\pdfoutput=1  

\documentclass[12pt,letterpaper,rotate]{article}

\usepackage{amsmath}
\usepackage{amsfonts}
\usepackage{amssymb}
\usepackage{graphicx}
\usepackage[numbers,sort&compress]{natbib}

\usepackage[hang,small,bf]{caption}

\usepackage{rotating}

\def\be{\begin{equation}}
\def\ee{\end{equation}}
\def\bea{\begin{eqnarray}}
\def\eea{\end{eqnarray}}
\def\pl{{\rm Pl}}
\def\d{\rm d}

 \def\vx{{\bf  x}} 
\def\vk{{\bf k}}

\def\Omsd{\Omega_{\sigma,{\rm dec}}}

\def\H{\cal H}
\def\pa{\partial}

\def\s{\sigma}

\def\x{{\bf x}}
\def\k{{\bf k}}

\def\H{\,{\cal  H}}

\newcommand{\Aa}{\ensuremath{\frac{a^{\prime}}{a}}}
\newcommand{\Ab}{\ensuremath{\Big(\frac{a^{\prime}}{a}\Big)^2}}
\newcommand{\Ac}{\ensuremath{\frac{a^{\prime\prime}}{a}}}

\newcommand{\La}{\ensuremath{\partial_i \,\partial^i}}
\newcommand{\deu}[1]{\ensuremath{{\delta #1}^{}}}

\def\cmm2{{\,\rm cm^{-2}}}
\def\cm2{{\,{\rm cm}^2}}
\def\cmm3{{\,{\rm cm}^{-3}}}
\def\gcmm3{{\,{\rm g\,cm^{-3}}}}

\def\calp{{\cal P}}

\def\yrs{\rm \,yrs}

\def\la{\mathrel{\mathpalette\fun <}}
\def\ga{\mathrel{\mathpalette\fun >}}

\def\fun#1#2{\lower3.6pt\vbox{\baselineskip0pt\lineskip.9pt
  \ialign{$\mathsurround=0pt#1\hfil##\hfil$\crcr#2\crcr\sim\crcr}}}

\def\f{\frac{1}{2}}

\def\p{ {\bf p} }

\def\d{{\rm d}}

\def\G{\Gamma}

\def\na{\boldsymbol{\nabla}}

\def\bseq{\begin{subequation}}
\def\eseq{\end{subequation}}
\def\bsea{\begin{subeqnarray}}
\def\esea{\end{subeqnarray}}

\def\c2bb{\cos^2 2 \beta}

\def\s2bb{\sin^2 2 \beta}

\begin{document}


\vspace{4.5cm}

\begin{center}


 {\LARGE Inflation and the Theory of Cosmological Perturbations} 
 \\[2.0cm]

{\large A. Riotto}
\\[0.5cm]

{\small \sl INFN, Sezione di Padova, via Marzolo 8, I-35131, Padova, Italy.}

\end{center}
\vspace{2cm} \hrule \vspace{0.3cm}
{\small  \noindent \textbf{Abstract} \\[0.3cm]
\noindent
These lectures provide a pedagogical introduction to inflation and the 
theory of cosmological perturbations generated during inflation
which  are  thought to be 
the origin of structure in the universe.


\vspace{1.3cm}
\hrule 

\vskip 4cm

\begin{center}
{\it 
Lectures given at the:\\

Summer School on\\

Astroparticle Physics and Cosmology\\

Trieste, 17 June - 5 July 2002}
\end{center}

\newpage
\vspace{2cm} \hrule \vspace{0.3cm}
{\small  \noindent \textbf{Notation} \\[0.3cm]
\noindent
A few words on the metric notation. We will be using the convention $(-,+,+,+)$, even though we might
switch time to time to the other option $(+,-,-,-)$. This might happen
for our convenience, but also  for pedagogical reasons. Students should not be shielded too much against
the phenomenon of changes of convention and notation in books and articles. 
\vspace{0.5cm} \hrule \vspace{0.3cm}
{\small  \noindent \textbf{Units} \\[0.3cm]
\noindent
We will adopt natural, or high energy physics, units. There is only one fundamental
dimension, energy, after setting
$\hbar=c=k_{\rm b}=1$, 

$$
[{\rm Energy}]=[{\rm Mass}]=[{\rm Temperature}]=[{\rm Length}]^{-1}=[{\rm Time}]^{-1}\, .
$$
The most common conversion factors and quantities we will make use of are
\vskip 0.2cm
\noindent
1 GeV$^{-1}= 1.97\times 10^{-14}$ cm=$6.59\times 10^{-25}$ sec,
\vskip 0.2cm
\noindent
1 Mpc= 3.08$\times 10^{24}$ cm=1.56$\times 10^{38}$ GeV$^{-1}$,
\vskip 0.2cm
\noindent
$M_{\rm Pl}= 1.22\times 10^{19}$ GeV,
\vskip 0.2cm
\noindent
$H_0$= 100\,h\, Km sec$^{-1}$ Mpc$^{-1}$=2.1$\,h\,\times 10^{-42}$ GeV,
\vskip 0.2cm
\noindent
$\rho_{\rm c}= 1.87h^2\cdot 10^{-29}{\rm g}\,{\rm cm}^{-3}=
1.05 h^2\cdot 10^4\,{\rm eV}\,{\rm cm}^{-3}=
8.1 h^2\times 10^{-47}$ GeV$^{4}$,
\vskip 0.2cm
\noindent
$T_0=2.75$ K=2.3$\times 10^{-13}$ GeV,
\vskip 0.2cm
\noindent
$T_{\rm eq}=5.5(\Omega_0 h^2)$ eV,
\vskip 0.2cm
\noindent
$T_{\rm ls}=0.26\, (T_0/2.75\, {\rm K})$ eV.
\vskip 0.2cm
\noindent
$s_0=2969\,{\rm cm}^{-3}$.
 \vspace{0.5cm}  \hrule
\def\thefootnote{\arabic{footnote}}
\setcounter{footnote}{0}

\vspace{0.50cm}

\vfill \noindent
{\footnotesize email: {\tt antonio.riotto@pd.infn.it}\hfill 

\newpage

\newpage
\tableofcontents

\newpage
\part{Introduction}
Inflation is a beautiful theoretical paradigm which explains why our universe looks the way we see it. It assumes that 
in the early universe  an infinitesimally small patch underwent this period of rapid exponential expansion becoming the universe (or much larger portion of)
we observe today. 
The observed universe is  therefore so homogeneous and isotropic because inhomogeneities were wiped out
 \cite{guth}.  In fact, the main subject of these lectures regards another incredible gift inflation give us \cite{hawking,starob82,guthpi}.
 The inflationary expansion of the universe quantum-mechanically excite quantum fields and stretches their perturbations
 from microphysical to cosmological scales. These vacuum fluctuations become classical on large scales and induce energy density fluctuations.
 When, after inflation, these fluctuations re-enter the observable universe, the generate temperature and matter anisotropies.
It is believed that this is  responsible for the observed
cosmic microwave background (CMB) anisotropy and for the large-scale
distribution of galaxies and dark matter. 
Inflation brings another bonus: it sources 
 gravitational waves as a vacuum fluctuation, which may contribute
to  CMB anisotropy polarization and are now the subject of an intense search program.
Therefore, a prediction of inflation is that 
all of the
structure we see in the universe is a result of quantum-mechanical
fluctuations during the inflationary epoch.

The goal of these lectures is to provide a pedagogical introduction
to the inflationary paradigm and to the theory of cosmological
perturbations generated during inflation and they  are organized as follows.

The reader might  find  also useful consulting
the following textbooks \cite{bookKT,bookLL,bookD,Mukhanov,Lidsey,Stewart,lr} and review \cite{lr} for more material.

\part{Basics of the Big-Bang model}
Our current understanding of the evolution of the universe is based upon 
the Friedmann-Robertson-Walker (FRW) cosmological model, or the hot 
big bang model as it is usually called. The model is so successful that 
it has become known as the standard cosmology. The FRW cosmology is so robust that it is possible to 
make sensible speculations about the universe at times as early as $10^{-43}$  
sec after the Big Bang.

Our universe is, at least on  large scales,  homogeneous and
isotropic. This observation gave rise to the so-called
Cosmological Principle.
    \begin{figure}[h!]
    \centering
        \includegraphics[width=.55\textwidth]{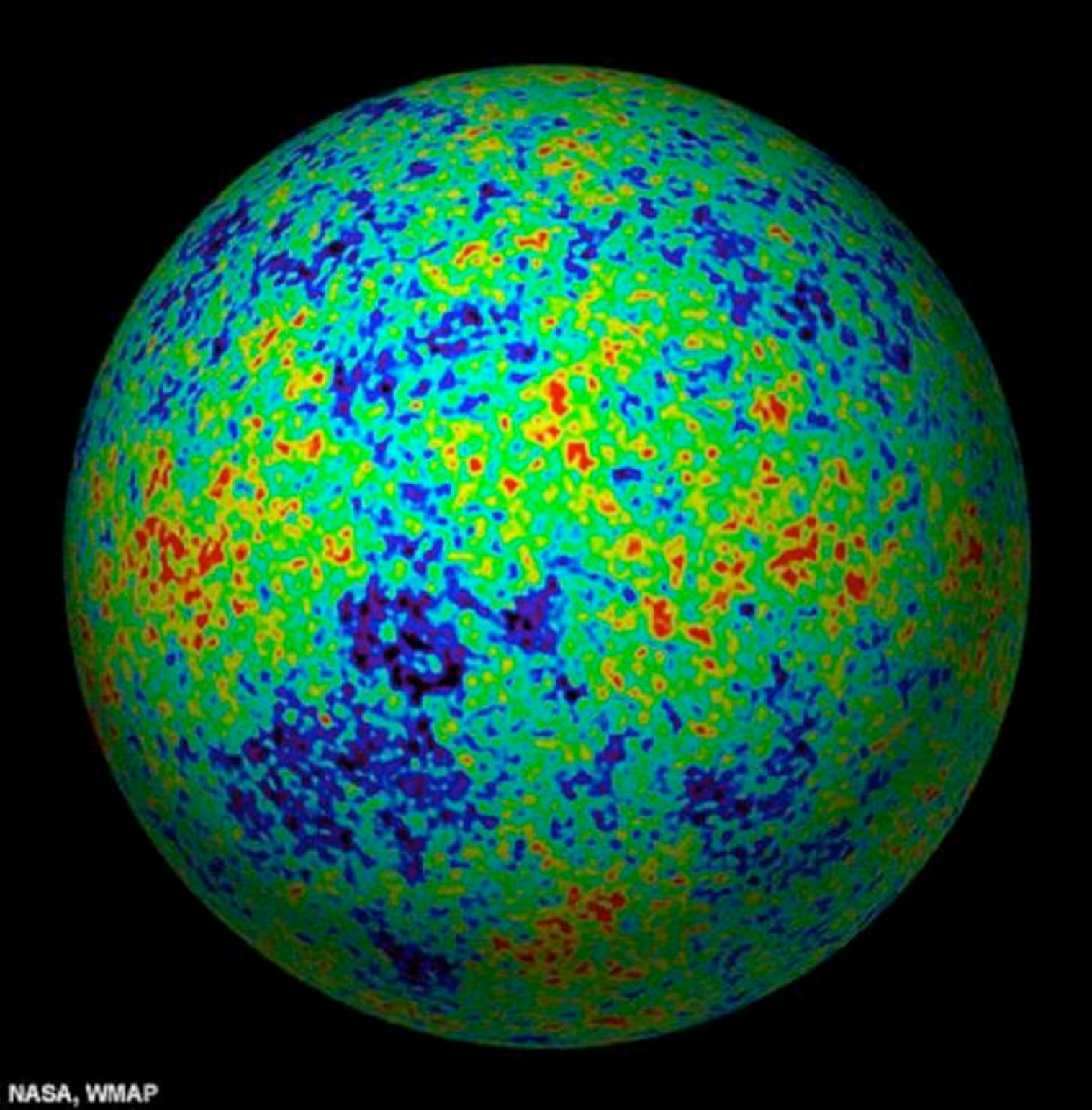}
\caption{The CMB radiation projected onto a sphere}
    \label{cmb3}
    \end{figure}
The best evidence for the isotropy of the observed universe is the 
uniformity of the temperature of the cosmic microwave background (CMB) radiation: intrinsic temperature anisotropies is smaller than about 
one part in $10^5$.  
This uniformity signals that at the epoch of 
last scattering for the CMB radiation  (about 200,000 yr after the bang) the  
universe was to a high degree of precision (order of $10^{-5}$ or so) isotropic and 
homogeneous. 
Homogeneity  and isotropy is of course true if the universe is observed at sufficiently large scales. Indeed, 
our  observed universe has a current size of of about  3000 Mpc. The inflationary theory, as we shall see, suggests that the universe continues to be homogeneous and isotropic
over distances larger than 3000 Mpc.

From now on we will work under the assumption that  our 
 observable universe is homogeneous and isotropic on large scales and that our spacetime is  maximally symmetric space satisfying the Cosmological Principle.  This is the so-called   Friedmann-Robertson-Walker  metric, which can be 
written in the form

\be
\label{FRW}
\d s^2=-\d t^2+a^2(t)\left[\frac{\d r^2}{1-k r^2}+r^2\d \theta^2+r^2\sin^2\theta\d\phi^2\right],
\ee
where $(t,r,\theta,\phi)$ are  comoving coordinates,  
$a(t)$ is the cosmic scale factor and  $k$ can be chosen to be $+1, -1$, or $0$ by rescaling the coordinates for spaces of constant 
positive, negative, or zero spatial curvature, respectively. The coordinate 
$r$ is dimensionless, {\it i.e.} $a(t)$ has dimensions of length and only relative ratios are physical, and $r$ ranges 
from 0 to 1 for $k = +1$. 
The time coordinate is just the proper (or clock) time measured 
by an observer at rest in the comoving frame, {\it i.e.}, $(r, \theta,\phi)$=constant.   Let us summarize here the main passages leading to the
FRW metric.

\begin{enumerate}
\item First of all, we write the FRW metric as
\be
\d s^2=-\d t^2+a^2(t)\widetilde{g}_{ij}\d x^i\d x^j.
\ee
From now on, all objects with a tilde will refer to three-dimensional quantities
calculated with the metric $\widetilde{g}_{ij}$.

\item One can then calculate the Christoffel symbols in terms of $a(t)$ and $\widetilde{\G}^i_{\,\,j k}$. 
Recalling from the GR course that the  Christoffel symbols are 

\be
\label{defGamma}
\fbox{$\displaystyle
 \G^{\mu}_{\,\,\nu\lambda}=\frac{1}{2}g^{\mu\rho}\left(\frac{\partial g_{\rho\nu}}{\partial x^\lambda}+\frac{\partial g_{\rho\lambda}}{\partial x^\nu}-\frac{\partial g_{\nu\lambda}}{\partial x^\rho}\right)$}\, ,
 \ee
we may compute the  non vanishing components 

\bea
\label{cfrw}
\G^{i}_{\,\, jk}&=& \widetilde{\G}^{i}_{\,\, jk},\nonumber\\
\G^{i}_{\,\, j0}&=&\frac{\dot a}{a}\,\delta^i_{\, j}, \nonumber\\
\G^{0}_{\,\, ij}&=&\frac{\dot{a}}{a} g_{ij}= \dot{a}a \widetilde{g}_{ij}.
\eea

\item 
The relevant components of the Riemann tensor 

\be
\label{riemann}
\fbox{$\displaystyle
R^\lambda_{\,\,\sigma\mu\nu}=\partial_\mu\G^{\lambda}_{\,\,\sigma\nu}-\partial_\nu\G^{\lambda}_{\,\,\sigma\mu}+\G^{\lambda}_{\,\,\mu\rho}\G^{\rho}_{\,\,\nu\sigma}-\G^{\lambda}_{\,\,\nu\rho}\G^{\rho}_{\,\,\mu\sigma}$}\, 
\ee
for the FRW metric are

\bea
R^i_{\,\,0j0}&=&-\frac{\ddot a}{a}\delta^i_{\, j},\nonumber\\
R^0_{\,\,i0j}&=&\ddot{a} a \widetilde{g}_{ij},\nonumber\\
R^k_{\,\,ikj}&=&\widetilde{R}_{ij}+2\dot{a}^2  \widetilde{g}_{ij}.
\eea

\item Now we can use $\widetilde{R}_{ij}=2 k \widetilde{g}_{ij}$ (as a consequence of the maximally symmetry of $\widetilde{g}_{ij}$) to calculate $R_{\mu\nu}$. The nonzero components are

\bea
\label{rfrw}
R_{00}&=&-3\frac{\ddot a}{a},\nonumber\\
R_{ij}&=&\left(a\ddot{a}+2\dot{a}^2+2 k\right)\widetilde{g}_{ij},\nonumber\\
&=&\left(\frac{\ddot{a}}{a}+2\frac{\dot{a}^2}{a^2}+2 \frac{k}{a^2}\right)g_{ij}.
\eea

\item  The Ricci scalar is

\be
\label{ricci}
R=\frac{6}{a^2}\left(a\ddot{a}+\dot{a}^2+k\right),
\ee
and

\item the Einstein tensor $G_{\mu\nu}=R_{\mu\nu}-\frac{1}{2}g_{\mu\nu} R$ has the components

\bea
\label{comp}
G_{00}&=&3 \left(\frac{\dot{a}^2}{a^2}+ \frac{k}{a^2}\right),\nonumber\\
G_{0i}&=&0,\nonumber\\
G_{ij}&=&-\left(2\frac{\ddot{a}}{a}+\frac{\dot{a}^2}{a^2}+ \frac{k}{a^2}\right)g_{ij}.
\eea
\end{enumerate}

\section{Standard cosmology}
 The dynamics of 
the expanding universe only appeared implicitly in the time dependence 
of the scale factor $a(t)$. To make this time dependence explicit, one must 
solve for the evolution of the scale factor using the Einstein equations

\be
\fbox{$\displaystyle
R_{\mu\nu}-\frac{1}{2}g_{\rm \mu\nu}R=8\pi G_{\rm N}T_{\mu\nu}-\Lambda g_{\rm \mu\nu},$}
\ee
where $T_{\mu\nu}$ is the stress-energy tensor for all the fields present  (matter, radiation, and so on) and we have also included the presence of a cosmological constant. With very minimal assumptions about the  right-hand side of the 
Einstein equations, it is possible to proceed without detailed knowledge 
of the properties of the fundamental fields that contribute to the stress 
tensor $T_{\mu\nu}$. 

\subsection{The stress-energy momentum tensor}
To be consistent with the symmetries of the metric, the total 
stress-energy tensor tensor must be diagonal, and by isotropy the spatial 
components must be equal. The simplest realization of such a 
stress-energy tensor is that of a perfect fluid characterized by a time-dependent 
energy density $\rho(t)$ and pressure $P(t)$

\be
\label{pf}
\fbox{$\displaystyle
T^{\mu}_{\,\nu}=(\rho+P)u^\mu u_\nu+P \delta^{\mu}_{\,\nu}={\rm diag}(-\rho,P,P,P),$}
\ee
where $u^\mu=(1,0,0,0)$ in a comoving coordinate system. This is precisely the energy-momentum
tensor of a perfect fluid. The four-vector $u_\mu$ is known as the velocity field of the fluid, and
the comoving coordinates are those with respect to which the fluid is at rest. 
In general, this matter content has to be supplemented by an equation of state. This is
usually assumed to be that of a barytropic fluid, {\it i.e.} one whose pressure depends only
on its density, $P=P(\rho)$. The most useful toy-models of cosmological fluids arise from
considering a linear relationship between $P$ and $\rho$ of the type

\be
\label{pf}
\fbox{$\displaystyle
P=w\rho$},
\ee
where $w$ is known as the equation of state parameter. Occasionally also more exotic
equations of state are considered.
For non-relativistic particles (NR) particles, there is no pressure, $p_{\rm NR} = 0$, {\it i.e.}  $w_{\rm NR}  = 0$, and such matter is
usually referred to as dust.
The trace of the energy-momentum tensor is

\be
T^\mu_{\, \mu}=-\rho+3 P.
\ee
For relativistic particles, radiation for example, the energy-momentum tensor is (like that of Maxwell theory)
traceless, and hence relativistic particles  have the equation of state

\be
P_{\rm r} =\frac{1}{3}\rho_{\rm r},
\ee
and thus $w_{\rm r} =1/3$. For physical (gravitating instead of anti-gravitating) matter one usually requires $\rho>0$
(positive energy) and either $P > 0$, corresponding to $w > 0$ or, at least,  $(\rho + 3P) > 0$,
corresponding to the weaker condition $w > -1/3$. A cosmological constant, on the other hand, corresponds, as we will see, to a matter
contribution with $w_\Lambda = -1$ and thus violates either $\rho > 0$ or $(\rho  + 3P) > 0$.

Let us now turn to the conservation laws associated with the energy-momentum tensor,

\be
\nabla_\mu T^{\mu\nu}=0.
\ee
The spatial components of this conservation law give

\be
\nabla_\mu T^{\mu i}=
\nabla_0 T^{0i}+ \nabla_j T^{ji}=0+ \nabla_j T^{ji}=P  \nabla_j g^{ij}=0,
\ee
where the last passage has been made because the metric is covariantly conserved. The only interesting conservation law is thus the zero-component

\be
\nabla_\mu T^{\mu 0}=\pa_\mu T^{\mu 0}+\G^\mu_{\,\,\mu\nu} T^{\nu 0}+\G^0_{\,\, \mu\nu} T^{\mu\nu}=0, 
\ee
which for a perfect fluid becomes

\be
\dot{\rho}+\G^\mu_{\,\,\mu 0}\rho+\G^0_{\,\, 00}\rho+\G^0_{\,\, ij} T^{ij}=0.
\ee
Using the  Christoffel symbols previously computed, see Eq. (\ref{cfrw}), we get

\be
\label{F3}
\fbox{$\displaystyle
\dot{\rho}+3H(\rho+P)=0$}.
\ee
For instance, when the pressure of the cosmic matter is negligible, like in the universe
today, and we can treat the galaxies (without disrespect) as dust, then one has

\be
\fbox{$\displaystyle
\rho_{\rm NR} \, a^3={\rm constant}\,\,\,\,{\rm (MATTER)}$}.
\ee
The energy (number) density scales like the inverse of the volume whose size is $\sim a^3$
On the other hand, if the universe is dominated by, say, radiation, then one has the
equation of state $P=\rho/3$, then

\be
\fbox{$\displaystyle
\rho_{\rm r} \, a^4={\rm constant} \,\,\,\,{\rm (RADIATION)}.$}
\ee
The energy density scales the like the inverse of the volume (whose size is $\sim a^3$) and the energy which scales like $1/a$ because of the red-shift: photon energies scale like the inverse
of their wavelengths which in turn scale like $1/a$. More generally, for matter with equation of state parameter $w$, one finds

\be
\fbox{$\displaystyle
\rho \, a^{3(1+w)}={\rm constant}.$}
\ee
In particular, for $w=-1$, $\rho$ is constant and corresponds, as we will see more explicitly
below, to a cosmological constant vacuum energy

\be
\fbox{$\displaystyle
\rho_{\Lambda}={\rm constant} \,\,\,\,{\rm (VACUUM}\,{\rm ENERGY)}.$}
\ee
The early universe was radiation dominated, the adolescent universe 
was matter dominated and the adult universe is dominated  by the cosmological constant. If the universe underwent inflation, there was again a very early 
period when the stress-energy was dominated by vacuum energy. As we 
shall see next, once we know the evolution of $\rho$ and $P$  in terms of the scale 
factor $a(t)$,  it is straightforward to solve for $a(t)$. 
Before going on, we want to emphasize the utility of describing the 
stress energy in the universe by the simple equation of state $P = w\rho$. This 
is the most general form for the stress energy in a FRW space-time and 
the observational evidence indicates that on large scales the FRW metric is 
quite a good approximation to the space-time within our Hubble volume. 
This simple, but often 
very accurate, approximation will allow us to explore many early universe 
phenomena with a single parameter.

\section{The Friedmann equations}
After these preliminaries, we are now prepared to tackle the Einstein equations. We
allow for the presence of a cosmological constant and thus consider the equations

\be
G_{\mu\nu}+\Lambda g_{\mu\nu}=8\pi G_{\rm N} T_{\mu\nu}.
\ee
It will be convenient to rewrite these equations in the form

\be
R_{\mu\nu}=8\pi G_{\rm N} \left(T_{\mu\nu}-\f g_{\mu\nu} T^\lambda_{\,\lambda}\right)+\Lambda g_{\mu\nu}.
\ee
Because of isotropy, there are only two independent equations, namely the $00$-component
and any one of the non-zero $ij$-components. Using 
Eqs. (\ref{rfrw}) we find

\bea
-3\frac{\ddot a}{a}&=&4\pi G_{\rm N}(\rho+3P)-\Lambda\, ,\nonumber\\
\frac{\ddot a}{a}+2\frac{\dot{a}^2}{a^2}+2\frac{k}{a^2}&=&4\pi G_{\rm N}(\rho-P)+\Lambda.
\eea
Using the first equation to eliminate $\ddot{a}$ from the second, one obtains the set of equations for the Hubble rate

\be
\label{F1}
\fbox{$\displaystyle
H^2+\frac{k}{a^2}=\frac{8\pi G_{\rm N}}{3}\rho+\frac{\Lambda}{3}$}
\ee
and for for the acceleration

\be
\label{F2}
\fbox{$\displaystyle
\frac{\ddot a}{a}=-\frac{4\pi G_{\rm N}}{3}(\rho+3P)+\frac{\Lambda}{3}$}.
\ee
Together, this set of equation is known as the Friedman equations. They are 
supplemented this by the conservation equation (\ref{F3}).
Note that because of the Bianchi identities, the Einstein equations and the conservation
equations should not be independent, and indeed they are not. It is easy to see that
(\ref{F1}) and (\ref{F3}) imply the second order equation ({\ref{F2}) so that, a pleasant simplification,
in practice one only has to deal with the two first order equations (\ref{F1}) and (\ref{F3}).
Sometimes, however, (\ref{F2}) is easier to solve than (\ref{F1}), because it is linear in $a(t)$, and
then (\ref{F1}) is just used to fix one constant of integration.

Notice that Eqs. (\ref{F1}) and (\ref{F2}) can be obtained, in the non-relativistic limit $P=0$ from Newtonian physics. Imagine that the distribution of matter is uniform and its matter density is $\rho$. Put a test particle with mass $m$ on a surface of a sphere of radius $a$ and let gravity act. The total energy is constant and therefore

\be
E_{\rm kin}+E_{\rm pot}=\frac{1}{2}m\dot{a}^2-G_{\rm N}\frac{m M}{a}=\kappa={\rm constant}.
\ee
Since the mass $M$ contained in a sphere of radius $a$ is $M=(4\pi\rho a^3/3)$, we obtain

\be
\frac{1}{2}m\dot{a}^2-\frac{4\pi G_{\rm N}}{3}m\rho a^2=\kappa={\rm constant}.
\ee
By dividing everything by $(m a^2/2)$ we obtain Eq. (\ref{F1}) with of course no cosmological constant and after setting $k=2\kappa/m$. Eq. (\ref{F2}) can be analogously obtained from Newton's law
relating the gravitational force and the acceleration (but still with $P=0$).

The expansion rate of the universe is determined by the Hubble rate $H$ which is not a constant
and generically scales like $t^{-1}$. The Friedmann equation (\ref{F1}) can be recast
as

\be
\label{omega}
\fbox{$\displaystyle
\Omega-1=\frac{\rho}{3 H^2/8\pi G_{\rm N}}=\frac{k}{a^2 H^2},$}
\ee
where we have defined the parameter $\Omega$ as the ratio between the 
energy density $\rho$ and the critical energy density $\rho_{\rm c}$

\be
\fbox{$\displaystyle
\Omega=\frac{\rho}{\rho_{\rm c}},\,\,\,\,\rho_{\rm c}=\frac{3 H^2}{8\pi G_{\rm N}}.$}
\ee
Since $a^2 H^2>0$, there is a correspondence between the sign of $k$ and the sign of $(\Omega-1)$

\be
\begin{array}{cccc}
k=+1&\Rightarrow &\Omega>1 & {\rm CLOSED},\\
k=0&\Rightarrow &\Omega=1 & {\rm FLAT},\\
k=-1&\Rightarrow &\Omega<1 & {\rm  OPEN}.\\
\end{array}
\ee
Eq. (\ref{omega}) is valid at all times, note also that both $\Omega$ and $\rho_{\rm c}$ are not constant in time. At early times once has a radiation-dominated (RD) phase radiation  and $H^2\sim a^{-4}$ with  $(\Omega-1)\sim a^2$; during the matter-dominated phase (MD) one finds  $H^2\sim a^{-3}$ with  $(\Omega-1)\sim a$. These relations will be crucial when we will study the inflationary universe. 
From the FRW metric it is also clear that 
 the effect of the curvature becomes important only at a comoving  radius
$r\sim |k|^{-1/2}$. So we define the physical radius of curvature of the universe $R_{\rm curv}=a(t)|k|^{-1/2}=(6/|^3{\cal R}|)^{1/2}$, related to the Hubble radius $H^{-1}$ by

\be
R_{\rm curv}=\frac{H^{-1}}{|\Omega-1|^{1/2}}.
\ee
When $|\Omega-1|\ll 1$, such a curvature radius turns out to be much larger than the Hubble radius and we can safely neglect the effect of curvature in the universe. Note also that for closed universes, $k=+1$,
$R_{\rm curv}$ is just the physical radius of the three-sphere.

\section{Equilibrium thermodynamics of the early universe}
Because the early universe was to a good approximation in thermal equilibrium at very early epochs, we can assume that it was characterized by a  RD phase we will quickly review some basic thermodynamics of.

The number density $n$, energy density $\rho$ and pressure $P$ of a dilute, weakly interacting gas of particles with $g$ internal degrees of freedom can be written in terms of its
phase space distribution function $f(\p)$

\bea
n&=&\frac{g}{(2\pi)^3}\int\d^3p\, f(\p),\nonumber\\
\rho&=&\frac{g}{(2\pi)^3}\int\d^3p\, E(\p)\,f(\p),\nonumber\\
P&=&\frac{g}{(2\pi)^3}\int\d^3p\,\frac{|\p|^2}{3E(\p)} f(\p),
\eea
where $E^2=|\p|^2+m^2$.The phase space  $f$ is given by the familiar Fermi-Dirac or Bose-Einstein distributions

\be
\fbox{$\displaystyle
f(\p)=\left[{\rm exp}(E/T)\pm 1\right]^{-1},$}
\ee
where we have neglected a possible  chemical potential,  $+1$ refers to Fermi-Dirac species and $-1$ to Bose-Einstein species.
From the equilibrium distributions, ¥ the number density $n$, energy density $\rho$ and pressure $P$  of a species of mass $m$, and temperature $T$ are

\bea
n&=&\frac{g}{2\pi^2}\int_m^\infty\,\d E\, E\,\frac{(E^2-m^2)^{1/2}}{{\rm exp}(E/T)\pm 1},\nonumber\\
\rho&=&\frac{g}{2\pi^2}\int_m^\infty\,\d E\, E^2\, \frac{(E^2-m^2)^{1/2}}{{\rm exp}(E/T)\pm 1},\nonumber\\
P&=&\frac{g}{6\pi^2}\int_m^\infty\,\d E\frac{(E^2-m^2)^{3/2}}{{\rm exp}(E/T)\pm 1}.
\eea
In the relativistic limit $T\gg m$ we obtain

\bea
\rho&=&\left\{
\begin{array}{cc}
(\pi^2/30)gT^4 & ({\rm BOSE})\\
(7/8)(\pi^2/30)gT^4 & ({\rm FERMI})
\end{array}\right.\nonumber\\
n&=&\left\{
\begin{array}{cc}
(\zeta(3)/\pi^2)gT^3 & ({\rm BOSE})\\
(3/4)(\zeta(3)/\pi^2)gT^3 & ({\rm FERMI})
\end{array}\right.\nonumber\\
P&=&\rho/3.
\eea
Here $\zeta(3)\simeq 1.2$ is the Riemann zeta function of three. The total energy density and pressure of all species in equilibrium can be expressed in terms of the photon temperature $T$

\bea
\rho_{\rm r}&=&T^4\sum_{{\rm all}\,{\rm species}}\left(\frac{T_i}{T}\right)^4
\frac{g_i}{2\pi^2}\int_{x_i}^\infty\,\d u\, \frac{(u^2-x_i^2)^{1/2} u^2}{\exp(u)\pm 1},\nonumber\\
P_{\rm r}&=&T^4\sum_{{\rm all}\,{\rm species}}\left(\frac{T_i}{T}\right)^4
\frac{g_i}{6\pi^2}\int_{x_i}^\infty\,\d u\, \frac{(u^2-x_i^2)^{3/2}}{\exp(u)\pm 1},
\eea
where $x_i=m_i/T$ and we have taken into account the possibility that 
the species have a different temperature than the photons. 

Since the energy density and pressure of non-relativistic species is exponentially smaller than that of relativistic species, it is a very good approximation to include only the relativistic species
in the sums and we obtain

\be
\fbox{$\displaystyle
\rho_{\rm r}=3 P_{\rm r}=\frac{\pi^2}{30}g_*(T)\,T^4,$}
\ee
where

\be
\fbox{$\displaystyle
g_*(T)=\sum_{\rm bosons} g_i \left(\frac{T_i}{T}\right)^4+\frac{7}{8}\sum_{\rm fermions} g_i \left(\frac{T_i}{T}\right)^4,$}
\ee
counts the effective total number of relativistic degrees of freedom in the plasma. 
For instance, for $T\ll$ MeV, the only relativistic species are the three neutrinos
with $T_\nu=(4/11)^{1/3} T_\gamma$ and the photons. This gives 

\be
g_*(\ll\,{\rm MeV})=\frac{7}{8}\cdot 3\,(={\rm families})\cdot 2\,(={\rm Weyl}\,\,{\rm d.o.f.})\cdot\left(\frac{4}{11}\right)^{4/3}+2\,(={\rm photon}\,\,{\rm d.o.f.})\simeq 3.36.
\ee
For 100 MeV $\ga T\ga$ 1 MeV, the electron and positron are additional relativistic degrees of freedom  and $T_\nu=T_\gamma$. We thus find

\begin{eqnarray}
g_*(100\,{\rm MeV}\ga T \ga 1\, {\rm MeV})&=&\frac{7}{8}\cdot 3\,(={\rm families})\cdot 2\,(={\rm Weyl}\,\,{\rm d.o.f.})+2\,(={\rm photon}\,\,{\rm d.o.f.})\nonumber\\
&+&
\frac{7}{8}\cdot 4\,(={\rm Dirac}\,\,{\rm d.o.f.})
\simeq 10.75.
\end{eqnarray}
 For $T\ga $ 300 GeV, all the species of the Standard Model (SM) are in equilibrium: 8 gluons, $W^\pm$, $Z$, three generations of quarks and leptons and one complex Higgs field and $g_*\simeq 106.75$. 

During early RD phase when $\rho\simeq \rho_{\rm r}$ and supposing that $g_*\simeq $ constant, we have that the Hubble rate is

\be
H^2=\frac{8\pi G_{\rm N}}{3}\rho_{\rm r}=\frac{8\pi}{3}\frac{\rho_{\rm r}}{M_\pl^2}=\frac{8\pi}{3M_\pl^2}\frac{\pi^2}{30}g_*(T)\,T^4,
\ee
where $ G_{\rm N}=1/M_\pl^2$ and $M_\pl\simeq 1.2\cdot 10^{19}$ GeV is the  Planck mass. We  get therefore

\be
\fbox{$\displaystyle
H\simeq 1.66\,g_*^{1/2}\,\frac{T^2}{M_\pl}$}
\ee
and the corresponding time is, being $H\simeq 1/2t^{-1}$, 

\be
\fbox{$\displaystyle
t\simeq 0.3\, g_*^{-1/2}\,\frac{M_\pl}{T^2}\simeq\left(\frac{T}{{\rm MeV}}\right)^{-2}\,{\rm sec}.$}
\ee
In thermal equilibrium the entropy per comoving volume $V$, $S$, is conserved. It is useful to define the entropy density $s$ as

\be
s=\frac{S}{V}=\frac{\rho+P}{T}.
\ee
It is dominated by the relativistic degrees of freedom and to a very good approximation

\be
\fbox{$\displaystyle
s=\frac{2\pi^2}{45}g_{*S}T^3,$}
\ee
where

\be
\fbox{$\displaystyle
g_{*S}(T)=\sum_{\rm bosons} g_i \left(\frac{T_i}{T}\right)^3+\frac{7}{8}\sum_{\rm fermions} g_i \left(\frac{T_i}{T}\right)^3,$}
\ee
For most of the history of the universe, all particles had the same temperature and one replace therefore $g_{*S}$ with $g_*$. Note also that $s$ is proportional to the number density of relativistic degrees of freedom and in particular it can be related to the photon number density
$n_\gamma$
\be
\fbox{$\displaystyle
s\simeq 1.8 g_{*S} n_\gamma.$}
\ee
Today $s_0\simeq 7.04 n_{\gamma,0}$. The conservation of $S$ implies that $s\sim a^{-3}$ and therefore 

\be
\fbox{$\displaystyle
g_{*S}T^3 a^3={\rm constant}$}
\ee
during the evolution of the universe we are concerned with.  
The fact that $S=g_{*S}T^3 a^3=$ constant, implies that the temperature of the universe evolves as 

\be
\fbox{$\displaystyle
T\sim g_{*S}^{-1/3} a^{-1}.$}
\ee
When $g_{*S}$ is constant one gets the familiar result $T\sim a^{-1}$. The factor $
g_{*S}^{-1/3}$ enters because whenever a particle species becomes non-relativistic and disappears from the plasma, its entropy is transferred to the other relativistic particles in the thermal plasma
causing $T$ to decrease slightly less slowly (sometimes it is said, but in a wrong way, that the universe slightly reheats up).

Lastly, an important quantity is
the entropy within a horizon volume:  $S_{\rm HOR}
\sim H^{-3}T^3$; during the RD epoch
$H\sim T^2/M_{\rm Pl}$, so that
\begin{equation}
S_{\rm HOR} \sim \left( {M_{\rm Pl}\over T} \right)^3.
\end{equation}
From this we will shortly conclude that at early times the comoving
volume that encompasses all that we can see today (that is a region as larger as our present horizon)  was comprised
of a very large number of causally disconnected regions.

\section{The particle horizon and the Hubble radius}
A fundamental question in cosmology that one might ask is: what 
fraction of the universe is in causal contact? More precisely, for a comoving 
observer with coordinates $(r_0, \theta_0, \phi_0)$, for what values of ($r,\theta,\phi$) would 
a light signal emitted at $t = 0$ reach the observer at, or before, time $t$? 
This can be calculated directly in terms of the FRW metric. A light signal 
satisfies the geodesic equation $\d s^2=0$. Because of the homogeneity of 
space, without loss of generality we may choose $r_0 = 0$. Geodesics passing 
through $r = 0$ are lines of constant $\theta$ and $\phi$, just as great circles passing 
from the poles of a two sphere are lines of constant $\theta$ ({\it i.e.}, constant  
longitude), so $\d\theta=\d\phi=0$. Of course, the isotropy of space makes the 
choice of direction $(\theta_0,\phi_0)$ irrelevant. Thus, a light signal emitted from 
coordinate position $(r_{\rm H},\theta_0,\phi_0)$ at time $t = 0$ will reach $r_0 = 0$ in a time 
$t$ determined by 

\be
\int_0^t\frac{\d t'}{a(t')}=\int_0^{r_{\rm H}}\frac{\d r'}{\sqrt{1- kr^{'2}}}.
\ee
The proper distance to the horizon measured at time $t$ is

\be
\fbox{$\displaystyle
R_{\rm H}(t)=a(t)\int_0^t\frac{\d t'}{a(t')}=a(t)\int_0^a\frac{\d a'}{a'}\frac{1}{a'H(a')}=
a(t)\int_0^{r_{\rm H}}\frac{\d r'}{\sqrt{1- kr^{'2}}}\,\,\,({\rm PARTICLE}\,\,{\rm HORIZON}).$}
\ee
If $R_{\rm H}(t)$ is finite, it  sets 
 the boundary between the visible universe and the part of the 
universe from which light signals have not reached us. The behavior of 
$a(t)$ near the singularity will determine whether or not the particle horizon is finite. We will see that in the standard cosmology $R_{\rm H}(t)\sim t$, that is the particle horizon is finite. 
The particle horizon should not be confused with the notion of the Hubble radius 

\be
\fbox{$\displaystyle
\frac{1}{H}=\frac{a}{\dot a}\,\,\,({\rm HUBBLE}\,\,{\rm RADIUS}).$}
\ee
The Hubble radius has the following meaning: it is the distance  travelled by particles in the course of one expansion time,  roughly the time which takes the  scale factor to double (think of the distance as $\d t\sim (\d a/a)H^{-1}$) \cite{bookD}. So the Hubble radius is different  way of measuring whether particles
are causally connected with each other. If they are separated by distances larger than the Hubble radius, they cannot currently communicate. 
Let us emphasize that the  particle horizon and the  Hubble radius are different quantities:  
particles   separated by distances greater than $R_H(t)$ have never communicated
with one another; on the contrary, if they are separated by an amount larger than  Hubble radius $H^{-1}$, this means that
they cannot communicate  at the given  time $t$.

We shall see that the standard cosmology the distance to the
horizon is finite, and up to numerical factors,
equal to the Hubble radius, $H^{-1}$, but during inflation, for instance, they are drastically different.
One can also define
a comoving particle horizon distance

\be
\label{horcom}
\fbox{$\displaystyle
\tau_{\rm H}=\int_0^t\,\frac{\d t'}{a(t')}=\int_0^a\frac{\d a'}{H(a') a^{'2}}=\int_0^a\d\ln a'\left(\frac{1}{Ha'}\right)
\,\,\,({\rm COMOVING}\,\,{\rm PARTICLE }\,\,{\rm HORIZON}).$}
\ee
Here, we have expressed the comoving horizon as the logarithmic  integral of the comoving Hubble radius $(aH)^{-1}$

\be
\label{radcom}
\fbox{$\displaystyle
\frac{1}{aH}\,\,\,({\rm COMOVING}\,\,{\rm HUBBLE}\,\,{\rm RADIUS}),$}
\ee
which will play a crucial role in inflation. 
 Let us
reiterate that  the  comoving horizon $\tau_{\rm H}$  and the comoving Hubble radius 
$(aH)^{-1}$ are different quantities. Particles separated by comoving distances greater than $\tau_{\rm H}$, they never 
talked to  one another; if separated by distances greater than 
$(aH)^{-1}$, they are not talking at  some time $\tau$. It is therefore possible that $\tau_{\rm H}$ could 
be much larger the comoving Hubble radius at the present epoch, so that there could not be any  communication  today but 
there was at earlier epochs. As we shall see, this might happen if the comoving Hubble radius 
early on was much larger than it is now so that  $\tau_{\rm H}$  got most of its contribution from 
early times. We will see that this could happen, but it does not happen during matter-dominated  or radiation-dominated 
 epochs. In those cases, the comoving Hubble radius increases with time, 
so typically we expect the largest contribution to $\tau_{\rm H}$ to come from the most recent 
times. 

Recall that in a universe dominated by a fluid with equation of state $P=w/\rho$ we have $n=2/3(1+w)$.
The  comoving Hubble radius goes like

\be
{\rm COMOVING}\,\,\,{\rm HUBBLE}\,\,\,{\rm RADIUS}=\frac{1}{aH}\sim \frac{t}{t^{n}}=t^{1-n}
\ee
In particular, for a MD universe $w=0$ and $n=2/3$, while for a RD universe $w=1/3$ and $n=1/2$. In both cases the comoving Hubble radius increases with time. 
We see that  in the standard cosmology the particle
horizon is finite, and up to numerical factors,
equal to  the Hubble radius, $H^{-1}$.
For this reason, one can use  the words  horizon and Hubble radius
interchangeably for standard cosmology. 
As we shall see, in inflationary models
the horizon and Hubble radius are drastically different
as the horizon distance grows exponentially relative
to the Hubble radius; in fact, at the end of inflation
they differ by $e^N$, where $N$ is the number of
e-folds of inflation. The horizon sets the length scale for which two points separated by
a distance larger than $R_H(t)$ they could never communicate, while the Hubble radius sets the scale
at which these two points could not communicate at the time $t$. 

Note also that a physical length scale $\lambda$ is within the Hubble radius
if $\lambda<H^{-1}$. Since we can identify the length scale
$\lambda$ with its wavenumber $k$, $\lambda=2\pi a/k$, 
we will have the following
rule
\begin{center}
\begin{tabular}{|p{13 cm}|}
\hline
\bea
\frac{k}{aH}&\ll& 1 \Longrightarrow {\rm SCALE}~~\lambda~~
{\rm OUTSIDE}~~{\rm THE}~~{\rm HUBBLE }~~{\rm RADIUS}\nonumber\\
\frac{k}{aH}&\gg& 1 \Longrightarrow {\rm SCALE}~~\lambda~~
{\rm WITHIN}~~{\rm THE}~~{\rm HUBBLE}~~{\rm RADIUS}\nonumber
\eea
\\
\hline
\end{tabular}
\end{center}
Notice that in standard cosmology

\be
\frac{\lambda}{{\rm PARTICLE}\,\,\,{\rm HORIZON}}=
\frac{\lambda}{R_H}=\lambda\, H\sim \frac{a H}{k}.
\ee
This shows once more that  Hubble radius and particle horizon can be used interchangeably in standard cosmology.

\section{Some conformalities}
\noindent
Before launching ourselves into the description of the shortcomings of the Big-Bang  model and  inflation,
we would like to go back to the concept of conformal time which will
be useful in the next sections.
The conformal time $\tau$ is defined through the following relation
\begin{equation}
\d\tau=\frac{\d t}{a}.
\label{jj}
\end{equation}
The metric $\d s^2=-\d t^2+a^2(t)\d {\bf x}^2$ then becomes
\begin{equation}
\d s^2 = a^2(\tau)\left[-\d\tau^2 +\d {\bf x}^2 \right].
\label{metricconf}
\end{equation}
The reason why $\tau$ is called conformal is manifest from Eq. 
(\ref{metricconf}): the corresponding FRW line element is conformal
to the Minkowski line element describing a static four dimensional
hypersurface.

Any function $f(t)$ satisfies the rule
\bea
\dot{f}(t)&=&\frac{f^\prime(\tau)}{a(\tau)},\\
\ddot{f}(t)&=&\frac{f^{\prime\prime}(\tau)}{a^2(\tau)}-
\H\frac{f^\prime(\tau)}{a^2(\tau)},
\label{rule1}
\eea
where a prime now indicates differentiation wrt to the conformal time
$\tau$ and
\be
\fbox{$\displaystyle
\H=\frac{a^\prime}{a}$}.
\ee
In particular we can set the following rules
\begin{center}
\begin{tabular}{|p{13 cm}|}
\hline
\bea
H&=&\frac{\dot a}{a}=\frac{a^\prime}{a^2}=\frac{\H}{a},\nonumber\\
\ddot a&=&\frac{a^{\prime\prime}}{a^2}-\frac{\H^2}{a},\nonumber\\
\dot H&=&\frac{\H^\prime}{a^2}-\frac{\H^2}{a^2},\nonumber\\
H^2&=&{8\pi G \rho \over 3} - {k\over a^2}\Longrightarrow
\H^2={8\pi G \rho a^2 \over 3}  -k\,\nonumber\\
\dot H&=&-4\pi G\left(\rho+P\right)\Longrightarrow 
\H^\prime=-\frac{4\pi G}{3}\left(\rho+3P\right)a^2,\nonumber\\
\dot{\rho} &+&3H(\rho +P) =0\Longrightarrow \rho^\prime +3\H(\rho +P)=0
\nonumber
\label{rules}
\eea
\\
\hline
\end{tabular}
\end{center}
Finally, 
if the scale factor $a(t)$ scales like $a\sim t^n$, solving the relation
(\ref{jj}) we find
\be
a\sim t^n\Longrightarrow a(\tau)\sim \tau^{\frac{n}{1-n}}.
\label{rule}
\ee
Therefore, for a RD era $a(t)\sim t^{1/2}$ one has $a(\tau)\sim\tau$ and for
a MD era $a(t)\sim t^{2/3}$, that is $a(\tau)\sim\tau^2$.

\part{The shortcomings of the standard Big-Bang Theory}
\noindent
This section is dedicated to a description of the drawbacks present in the cosmological Big-Bang theory. They are the origin for understanding why inflation is needed.
\subsection{The Flatness Problem}
\noindent
Let us make a tremendous extrapolation and assume  that Einstein
equations are valid until the Plank era, when the temperature of
the universe is $T_{\rm Pl}\sim  10^{19}$ GeV.
From the equation for the curvature

\be
\Omega-1=\frac{k}{H^2 a^2},
\ee
 we read that 
if the universe is perfectly flat, then $(\Omega=1)$ at all times.
On the other hand, if there is even a small curvature term, the time
dependence of $\left(\Omega-1\right)$ is quite different.

During a RD period, we have that $H^2\propto
\rho_{\rm r}\propto a^{-4}$ and
\begin{equation}
\Omega -1 \propto \frac{1}{a^2 a^{-4}}\propto a^2.
\end{equation}
During MD, $\rho_{\rm NR}\propto a^{-3}$ and   
\begin{equation}
\Omega -1 \propto \frac{1}{a^2 a^{-3}}\propto a.
\end{equation}
In both cases $(\Omega-1)$ decreases going backwards with time. 
Since we know that today $(\Omega_0-1)$ is of order unity at present, 
we can deduce its value at $t_{\pl}$ (the time at which the temperature
of the universe is $T_{\pl}\sim 10^{19}$ GeV)
\begin{equation}
\label{yxc}
\frac{\mid \Omega -1 \mid_{T=T_{\pl}}}{\mid \Omega -1
\mid_{T=T_{0}}} \approx \left(\frac{a^2_{\pl}}{a_{0}^2}\right) \approx
\left(\frac{T_0^2}{T^2_{\pl}}\right)  \approx \mathcal{O}(10^{-64}).
\end{equation}
where  $0$ stands for the present epoch, and $T_0\sim
10^{-13}$ GeV is the present-day
temperature of the CMB radiation. If we are not so brave and 
go back simply to the epoch of
nucleosynthesis when light elements abundances were formed, at
$T_N\sim$ 1 MeV,
we get
\begin{figure}[h!]
    \centering
        \includegraphics[width=.55\textwidth]{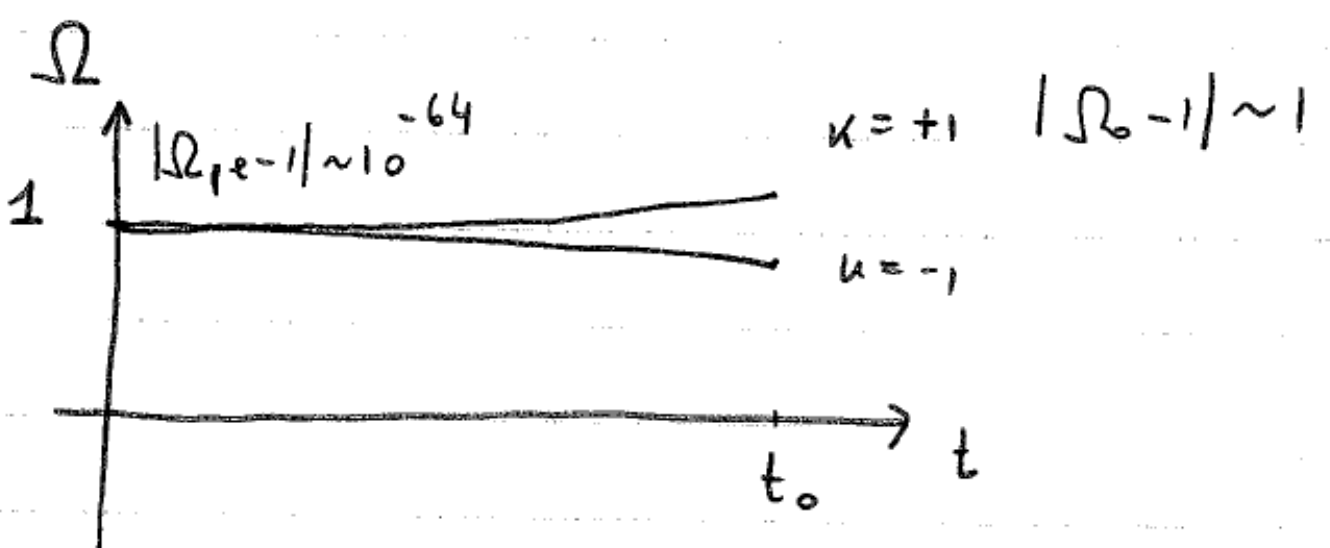}
\caption{Illustration of the flatness problem in standard cosmology.}
    \label{hor1}
\end{figure}
\begin{equation}
\frac{\mid \Omega -1 \mid_{T=T_N}}{\mid \Omega -1
\mid_{T=T_{0}}} \approx \left(\frac{a^2_{N}}{a_{0}^2}\right) \approx
\left(\frac{T_0^2}{T^2_N}\right)  \approx \mathcal{O}(10^{-16}).
\end{equation}
In order to get the correct value of $(\Omega_0-1)\sim 1$ at present,
the value of $(\Omega-1)$ at early times have to be fine-tuned to values
amazingly close to zero, but without being exactly zero. This is the reason why
the flatness problem is also dubbed the `fine-tuning problem'.\\

\subsection{The Entropy Problem}
\noindent
Let us now see how the hypothesis of adiabatic expansion of the
universe is connected with the flatness problem. 
From the Friedman equations  we know that  during a 
RD period
\begin{equation}
H^2\simeq \rho_{\rm r}\simeq \frac{T^4}{M_{\rm Pl}^2},
\end{equation}
from which we deduce
\begin{equation}
\Omega-1=\frac{k M_{\rm Pl}^2}{a^2 T^4}=\frac{k M_{\rm Pl}^2}{S^{\frac{2}{3}} T^2}.
\end{equation}
Under the hypothesis of adiabaticity, $S$ is constant over the
evolution of the universe and therefore
\be
\left|\Omega -1\right|_{t=t_{\pl}}
=\frac{M_{\rm Pl}^2}{T_\pl^2}\frac{1}{S_{\rm U}^{2/3}}=\frac{1}{S_{\rm U}^{2/3}}\approx
10^{-60},
\label{con}
\ee
where we have used the fact that the present horizon contains a total entropy

\be
S_{\rm U} = {4\pi\over 3}H_0^{-3}s={4\pi\over 3}H_0^{-3}\frac{2\pi^2 g_*(T_0) T_0^3}{45} \simeq 10^{90}.
\ee
We have discovered that  $(\Omega-1)$ is so close to
zero at early epochs because the total entropy of our universe
is so incredibly large.
The flatness problem is therefore a problem of understanding why the 
(classical) initial conditions corresponded to a universe that was so close 
to spatial flatness. One would have indeed expected the most natural number for the total entropy of the universe to be
of the order of unity at the Planckian temperature, when the horizon itself was of the order of the Planckian length.
In a sense, the problem is one of fine--tuning and 
although such a balance is possible in principle, one nevertheless feels 
that it is unlikely. On the other hand, the flatness problem arises because 
the entropy in a comoving volume is conserved. It is possible, therefore,  
that the problem could be resolved if the cosmic expansion was 
non--adiabatic for some finite time interval 
during the early history of the universe. \\

\subsection{The horizon problem}
\noindent
According to the standard cosmology,
photons decoupled from the rest of the components (electrons and baryons)
 at a temperature of the order of 0.3 eV. This  corresponds to the
so-called  surface of `last-scattering' at a red shift of about $1100$ and
an age of about $180,000\,(\Omega_0 h^2)^{-1/2}\yrs$.

From the epoch of last-scattering onwards, photons free-stream
and reach us basically untouched. Detecting primordial photons
is therefore equivalent to take a picture of the universe when the
latter was about 300,000 $\yrs$ old.
The spectrum of the cosmic background radiation is consistent
that of a black body at temperature 2.73 K over more than three
decades in wavelength; see Fig. \ref{fig:spectrum}.
The length corresponding to our present Hubble radius (which is
approximately the radius of our observable universe) at the time
of last-scattering was
$$
\lambda_{\rm H}(t_{\rm ls})=R_{\rm H}(t_0) \left(\frac{a_{\rm ls}}{a_0}
\right)=R_{\rm H}(t_0) \left(\frac{T_{0}}{T_{\rm ls}}\right).
$$
On the other hand, during the MD period, 
the Hubble length has decreased with a different law
$$
H^2\propto \rho_{\rm NR} \propto a^{-3} \propto T^{3}.
$$
At last-scattering
$$
H_{\rm ls}^{-1}=R_{\rm H}(t_0)\left( \frac{T_{\rm ls}}{T_0} \right)^{-3/2}\ll R_{\rm H}(t_0).
$$
The length corresponding to our present Hubble radius was much
larger that the horizon at that time. This can be shown comparing
the volumes corresponding to these two scales

\begin{equation}
\frac{\lambda^3_{H}(T_{\rm ls})}{H_{\rm ls}^{-3}}=
\left(\frac{T_0}{T_{\rm ls}}\right)^{-\frac{3}{2}}\approx 10^6.
\label{ppz}
\end{equation}
\begin{figure}[h!]
    \centering
        \includegraphics[width=.85\textwidth]{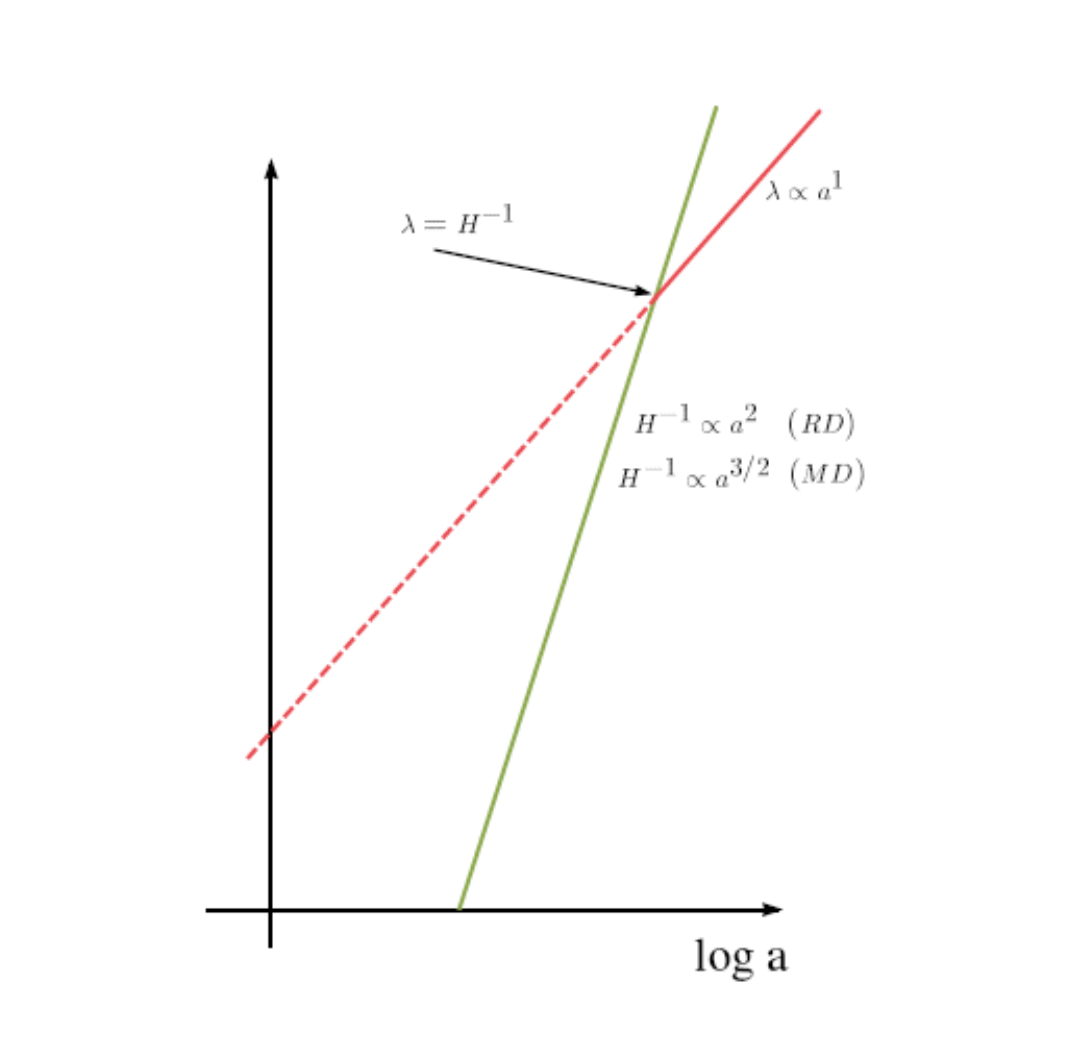}
\caption{The horizon scale (green line) and a physical scale $\lambda$
(red line) as function of the scale factor $a$ From Ref. \cite{kolbreview}.}
    \label{fig:normal}
\end{figure}
There were $\sim 10^6$ casually
disconnected regions within the volume that now corresponds to our
horizon! 
It is difficult to
come up with a process other
than an early hot and dense phase in the history
of the universe that would lead to a precise
black body  \cite{dnsnature} for a bath of photons which
were causally disconnected the last time  they interacted with the
surrounding plasma.

The horizon problem is well represented by Fig. \ref{fig:normal}
where the green line indicates the horizon scale and the red line any
generic physical length scale $\lambda$. Suppose, indeed that $\lambda$
indicates the distance between two photons we detect today. From
Eq. (\ref{ppz}) we discover that at the time of emission (last-scattering)
the two photons could not talk to each other, the red line is above the
green line.

There is another aspect of the horizon problem which is related to the
problem of initial conditions for the cosmological perturbations.

  The temperature difference measured
between two points separated by a large angle ($\ga 1^\circ$)
is due to the so-called Sachs-Wolfe effect and is caused by the fact that these two points had a different value of the gravitational potential
associated to it at 
the last-scattering surface. 
\begin{figure}[h!]
    \centering
        \includegraphics[width=.75\textwidth]{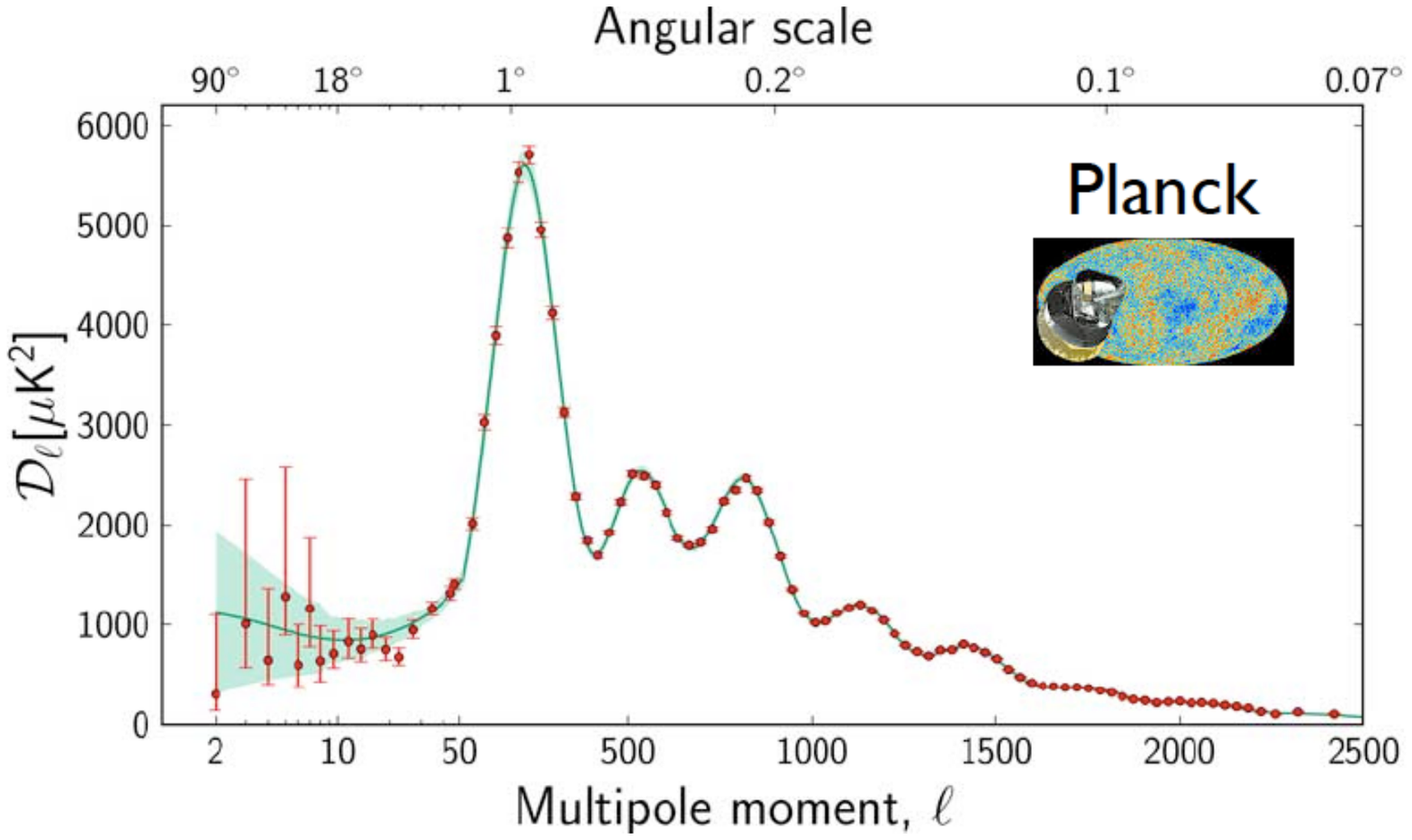}
\caption{The CMBR anisotropy as function of $\ell$ from the recent Planck satellite data.}
    \label{cobe}
\end{figure}
The
temperature anisotropy is commonly expanded  in spherical harmonics
\begin{equation}
\frac{\Delta T}{T}(x_0,\tau_0,{\bf n})=\sum_{\ell m}
a_{\ell m}(x_0)Y_{\ell m}({\bf n}),
\end{equation}
where $x_0$ and $\tau_0$ are our position and the present time, respectively,
 ${\bf n}$ is the
direction of observation, $\ell'$s are the
different multipoles  and\footnote{An alternative definition is $C_\ell=
\langle \left|a_{\ell m}\right|^2\rangle=\frac{1}{2\ell +1}
\sum_{m=-\ell}^{\ell}
\left|a_{\ell m}\right|^2$.}
\begin{equation}
\langle a_{\ell m}a^*_{\ell'm'}\rangle=\delta_{\ell\ell'}\delta_{mm'} C_\ell,
\end{equation}
where the deltas are due to the fact that the process that created
the anisotropy is statistically isotropic. 
The $C_\ell$ are the so-called CMB power spectrum  \cite{hu}.
For homogeneity and isotropy, the $C_\ell$'s are neither a function
of $x_0$, nor of $m$.
The two-point-correlation function is related to the $C_\ell$'s in
the following way
\begin{eqnarray}
\Big<\frac{\delta T({\bf n})}{T}\frac{\delta T({\bf n}')}{T}\Big>&=&
\sum_{\ell\ell'mm'}\langle a_{\ell m}a^*_{\ell'm'}\rangle
Y_{\ell m}({\bf n})Y^*_{\ell'm'}({\bf n}')\nonumber\\
&=&\sum_\ell C_\ell \sum_m
Y_{\ell m}({\bf n})Y^*_{\ell m}({\bf n}')=\frac{1}{4\pi}\sum_\ell (2\ell+1) 
C_\ell
P_\ell(\mu={\bf n}\cdot{\bf n}')
\label{jas}
\end{eqnarray}
where we have used the addition theorem for the spherical
harmonics, and $P_\ell$ is the Legendre polynomial of order $\ell$. 
In expression (\ref{jas}) the expectation value is an ensemble average. It 
can be regarded as an average over the possible observer positions, but
not in general as an average over the single sky we observe, because of the
cosmic variance\footnote{The usual hypothesis is that we observe a typical
realization of the ensemble. This means that we expect  the difference 
between the observed values $|a_{\ell m}|^2$ and the 
ensemble averages $C_\ell$ to be of the order of the mean-square deviation
of  $|a_{\ell m}|^2$ from $C_\ell$. The latter is called
cosmic variance and, because we are dealing with a Gaussian distribution, it is
equal to $2C_\ell$ for each multipole $\ell$. For a single $\ell$, averaging 
over the $(2\ell +1)$ values of $m$ reduces the cosmic variance
by a factor $(2\ell +1)$, but it remains a serious limitation for low 
multipoles.}.

Let us now consider the last-scattering surface. In comoving coordinates
the latter is `far' from us a distance equal to
\be
\int_{t_{\rm ls}}^{t_0}\,\frac{\d t}{a}=\int_{\tau_{\rm ls}}^{\tau_0}\,
\d\tau=\left(\tau_0-\tau_{\rm ls}\right).
\ee
A given comoving 
scale $\lambda$  is therefore projected on the last-scattering surface
sky on an angular scale
\begin{equation}
\theta \simeq \frac{\lambda}{\left(\tau_0-\tau_{\rm ls}\right)},
\end{equation}
where we have neglected tiny curvature effects.
Consider now
that the scale $\lambda$ is of the order of the comoving sound horizon at the 
time of last-scattering, $\lambda\sim c_s\tau_{\rm ls}$, where
$c_s\simeq 1/\sqrt{3}$ is the sound velocity at which photons
propagate in the plasma at the last-scattering.
This corresponds
to an angle 
\be
\theta\simeq c_s\frac{\tau_{\rm ls}}{\left(\tau_0-\tau_{\rm ls}\right)}\simeq
c_s\frac{\tau_{\rm ls}}{\tau_0},
\ee
where the last passage has been performed knowing that $\tau_0\gg
\tau_{\rm ls}$. Since the universe is MD from the
time of last-scattering onwards, the scale factor has the following
behavior:  $a\sim T^{-1}\sim 
t^{2/3}\sim \tau^2$, where we have made use of the relation (\ref{rule}).
The angle $\theta_{\rm HOR}$ 
subtended by the sound horizon on the last-scattering
surface then becomes
\be
\theta_{\rm HOR}
\simeq c_s\left(\frac{T_0}{T_{\rm ls}}\right)^{1/2}\sim 1^\circ,
\ee
where we have used $T_{\rm ls}\simeq 0.3$ eV and $T_0\sim 10^{-13}$ GeV.
This corresponds to a multipole $\ell_{\rm HOR}$
 
\begin{equation}
\ell_{\rm HOR}=\frac{\pi}{\theta_{\rm HOR}}\simeq 200.
\end{equation}
From these estimates we conclude that  
two photons which on the last-scattering surface were separated
by an angle larger than $\theta_{\rm HOR}$, corresponding to
multipoles smaller than $\ell_{\rm HOR}\sim 200$ were not in causal
contact. 
On the other hand, 
from Fig. \ref{cobe} it is clear that small anisotropies, of the 
{\it same} order of magnitude $\delta T/T\sim 10^{-5}$ are present at $\ell\ll
200$. We conclude that one of the striking features of the CMB 
fluctuations is that they appear to be non causal.
Photons at the last-scattering surface which were causally disconnected
have the same small anisotropies!
The existence of particle
horizons in the standard cosmology precludes explaining
the smoothness as a result of microphysical events:  the
horizon at decoupling, the last time one could imagine
temperature fluctuations being smoothed by particle interactions,
corresponds to an angular scale on the sky of about
$1^\circ$, which precludes temperature variations on larger scales
from being erased.

From the considerations made so far, it appears that solving the
shortcomings of the standard Big Bang theory requires two basic
modifications of the assumptions made so far:

\begin{itemize}

\item { The universe has to go through a non-adiabatic period.
This is necessary to solve the entropy and the flatness problem. A 
non-adiabatic phase may give rise to the large entropy $S_{\rm U}$ we observe
today.}

\item{The universe has to go through a primordial period during which
the physical scales $\lambda$ evolve faster than the Hubble radius  $H^{-1}$.}
\end{itemize}
The second  condition is obvious from Fig. \ref{inflation}.
\begin{figure}[h!]
    \centering
        \includegraphics[width=.85\textwidth]{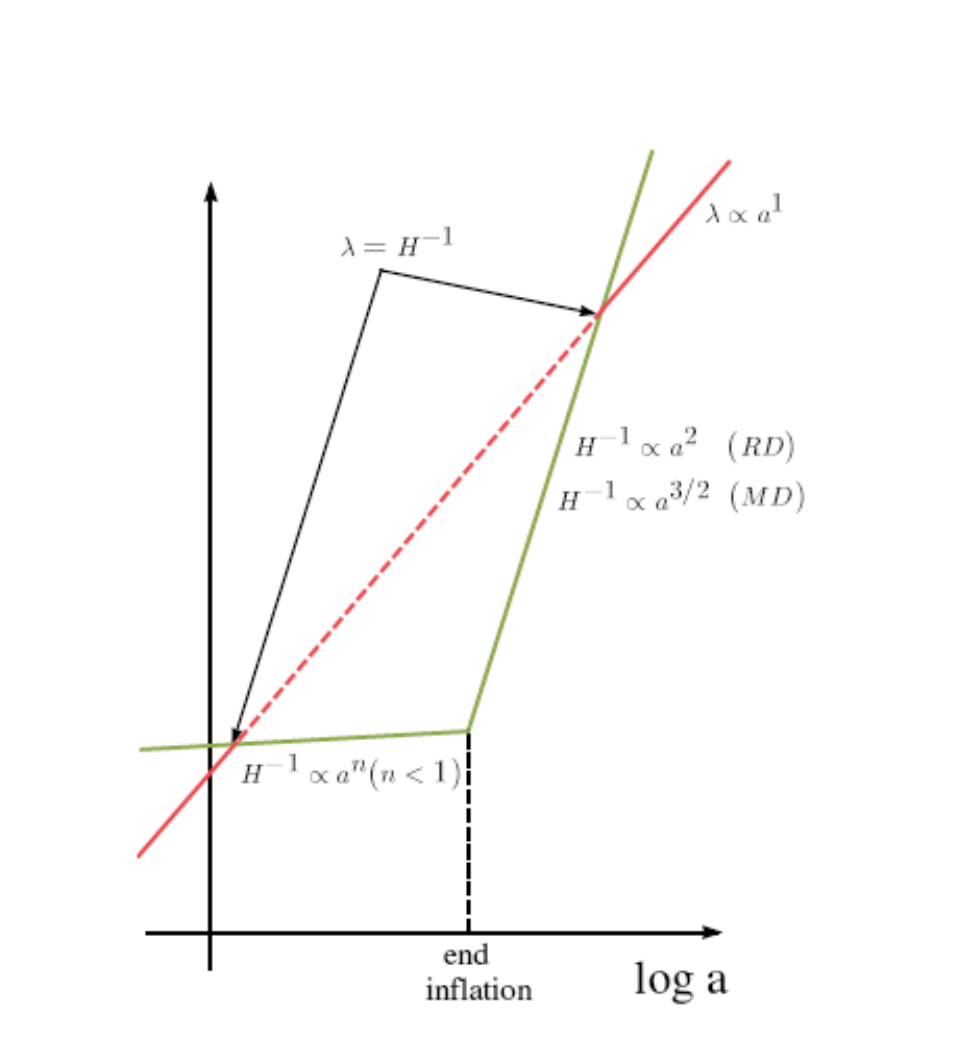}
\caption{The behavior of a generic scale $\lambda$ and the Hubble radius
$H^{-1}$ in the standard inflationary model. From Ref. \cite{kolbreview}.}
    \label{inflation}
\end{figure}
If there is period during which physical length scales grow faster
than the Hubble radius $H^{-1}$, length scales $\lambda$ 
which are within the horizon today, $\lambda<H^{-1}$ (such
as the distance between two detected photons) and 
were outside the
Hubble radius  at some period,  $\lambda>H^{-1}$ (for 
instance at the time of last-scattering when the
two photons were emitted), had a chance to be within the Hubble radius
at some primordial epoch,  $\lambda<H^{-1}$ again. If this happens,
the homogeneity and the isotropy of the CMB can be easily explained:
photons that we receive today and were emitted from the last-scattering 
surface from causally disconnected regions have the same temperature
because they had a chance to talk to each other at some primordial
stage of the evolution of the universe.
The solution to the horizon is based on the difference  between the (comoving) particle horizon and the (comoving) Hubble radius:  $R_H$ is bigger  than the Hubble radius
now, so that particles are in causal contact early on, but not at later epochs.

The second condition 
can be easily expressed as a condition on the
 scale factor $a$. Since a given scale $\lambda$ scales like
$\lambda\sim a$ and the Hubble radius $H^{-1}=a/\dot a$, we need to impose
that there is a period during which
\be
\left(\frac{\lambda}{H^{-1}}\right)^{\cdot}>0\Rightarrow \ddot a>0.
\label{fund}
\ee
Notice that is equivalent to require that the ratio between the comoving length scales $\lambda/a$
 and the comoving Hubble radius 
 
 \be
\left(\frac{\lambda}{H^{-1}}\right)^{\cdot}=\left(\frac{\lambda/a}{H^{-1}/a}\right)^{\cdot}=
\left(\frac{\lambda/a}{1/aH}\right)^{\cdot}>0
\label{fund}
\ee
increases with time. 
We can therefore introduced
the following rigorous definition: an  inflationary stage  is 
a period of the universe during which the latter accelerates
\begin{center}
\begin{tabular}{|p{13.0 cm}|}
\hline
$$
{\rm INFLATION}~~~\Longleftrightarrow~~~\ddot a>0.
$$
\\
\hline
\end{tabular}
\end{center}
\vskip 0.2cm

{\it \underline{Comment}}: Let us stress that during such a accelerating  phase 
the universe expands {\it adiabatically}. This means that during inflation
one can exploit the usual FRW equations. 
It must be 
clear therefore that the non-adiabaticity condition is satisfied not during 
inflation, but during the phase transition between the end of inflation
and the beginning of the RD phase. At this transition phase
a large entropy is generated under the form of relativistic degrees of freedom:
the Big Bang has taken place.

\part{The standard inflationary universe}
\noindent
From the previous section we have learned that 
an accelerating stage during the primordial phases  of the evolution of
the universe
might be able to solve the horizon problem. Therefore 
we learn that 
$$
\ddot a>0 \Longleftrightarrow (\rho +3 P)<0.
\label{gg}
$$
An accelerating period is obtainable only if the overall pressure $P$ 
of the universe is negative: $P<-\rho/3$. Neither 
a RD
phase nor a MD phase (for which $P=\rho/3$ and $P=0$, 
respectively) satisfy such a condition. Let us postpone for the time being
the problem of finding a `candidate' able to provide the condition
$P<-\rho/3$. For sure, 
inflation  is a phase of the history of the universe
occurring before the era of nucleosynthesis ($t \approx 1$ sec, $T
\approx 1$ MeV) during which the light elements abundances were
formed. This is because nucleosynthesis 
is the earliest epoch  we have
experimental data from and there is agreement 
with what the  standard Big-Bang theory predicts. However,  
the thermal history of the universe before 
the epoch of nucleosynthesis is unknown. 

In order to study the properties of the period of inflation, we assume the
extreme condition $P=-\rho$ which considerably simplifies the
analysis. A period of the universe during which
$P=-\rho$ is called  {\it de Sitter} stage. 
By inspecting the FRW equations and the energy conservation equation, 
we learn that during the de Sitter phase 
\bea
\rho&=&~~{\rm constant},\nonumber\\
H_{\rm I}&=&~~{\rm constant},\nonumber
\eea
where we have indicated by $H_{\rm I}$ the value of the Hubble rate during inflation.
Correspondingly, we obtain 
\be
a=a_{\rm I}\, e^{H_{\rm I}(t-t_{\rm I})},
\ee
where $t_{\rm I}$ denotes the time at which inflation starts.
Let us now see how such a period of exponential expansion
takes care of the shortcomings of the standard Big Bang Theory.\footnote{
Despite the fact that the growth of the scale factor is exponential
and the expansion is {\it superluminal}, this is not
in contradiction with what dictated by General Relativity. Indeed, it is the
space-time itself which is propagating so fast and not a light signal in it.}

\subsection{Inflation and the horizon Problem}
\noindent
During the  inflationary (de Sitter) epoch the Hubble radius $H_{\rm I}^{-1}$  is
constant. If inflation lasts long enough, all the physical scales
that have left  the Hubble radius  during the RD or 
MD phase can
re-enter the Hubble radius in the past: this is 
because such scales  are exponentially reduced.
Indeed,  during inflation the particle horizon grows exponential

\be
R_H(t)=a(t)\int_{t_{\rm I}}^t\frac{\d t'}{a(t')}=a_{\rm I}\, e^{H_{\rm I}(t-t_{\rm I})}\left(-\frac{1}{H_{\rm I}}\right)
\left[e^{-H_{\rm I}(t-t_{\rm I})}\right]_{t_{\rm I}}^t\simeq \frac{a(t)}{H_{\rm I}}, 
\ee
while the Hubble radius remains constant

\be
{\rm HUBBLE}\,\,\,{\rm RADIUS}=\frac{a}{\dot a}=H_{\rm I}^{-1},
\ee
and points that our causally disconnected today could have been in contact during inflation. 
Notice that in comoving coordinates the comoving Hubble radius shrink exponentially

\be
{\rm COMOVING }\,\,\,{\rm HUBBLE}\,\,\,{\rm RADIUS}=H_{\rm I}^{-1}e^{-H_{\rm I}(t-t_{\rm I})},
\ee
while comoving length scales remain constant. 
As we have seen in the previous section, 
this explains both the problem of the homogeneity of CMB
and the initial condition problem of small
cosmological perturbations.
Once the physical length is within the horizon,
microphysics can act, the universe can be made
approximately homogeneous and the primeval inhomogeneities can
be created. 

Let us see how long inflation must
be sustained in order to solve the horizon problem.
Let $t_{\rm I}$ and $t_{\rm f}$ be, respectively, the time of beginning and
end of inflation. We can define the corresponding number of e-foldings $N$
\be
N={\rm ln}\left[H_{\rm I}(t_{\rm e}-t_{\rm I})\right].
\ee
A necessary condition to solve the horizon problem is that the
largest scale we observe today, the present horizon $H_0^{-1}$, was
reduced  during inflation to a value $\lambda_{H_0}(t_{\rm I})$ 
smaller than the value of
Hubble radius $H_{\rm I}^{-1}$ during inflation.
This gives
$$
\lambda_{H_0}(t_{\rm I})=H^{-1}_0 \left(\frac{a_{t_{\rm f}}}{a_{t_0}}
\right) \left(\frac{a_{t_{\rm I}}}{a_{t_{\rm f}}}\right)=
H_0^{-1} \left(\frac{T_0}{T_{\rm f}}
\right) e^{-N}\la H_{\rm I}^{-1},
$$
where we have neglected for simplicity the short period of MD
and we have called $T_{\rm f}$ the temperature at the end of inflation (to be 
identified with  the reheating temperature $T_{\rm RH}$ at the beginning of 
the RD phase after inflation, see later).  
We get 
$$
N \ga \ln\left( \frac{T_0}{H_0} \right) - \ln\left( \frac{T_{\rm f}}{H_{\rm I}}
\right) \approx 67 + \ln\left( \frac{T_{\rm f}}{H_{\rm I}} \right).
$$
Apart from the logarithmic dependence, we obtain  $N \ga 70$.

\subsection{Inflation and the flatness problem}
\noindent
Inflation solves elegantly also the flatness problem. Since during inflation the
Hubble rate is constant
$$
\Omega -1 = \frac{k}{a^2H^2}\propto \frac{1}{a^2}.
$$
On the other end the condition (\ref{yxc}) tells us that
to reproduce a value of $(\Omega_0-1)$ of order of unity today
the initial value of $(\Omega-1)$ at the beginning of the
RD phase must be $\left|\Omega-1\right|\sim 10^{-60}$.
Since we identify the beginning of the RD phase
with the beginning of inflation, we require
$$
\left|\Omega -1\right|_{t=t_{\rm f}}\sim 10^{-60}.
$$
\begin{figure}[h!]
    \centering
        \includegraphics[width=.55\textwidth]{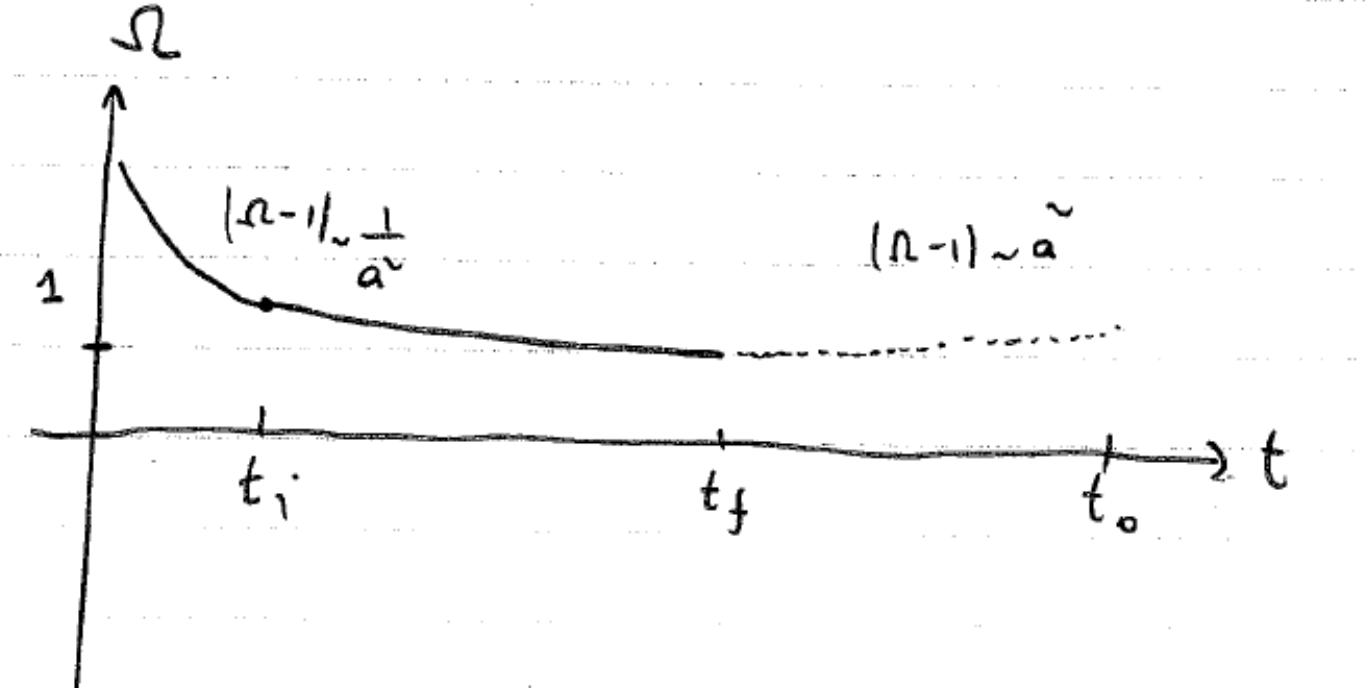}
\caption{Illustration of the solution of the flatness problem in standard inflationary cosmology.}
    \label{hor2}
\end{figure}
During inflation
\begin{equation}
\frac{\left|\Omega -1\right|_{t=t_{\rm f}}}{\left|
\Omega -1\right|_{t=t_{\rm I}}}= \left(\frac{a_{\rm I}}{a_{\rm f}}
   \right)^2 = e^{-2N}.
\label{fold}
  \end{equation}
Taking $\left|
\Omega -1\right|_{t=t_{\rm I}}$ of order unity, 
it is enough to
require that $N \approx 70$ to solve the flatness problem.

{\it 1. \underline{Comment}}:  In the previous section we have written that
the flatness problem can be also seen as a fine-tuning problem of one
part over $10^{60}$. Inflation ameliorates this fine-tuning problem, by
explaining a tiny number $\sim 10^{-60}$ with a number $N$ of the order
70.

{\it 2. \underline{Comment}}:  The number $N\simeq 70$ 
has been obtained requiring that
the present-day value of 
$(\Omega_0-1)$ is of order unity. For the expression (\ref{fold}), it
is clear that, if the period of inflation lasts longer than 70 e-foldings, 
the present-day value of $\Omega_0$ will be equal to unity with a great
precision. One can say that a generic prediction of inflation is that
\begin{center}
\begin{tabular}{|p{13.0 cm}|}
\hline
$$
{\rm INFLATION}~~~\Longrightarrow~~~\Omega_0=1.
$$
\\
\hline\end{tabular}
\end{center}
This statement, however, must be taken {\it cum grano salis} and
properly specified. Inflation does not change the global geometric properties
of the space-time. If the universe is open or closed, it will
always remain flat or closed, independently from inflation. 
What inflation does is to magnify the radius of curvature 
$R_{\rm curv}$  
so that locally
the universe is flat with a great precision. 
The current data on the
CMB anisotropies confirm this prediction.

\subsection{Inflation and the entropy problem}
\noindent
In the previous section, we have seen that the  
flatness problem arises because 
the entropy in a comoving volume is conserved. It is possible, therefore,  
that the problem could be resolved if the cosmic expansion was 
non-adiabatic for some finite time interval 
during the early history of the universe. We need to produce
a large amount of entropy $S_{\rm U}\sim 10^{90}$. Let us  
postulate that the entropy changed by an amount
\begin{equation}
S_{\rm f}=Z^3\,S_{\rm I}
\end{equation}
from the beginning to the end of the inflationary 
period, where $Z$ is a numerical factor.
It is very natural to assume that the total entropy of the universe
at the beginning of inflation was of order unity, one particle
per horizon. Since, from the end of inflation onwards, the universe expands
adiabatically, we have $S_{\rm f}=S_{\rm U}$. This gives $Z\sim 10^{30}$. On the other
hand, since $S_{\rm f}\sim \left(a_{\rm f}  T_{\rm f}\right)^3$ and 
$S_{\rm I}\sim \left(a_{\rm I}  T_{\rm I}\right)^3$, where $T_{\rm f}$ and $T_{\rm I}$ are the temperatures
of the universe at the end and at the beginning of inflation,
we get
\be
\left(\frac{a_{\rm f}}{a_{\rm I}}\right)=e^N\approx 10^{30}\left(\frac{T_{\rm I}}{T_{\rm f}}\right),
\ee
which gives again $N\sim 70$ up to the logarithmic factor ${\rm ln}
\left(T_{\rm I}/T_{\rm f}\right)$.
We stress again that such a large amount of entropy is not
produced during inflation, but  during the non-adiabatic 
phase transition which gives rise to the usual RD phase.

\subsection{Inflation and the inflaton}
\noindent
In the previous subsections we have described the various
advantages of having a period of accelerating phase. The latter
required $P<-\rho/3$. Now, we would like to show that this condition
can be attained by means of  a simple 
scalar field. We shall call  this field the {\it  inflaton} $\phi$.

The action of the inflaton field reads
\begin{equation}
S=\int \d^4x\, \sqrt{-g}\,\mathcal{L}=\int\, d^4x\, \sqrt{-g}\,
\left[-\frac{1}{2}
\partial_{\mu}\phi
\partial^{\mu}\phi -V(\phi)\right],
\end{equation}
where $\sqrt{-g}=a^3$ for the FRW metric.
From the Eulero-Lagrange equations
\begin{equation}
\partial^{\mu}\frac{\delta(\sqrt{-g}\mathcal{L})}{\delta\,
\partial^{\mu}\phi}- \frac{\delta(\sqrt{-g}\mathcal{L})}{\delta
\phi}=0, 
\end{equation}
we obtain
\begin{equation}
\fbox{$\displaystyle
\ddot{\phi}+ 3H\dot{\phi}-\frac{{\boldsymbol\nabla}^2\phi}{a^2}+V'(\phi)=0$},
\label{nabla}
\end{equation}
where $V'(\phi)=\left(\d V(\phi)/\d\phi\right)$. Note, in particular, the
appearance of the friction term $3H\dot{\phi}$: a scalar field
rolling down its potential suffers a friction due to the
expansion of the universe.

We can write the energy-momentum tensor of the scalar field
$$
T_{\mu\nu}=\partial_{\mu}\phi \partial_{\nu}\phi
+g_{\mu\nu}\, \mathcal{L}.
$$
The corresponding energy density $\rho_\phi$ and pressure density $P_\phi$ 
are
\begin{eqnarray}
\rho_{\phi}=T_{00}=\frac{\dot{\phi}^2}{2} + V(\phi)+ 
\frac{(\nabla \phi)^2}{2a^2},  \\
P_{\phi}=\frac{T^{i}_{\,\,i}}{3}=\frac{\dot{\phi}^2}{2} - V(\phi)- \frac{(\nabla
\phi)^2}{6a^2}.
\end{eqnarray}
Notice that, if
the gradient term were dominant, we would obtain
$P_\phi=-\rho_\phi/3$, not enough to drive inflation. 
We can now split the inflaton field in 
$$
\phi(t)=\phi_{0}(t)+\delta\phi({\bf x},t),
$$
where $\phi_{0}$ is the `classical' (infinite wavelength) field, that is 
the expectation value of the inflaton field 
on the initial isotropic and
homogeneous state, while $\delta\phi({\bf x},t)$ represents the quantum
fluctuation around $\phi_{0}$.
As for now, we will be only concerned with the evolution of the
classical field $\phi_0$. 
This separation is justified by the fact that quantum fluctuations are much
smaller than the classical value and therefore negligible when looking at the 
classical evolution.
The energy-momentum tensor becomes
\begin{eqnarray}
T_{00}=\rho_{\phi}=\frac{\dot{\phi_0}^2}{2} + V(\phi_0),\\
\frac{T^{i}_{\,\,i}}{3}=P_{\phi}=\frac{\dot{\phi}_0^2}{2} - V(\phi_0).
\end{eqnarray}
If
$$
V(\phi_0) \gg \dot{\phi}_0^2
$$
we obtain the following condition
$$
P_\phi\simeq -\rho_\phi.
$$
From this simple calculation, 
we realize that a scalar field whose energy is dominant
in the universe and whose potential energy  
dominates over the kinetic term drives inflation. Inflation
is driven by the vacuum energy of the inflaton field.

\subsection{Slow-roll conditions}

Let us now quantify  better under which circumstances a scalar field
may give rise to a period of inflation. 
The equation of motion of the classical value of the field is
 \begin{equation}
 \ddot{\phi}_0+3H\dot{\phi}_0+V'(\phi_0)=0.
\label{poi}
 \end{equation}
If we require that $\dot{\phi}_0^2\ll V(\phi_0)$, the scalar field 
is slowly rolling down
its potential. This is the reason why such  a period is called {\it slow-roll}.
We may also expect that, being the potential flat,  
$\dot{\phi}_0$ is negligible as well. We
 will assume that this is true and we will quantify this condition soon.
The FRW equation  becomes
\be
H^2\simeq \frac{8\pi G_{\rm N}}{3}\,V(\phi_0),
\ee
where we have assumed that the inflaton field dominates the
energy density of the universe.
The new equation of motion becomes
\be
 3H\dot{\phi}_0=-V'(\phi_0),
\label{friction}
\ee
which gives $\dot{\phi}_0$ as a function of $V'(\phi_0)$.
Using Eq. (\ref{friction}) slow-roll  conditions then require
$$
\fbox{$\displaystyle
\dot\phi_0^2 \ll  V(\phi_0)   \\  \Longrightarrow  \\  \frac{(V')^2}{V} \ll
H^2$} \label{slowroll1}
$$
and
$$
\fbox{$\displaystyle
|\ddot{\phi}_0| \ll |3H\dot{\phi}_0| \\  \Longrightarrow  \\  |V''| \ll H^2$}.  
\label{slowroll2}
$$
It is now useful to define the  slow-roll 
parameters, $\epsilon$ and $\eta$ in the following way
\begin{center}
\begin{tabular}{|p{13 cm}|}
\hline
\begin{eqnarray}
\epsilon&=&-\frac{\dot{H}}{H^2}=4\pi G_{\rm N}\frac{\dot{\phi}_0^2}{H^2}=\frac{\dot{\phi}_0^2}{2\overline{M}^2_\pl \,H^2}
=\frac{1}{16\pi
G_{\rm N}}\left(\frac{V'}{V}\right)^2, \label{epsilon slow roll}\nonumber\\
\eta&=&\frac{1}{8\pi G_{\rm N}} \left(\frac{V''}{V}\right)=\frac{1}{3}
\frac{V''}{H^2},\nonumber\\
\delta&=&\eta-\epsilon=-\frac{\ddot{\phi}_0}{H \dot{\phi}_0},\nonumber
\end{eqnarray}
\\
\hline
\end{tabular}
\end{center}
where we have indicated by $\overline{M}_{\rm Pl}$ the reduces Planck mass, 
\be
\overline{M}^2_{\rm Pl}=\frac{1}{8\pi G_{\rm N}}=\frac{{M_\pl}^2}{8\pi}.
\ee
It might be useful to have the same parameters expressed in terms of
conformal time
\begin{center}
\begin{tabular}{|p{13 cm}|}
\hline
\begin{eqnarray}
\epsilon&=&1-\frac{\H^\prime}{\H^2}=4\pi G_{\rm N}\frac{\phi_0^{\prime 2}}{\H^2},
\label{epsilon slow roll conf}\nonumber\\
\delta&=&\eta-\epsilon=1-\frac{\phi_0^{\prime\prime}}{\H \phi_0^\prime}
\, .\nonumber
\end{eqnarray}
\\
\hline
\end{tabular}
\end{center}
The parameter $\epsilon$ quantifies how much the
Hubble rate $H$ changes with time during inflation. Notice that, since
$$
\frac{\ddot a}{a}=\dot H+H^2=\left(1-\epsilon\right)H^2,
$$
inflation can be attained only if $\epsilon<1$:
\begin{center}
\begin{tabular}{|p{13 cm}|}
\hline
$$
{\rm INFLATION}~~~\Longleftrightarrow ~~~\epsilon <1.
$$
\\
\hline
\end{tabular}
\end{center}
As soon as this condition fails, inflation ends. In general, slow-roll 
inflation
is attained if $\epsilon\ll 1$ and $|\eta|\ll 1$. During inflation
the  slow-roll parameters $\epsilon$ and $\eta$ can be considered
to be approximately constant  since the potential $V(\phi)$
is very flat.

\vskip 0.2cm

{\it \underline{Comment}}:  In the following, we will work at {\it
first-order} perturbation in the slow-roll parameters, that is we will
take only the first power of them. Since, using their definition, it is
easy to see that $\dot\epsilon,\dot\eta={\cal O}
\left(\epsilon^2,\eta^2\right)$, this amounts to saying that we will
treat the slow-roll parameters as constant in time.
\vskip 0.2cm

Within these approximations, it is easy to compute the number of
e-foldings between the beginning and the end of inflation.
If we indicate by $\phi_{\rm i}$ and  $\phi_{\rm f}$ the values of the inflaton
field at the beginning and at the end of inflation, respectively,
we have that the {\it total} number of e-foldings is
\begin{eqnarray}
N&\equiv&\int_{t_{\rm i}}^{t_{\rm f}}\,H\,\d t\nonumber\\
&\simeq& \int^{\phi_{\rm f}}_{\phi_{\rm i}}
\d\phi_0\frac{H}{\dot{\phi}_0}\nonumber\\
&\simeq&-3 \int^{\phi_{\rm f}}_{\phi_{\rm i}}
\d\phi_0\frac{H^2}{V'}\nonumber\\
&\simeq& -8\pi G_{\rm N} \int^{\phi_{\rm f}}_{\phi_{\rm i}} \frac{V}{V'}\,\d\phi_0\nonumber\\
&=&-\frac{1}{\overline{M}^2_\pl} \int^{\phi_{\rm f}}_{\phi_{\rm i}} \frac{V}{V'}\,\d\phi_0
 \end{eqnarray}
We may also compute the number of e-foldings $\Delta N$ which are left to go
to the end of inflation

\be
\label{togo}
\Delta N\simeq  8\pi G_{\rm N} \int^{\phi_{\Delta N}}_{\phi_{\rm f}}
\frac{V}{V'}\,\d\phi_0,
\ee
where $\phi_{\Delta N}$ is the value of the inflaton field
when there are $\Delta N$ e-foldings to the end of inflation.

{\it 1. \underline{Comment}}:  A given scale
length $\lambda=a/k$ leaves the Hubble radius  when $k=aH_{\bf k}$
where
$H_{\bf k}$ is the 
the value of the Hubble rate at that time. One can compute easily
the rate of change of $H^2_k$ as a function of $k$
\be
\frac{\d {\rm ln} \,H_{\bf k}^2}{\d {\rm ln} \,k}=
\left(\frac{\d {\rm ln} \,H_{\bf k}^2}{\d t}\right)\left(\frac{\d t}{\d {\rm ln} \,a}
\right)\left( \frac{\d {\rm ln} \,a}{\d {\rm ln} \,k}\right)=
2\frac{\dot H}{H}
\times \frac{1}{H}\times 
1=2\frac{\dot H}{H^2}=-2\epsilon.
\label{z}
\ee

{\it 2. \underline{Comment}}:  Take a  given physical scale $\lambda$ today 
which crossed the Hubble radius  during inflation. This happened when
$$
\lambda\left(\frac{a_{\rm f}}{a_0}\right)e^{-\Delta N_\lambda}=\lambda
\left(\frac{T_0}{T_{\rm f}}\right)e^{-\Delta N_\lambda}=H_{\rm I}^{-1},
$$
where $\Delta N_\lambda$ indicates the number of e-foldings from the time the
scale crossed the Hubble radius during inflation to the end of inflation.
This relation gives a way to determine the number of e-foldings 
to the end of inflation corresponding to a given scale
$$
\Delta N_\lambda\simeq 65 +{\rm ln}\left(\frac{\lambda}{3000\,\,{\rm Mpc}}
\right)+2\,{\rm ln}\left(\frac{V^{1/4}}{10^{14}\,\,{\rm GeV}}
\right)-{\rm ln}\left(\frac{T_{\rm f}}{10^{10}\,\,{\rm GeV}}
\right).
$$
Scales relevant  for the CMB anisotropies 
correspond  to $\Delta N\sim $60.

\subsection{The last stage of inflation and reheating}
\noindent
Inflation ended when the potential energy associated with the inflaton
field became smaller than the kinetic energy of the oscillating field.   The process by which the energy of the inflaton field is
transferred from the inflaton field to radiation has been dubbed
{\it reheating}. In the old theory of reheating \cite{dolgov,abbot}   the comoving energy
density in the zero mode of the inflaton decays into normal particles. The latter then 
 scatter and thermalize to form a thermal background. 
Of particular interest is a quantity known usually 
as the reheat temperature,
denoted as $T_{\rm RH}$.  It is calculated by assuming 
an instantaneous conversion of the energy density in the inflaton 
field into radiation. The decay happens when the width of the inflaton energy,
$\Gamma_\phi$, is equal to $H$, the expansion rate of the universe. 

The reheat temperature is calculated quite easily.   After inflation
the inflaton field executes coherent oscillations about the minimum
of the potential at some $\phi_0\simeq \phi_{\rm m}$

\be
V(\phi_0)\simeq \frac{1}{2}V''(\phi_{\rm m})(\phi_0-\phi_{\rm m})^2\equiv\frac{1}{2}m^2(\phi_0-\phi_{\rm m})^2.
\ee
Indeed, the equation of motion for $\phi_0$ is

\be
\ddot{\phi}_0
+ 3H\dot{\phi}_0
+m^2(\phi_0-\phi_{\rm m})=0,
\ee
whose solution is 

\be
\phi_0(t)=\phi_{\rm i}\left(\frac{a_{\rm i}}{a}\right)^{3/2}\,\cos\left[m(t-t_{\rm i})\right],
\ee
where now the label $_{\rm i}$ denotes here the beginning of the oscillations.
Since the period of the oscillation is much shorter than the Hubble time, $H\gg m$, we can 
compute the equation satisfied by the energy density stored in the oscillating field   averaged over many oscillations

\bea
\label{rho}
\langle\dot{\rho}_\phi\rangle&=&\Big<\frac{\d }{\d t}\left(\frac{1}{2}\dot{\phi}_0^2+V(\phi_0)\right)\Big>_{{\rm many}\,{\rm  oscillations}}\nonumber\\
&=&\Big<  \dot\phi_0\left(\ddot\phi_0+V'(\phi_0)\right)\Big>_{{\rm many}\,{\rm  oscillations}}\nonumber\\
&= &\Big<  \dot\phi_0\left(-3H\dot\phi_0\right)\Big>_{{\rm many}\,{\rm  oscillations}}\nonumber\\
&=&-3H \Big<  \dot{\phi}_0^2\Big>_{{\rm many}\,{\rm  oscillations}}\nonumber\\
&=&-3H \Big<  \rho_\phi\Big>_{{\rm many}\,{\rm  oscillations}},
\eea
where we have used the equipartition property of the energy density during the oscillations
$\langle\dot{\phi}_0^2/2\rangle=\langle V(\phi_0)\rangle=\langle\rho_\phi/2\rangle$ and Eq. (\ref{poi}). The
solution of Eq. (\ref{rho}) is (removing the symbol of averaging)

\be
\rho_\phi=\left(\rho_\phi\right)_{\rm i}\left(\frac{a_{\rm i}}{a}\right)^3.
\ee
The Hubble expansion rate as a function of $a$ is 
\begin{equation}
H^2(a) = \frac{8\pi}{3}\frac{(\rho_\phi)_{\rm i}}{M_{\rm Pl}^2}
	\left( \frac{a_{\rm i}}{a} \right)^3.
\end{equation}
Equating $H(a)$ and $\Gamma_\phi$ leads to an expression for $(a_{\rm i}/a)$.
Now if we assume that all available coherent energy density is
instantaneously converted into radiation at this value of $(a_{\rm i}/a)$, we
can find the reheat temperature by setting the coherent energy
density, $\rho_\phi=(\rho_\phi)_{\rm i}(a_{\rm i}/a)^3$, equal to the radiation energy
density, $\rho_{\rm r}=(\pi^2/30)g_*T_{\rm RH}^4$, where $g_*$ is the effective
number of relativistic degrees of freedom at temperature $T_{\rm RH}$.
The result is
\begin{equation}
\label{eq:TRH}
\fbox{$\displaystyle
T_{\rm RH} = \left( \frac{90}{8\pi^3g_*} \right)^{1/4}
	\sqrt{ \Gamma_\phi M_{\rm Pl} } \
       = 0.2 \left(\frac{200}{g_*}\right)^{1/4}
      \sqrt{ \Gamma_\phi M_{\rm Pl} }.$}
\end{equation}
In some models of inflation reheating can be anticipated by a period of
preheating  \cite{preheating}  when the 
the classical inflaton field  very rapidly decays into $\phi$-particles or 
into other bosons due to broad parametric resonance. 

In preheating 
there is a new decay channel that is non-perturbative: 
stimulated emissions of bosonic particles
into energy bands with large occupancy numbers are induced due to the coherent oscillations of the inflaton field. 
The oscillations of the inflaton field induce mixing of positive and
negative frequencies in the quantum state of the field it couples to because of the {\it time-dependent} mass of the quantum field. Let us take,  for the sake of  simplicity,  to the case of 
a massive inflaton $\phi$ with quadratic potential $V(\phi)=\frac{1}{2}m^2\phi^2$ and coupled
to a massless scalar field $\chi$ via the quartic coupling $g^2\phi_0^2\chi^2$. 

Neglecting the Hubble rate in the frequency term (being smaller than the time-dependent term), the evolution equation for the
Fourier modes of the $\chi$ field with momentum ${\bf k}$ is
\be
\ddot X_{\bf k} + \omega_k^2 X_{\bf k}=0,
\ee
with
\begin{eqnarray}
X_{\bf k}&=&a^{3/2}(t)\chi_{\bf k},\nonumber\\
\omega_k^2 &=& k^2/a^2(t) + g^2\phi_0^2(t).
\end{eqnarray}
This Klein-Gordon equation may be cast
in the form of a Mathieu equation
\be
X_{\bf k}'' + [A(k) - 2q\cos2z]X_{\bf k} =0,
\ee
where $z=m t$ and
\begin{eqnarray}
\label{pa}
A(k)&=&\frac{k^2}{a^2 m^2}+2q,\nonumber\\
q&=&g^2\frac{\Phi^2}{4 m^2},
\end{eqnarray} 
where $\Phi$ is the
amplitude and $m$ is the frequency of inflaton oscillations,
$\phi_0(t)=\Phi(t)\sin(m t)$. Notice that, at least initially, if  $\Phi\gg M_{\rm Pl}$ 
\be
g^2\frac{\Phi^2}{4 m^2}\gg  g^2 \frac{M_{\rm Pl}^2}{m^2}
\ee
can be extremely large. If so, 
 the resonance is broad. 
For certain values of the parameters $(A,q)$
there are exact solutions $X_{\bf k}$ and the corresponding number density $n_{\bf k}$  grows exponentially with time because they belong to an instability band of the Mathieu
equation 
\be
X_{\bf k}\propto e^{\mu_{\bf k} m t}\Rightarrow n_{\bf k}\propto e^{2\mu_{\bf k} m t},
\ee
where the parameter $\mu_{\bf k}$ depends upon the instability band and, in the broad resonance case, $q\gg 1$, it is $\sim 0.2$.

These instabilities can be interpreted as coherent
``particle'' production with large occupancy numbers. One way of
understanding this phenomenon is to consider the energy of these modes
as that of a harmonic oscillator, $E_{\bf k} = |\dot X_{\bf k}|^2/2 + \omega_{\bf k}^2
|X_{\bf k}|^2/2 =  \omega_{\bf k} (n_{\bf k} + 1/2)$.  The occupancy number of
level ${\bf k}$ can grow exponentially fast, $n_{\bf k}\sim\exp(2\mu_{\bf k} m t)\gg1$,
and these modes soon behave like classical waves.  The parameter $q$ during preheating determines
the strength of the resonance.
It is important to notice that, after the short period of preheating, the universe is likely to enter a long period of matter domination where the biggest contribution to the energy density  of the universe is provided by 
the residual small amplitude oscillations of the classical inflaton field and/or by the inflaton quanta produced during the back-reaction processes. This period will end when the age of the universe becomes of the order of the perturbative lifetime of the inflaton field, $t\sim \Gamma_\phi^{-1}$. At this point, the universe will be reheated up to a temperature $T_{\rm RH}$  obtained applying the old theory of reheating described previously.

\subsection{A brief survey of inflationary models}
\noindent
Even restricting ourselves to a simple single-field inflation scenario, the
number of models available to choose from is large.
 It is convenient to define a general classification
 scheme, or ``zoology'' , for 
models of inflation. We divide models into three general types: {\it 
large-field}, {\it small-field}, and {\it hybrid} \cite{dodelson97}. A generic single-field potential  can be characterized by two 
independent mass scales: a ``height'' $\Lambda^4$, corresponding to the vacuum 
energy density during inflation, and a ``width'' $\mu$, corresponding to the 
change in the field value $\Delta \phi$ during inflation:
\begin{equation}
V\left(\phi\right) = \Lambda^4 f\left({\phi \over \mu}\right).
\end{equation}
Different models have different forms for the function $f$.
Let us now briefly describe the different class of models.

\subsubsection{Large-field models}

Large-field models are potentials typical of the ``chaotic'' 
inflation 
scenario \cite{linde83}, 
in which the scalar field is displaced from the minimum 
of the potential by an amount usually of order the Planck mass. 
\begin{figure}[h!]
    \centering
        \includegraphics[width=.45\textwidth]{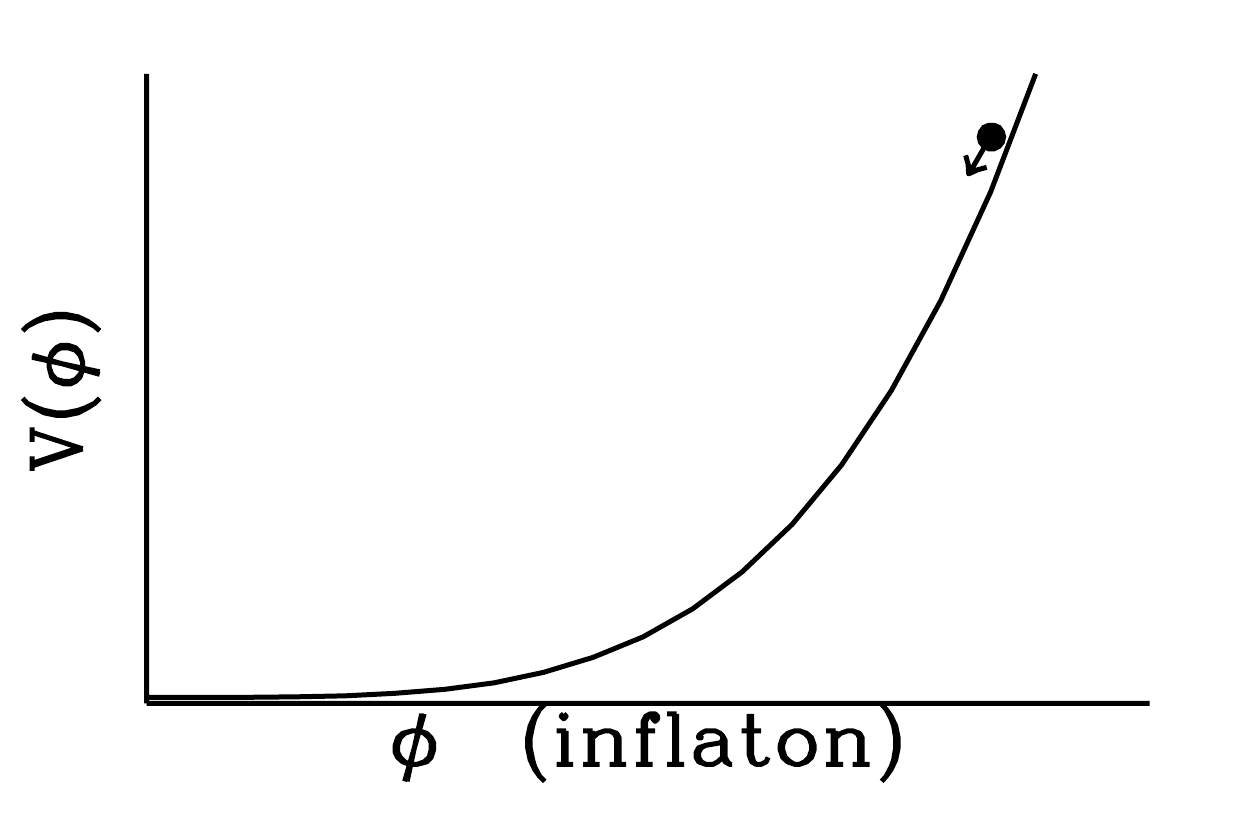}
\caption{Large field models of inflation. From Ref. \cite{kolbreview}. }
    \label{large}
\end{figure}
Such models are 
characterized by  $V''\left(\phi\right) > 0$, and $-\epsilon < \delta \leq 
\epsilon$. The generic large-field potentials we consider are polynomial 
potentials $V\left(\phi\right) = \Lambda^4
\left({\phi / \mu}\right)^p$,
and exponential potentials, $V\left(\phi\right) = \Lambda^4 \exp\left({\phi / 
\mu}\right)$. 
In the chaotic inflation scenario, it is assumed that the universe emerged 
from a quantum gravitational state with an energy density comparable to that 
of the Planck density. This implies that $V (\phi ) \approx M_{\rm Pl}^4$ 
and results in a large friction term in the Friedmann equation. 
Consequently, the inflaton will slowly roll  down its potential.
The condition 
for inflation is therefore satisfied and the scale factor grows as 
\begin{equation}
a(t) =a_{\rm I} e^{\left( \int^t_{t_{\rm I}} \d t' H(t') \right)}.
\end{equation}
The simplest chaotic inflation model is that of a free field with a 
quadratic potential, $V(\phi) =m^2 \phi^2/2$, where $m$ represents the mass 
of the inflaton. During inflation the scale factor grows as 
\begin{equation}
a(t) = a_{\rm I} e^{2\pi G_{\rm N} (\phi^2_{\rm I} - \phi^2 (t))}
\end{equation}
and inflation ends when $\phi = {\cal{O}}(M_{\rm Pl})$. If inflation 
begins when $V(\phi_{\rm i} ) \approx M_{\rm Pl}^4$, the scale factor grows 
by a factor $\exp( 4\pi M_{\rm Pl}^2/m^2)$ before the inflaton reaches the 
minimum of its potential. We will later  show that the mass 
of the field should be $m \approx 10^{-6}M_{\rm Pl}$ if the microwave 
background constraints are to be satisfied. This implies that the volume of 
the universe will increase by a factor of $Z^3 \approx e^{3 \times 
10^{12}}$ and this is more than enough inflation to solve the problems of 
the hot big bang model.

In the chaotic inflationary scenarios, the present-day universe is only
a small portion of the universe which suffered inflation.
Notice also that the typical values of the inflaton field
during inflation are of the order of $M_{\rm Pl}$, giving rise to the
possibility of testing planckian physics \cite{ckrt}.

\subsubsection{Small-field models}
\noindent
Small-field models are the type of potentials that arise naturally
 from spontaneous symmetry breaking (such as the original models of ``new'' 
inflation  \cite{linde82,albrecht82})  and from pseudo Nambu-Goldstone modes 
(natural inflation \cite{freese90}). The field starts from near an
 unstable equilibrium (taken to be at the origin) and rolls 
down the potential to a stable minimum. 
\begin{figure}[h!]
    \centering
        \includegraphics[width=.45\textwidth]{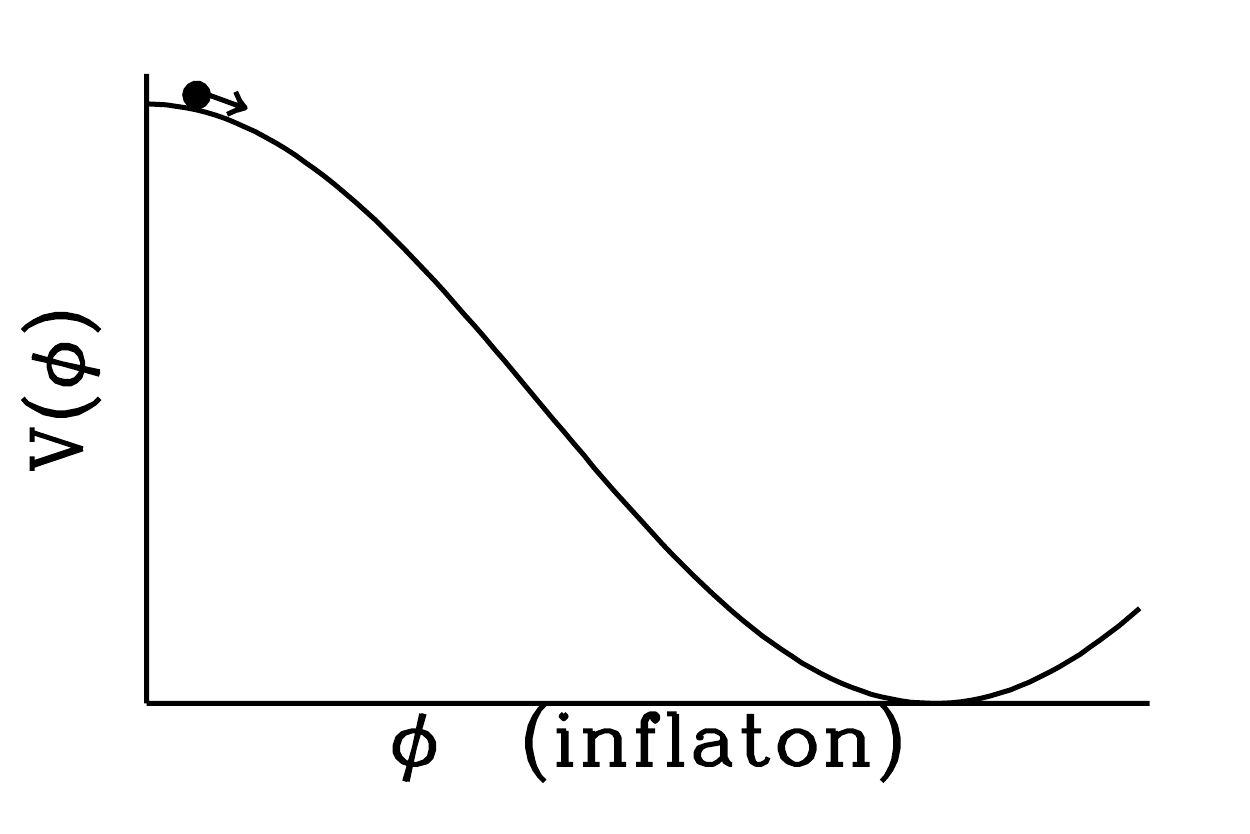}
\caption{Small field models of inflation.  From Ref. \cite{kolbreview}.}
    \label{small}
\end{figure}
Small-field models are 
characterized by $V''\left(\phi\right) < 0$ and $\eta < -\epsilon$. Typically 
$\epsilon$ is close to zero. 
The generic small-field potentials we consider are of the form 
$V\left(\phi\right) = \Lambda^4 \left[1 - \left({\phi / \mu}\right)^p\right]$,
 which can be viewed as a lowest-order Taylor expansion of an arbitrary
 potential about the origin. See, for instance, Ref. \cite{Dine:1997kf}.

\subsubsection{Hybrid models}
\noindent
The hybrid scenario \cite{linde91,linde94,copeland94}  frequently appears in 
models which incorporate inflation into supersymmetry \cite{Riotto:1997iv}
and supergravity  \cite{Linde:1997sj}. In a typical hybrid 
inflation model, the scalar field responsible
for inflation evolves toward a minimum with nonzero vacuum energy. The end of 
inflation arises as a
 result of instability in a second field. Such models are 
characterized by $V''\left(\phi\right) > 0$ and $0 < \epsilon < \delta$. We 
consider generic potentials for hybrid inflation of the form 
$V\left(\phi\right) 
= \Lambda^4 \left[1 + \left({\phi / \mu}\right)^p\right].$ The field value at 
the end of inflation is determined by some other physics, so there is a second 
free parameter characterizing the models. 
\begin{figure}[h!]
    \centering
        \includegraphics[width=.45\textwidth]{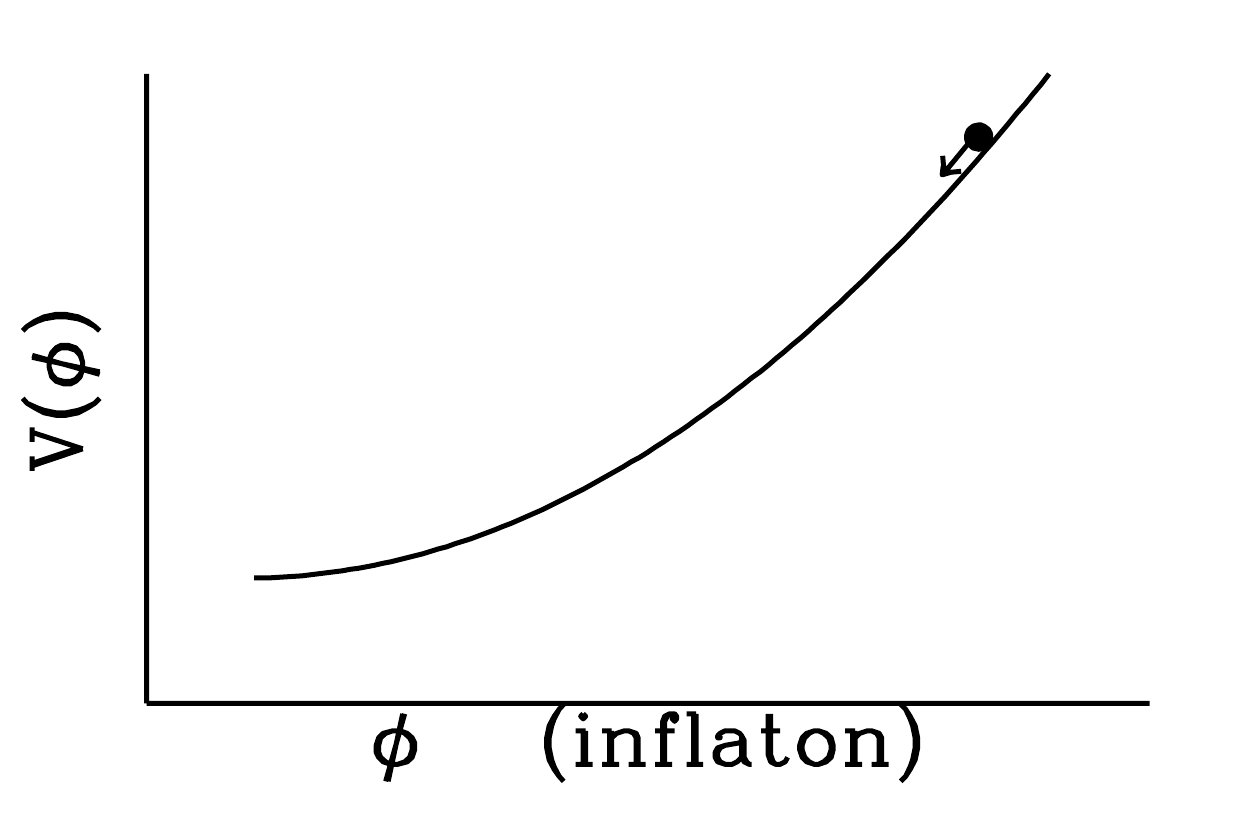}
\caption{Hybrid field models of inflation. From Ref. \cite{kolbreview}.}
    \label{hybrid}
\end{figure}

This enumeration of models is certainly not exhaustive. There are a number of
single-field models that do not fit well into this scheme, for example
logarithmic potentials $V\left(\phi\right) \propto
\ln\left(\phi\right)$ typical of supersymmetry \cite{lr,Halyo:1996pp,Binetruy:1996xj,Dvali:1997mh,
Lyth:1997pf,Riotto:1997wy,Espinosa:1998ks,King:1998uv}. 
Another example is potentials
with negative powers of the scalar field $V\left(\phi\right) \propto
\phi^{-p}$ \cite{barrow93} used in intermediate inflation 
 and dynamical
supersymmetric inflation  \cite{kinney97,kinney98}. 
Both of these cases require 
and auxiliary field to end inflation and are more properly categorized as 
hybrid models, but fall into the small-field class. 
However, the three classes
categorized by the relationship between the slow-roll parameters 
as $-\epsilon < 
\delta \leq \epsilon$ (large-field), $\delta 
\leq -\epsilon$ (small-field) 
and $0 < \epsilon < \delta$ (hybrid) seems to be good enough for
comparing theoretical expectations with experimental data.

\part{Inflation and the cosmological perturbations}

As we have seen in the previous section, 
the early universe was made very nearly uniform by a primordial
inflationary stage.  However, the important
caveat in that statement is the word `nearly'.  Our current
understanding of the origin of structure in the universe is that it
originated from small `seed' perturbations, which over time grew to
become all of the structure we observe.  
Once the universe becomes matter dominated ) some seeds of the density
inhomogeneities start growing 
thanks to the phenomenon of gravitational instabilities  thus 
forming   the structure we see today \cite{sf}. The gravitational instability is called  Jeans instability.

The presence of the primordial  inflationary seeds  is also confirmed by detailed measurements of the CMB anisotropies;   the   temperature
anisotropies at angular scales larger than $1^\circ$ are caused
by  some inflationary  inhomogeneities  since
causality prevents  microphysical processes
from producing anisotropies on angular scales larger
than about $1^\circ$, the angular size of the horizon
at last-scattering.

Our  best guess for the origin
of these perturbations is quantum fluctuations during an inflationary
era in the early universe.
Although originally introduced as a
possible solution to the  cosmological conundrums such as the
horizon, flatness and entropy problems, by far the most useful
property of inflation is that it generates spectra of both density
perturbations 
and gravitational waves. These perturbations extend from extremely
short scales to cosmological scales by the stretching of space during inflation.

Once inflation has ended,
however, the Hubble radius increases faster than the scale factor, so
the fluctuations eventually reenter the Hubble radius during the
radiation- or matter-dominated eras. The fluctuations that exit around
60 $e$-foldings or so before reheating reenter with physical
wavelengths in the range accessible to cosmological observations.
These spectra provide a distinctive signature of inflation. They can
be measured in a variety of different ways including the analysis of
microwave background anisotropies. 

The physical processes which give rise to the structures we observe today
are well-explained in Fig. \ref{pert}.

\begin{figure}[h!]
    \centering
        \includegraphics[width=.65\textwidth]{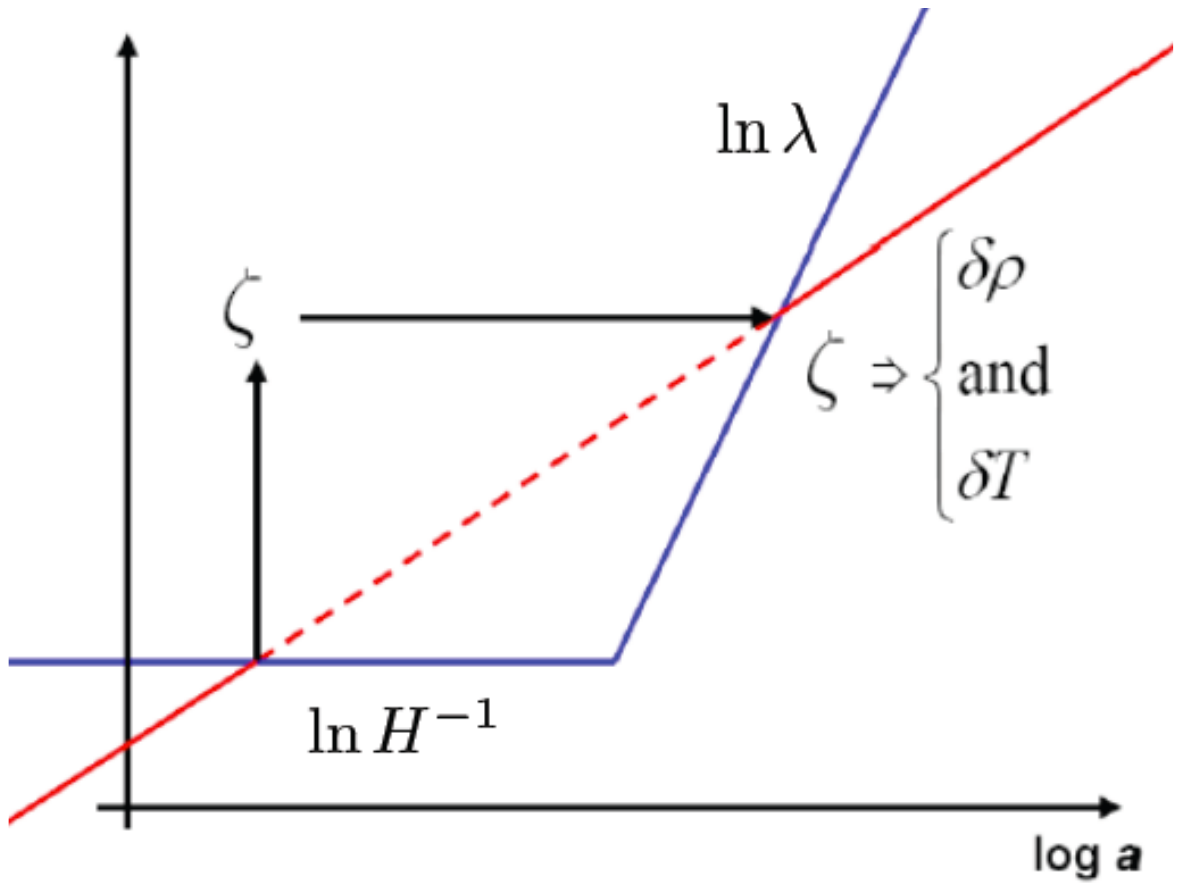}
\caption{A schematic representation of the generation of quantum fluctuations during inflation. From Ref. \cite{kolbreview}. }
    \label{pert}
\end{figure}
Since gravity talks to any component of the universe, small fluctuations
of the inflaton field are intimately related to fluctuations of the
space-time metric, giving rise to perturbations of the curvature
${\cal R}$ (which will be  defined in the following; the reader
may loosely think of it as a gravitational potential). The
wavelengths $\lambda$ of these  
perturbations grow exponentially and leave soon the
Hubble radius when $\lambda>H^{-1}$. On super-Hubble scales, curvature 
fluctuations are frozen in and may be considered as classical. Finally, 
when the
wavelength of these fluctuations reenters the horizon, at some
radiation- or matter-dominated epoch, the curvature (gravitational
potential) perturbations
of the space-time give rise to matter (and temperature) 
perturbations 
via the Poisson equation. These fluctuations will then start growing
giving rise to the structures we observe today.

In summary, two are the key ingredients for understanding 
the observed structures in the universe within the
inflationary scenario:

\begin{itemize}

\item Quantum fluctuations of the inflaton field 
are excited during inflation and stretched to cosmological scales. At the
same time, being the inflaton fluctuations connected
to the metric perturbations through  Einstein's equations, 
ripples on the metric are also excited and stretched to cosmological 
scales.

\item Gravity acts a   messenger since it communicates to baryons and photons
the small seed perturbations once a given wavelength becomes smaller
than the Hubble radius  after inflation.

\end{itemize}

Let us know see how quantum fluctuations are generated during inflation.
We will proceed by steps. First, we will consider the simplest
problem of studying the quantum fluctuations of a generic scalar field
during inflation: we will
learn how perturbations evolve as a function of time and compute their
spectrum. Then -- since a satisfactory description of the generation of
quantum fluctuations have to take both the inflaton and the metric
perturbations into account -- we will study the system composed by 
quantum
fluctuations of the inflaton field  and  quantum fluctuations
of the metric.

\section{Quantum fluctuations of a generic massless scalar field during 
inflation}

Let us first see how the
fluctuations  of a generic scalar field $\chi$, which is {\it not}
the inflaton field, behave during inflation. To warm up we first
consider a de Sitter epoch during which the Hubble rate is
constant.

\subsection{Quantum fluctuations of a generic massless scalar field during 
a de Sitter stage}

We assume this field to be massless. The massive case will be analyzed in 
the
next subsection. Expanding the scalar field $\chi$ in Fourier modes
$$
\delta\chi({\bf x},t)=\int\,\frac{\d^3 k}{(2\pi)^{3}}\,e^{i{\bf k}
\cdot{\bf x}}\,
\delta\chi_{{\bf k}}(t),
$$
we can write the equation for the
fluctuations as 
\be
\delta\ddot{\chi}_{\bf  k}
+3H\,\delta\dot{\chi}_{\bf k}+
\frac{k^2}{a^2}\,\delta\chi_{\bf k}=0.
\label{quantum}
\ee
Let us study the qualitative behavior of the solution to Eq. (\ref{quantum}).

\begin{itemize}

\item For wavelengths within the Hubble radius, $\lambda\ll H^{-1}$, the 
corresponding wavenumber satisfies the relation $k\gg a\,H$. In this
regime, we can neglect the friction term $3H\,\delta\dot{\chi}_{\bf k}$
and 
Eq. (\ref{quantum}) reduces to
\be
\delta\ddot{\chi}_{\bf k}+
\frac{k^2}{a^2}\,\delta\chi_{\bf k}=0,
\ee
which is -- basically -- the equation of motion of an harmonic oscillator.
Of course, the frequency term $k^2/a^2$ depends upon time because
the scale factor $a$ grows exponentially. On the qualitative level,
however, one expects that when the wavelength of the fluctuation is within
the horizon, the fluctuation oscillates.

\item For wavelengths above the Hubble radius, $\lambda\gg H^{-1}$, the 
corresponding wavenumber satisfies the relation $k\ll aH$
and the term $k^2/a^2$ can be safely neglected. Eq. (\ref{quantum}) reduces to
\be
\delta\ddot{\chi}_{\bf  k}
+3H\,\delta\dot{\chi}_{\bf k}=0,
\ee
which tells us that on super-Hubble scales $\delta\chi_{\bf k}$ remains 
constant.

\end{itemize}
We have therefore the following picture: take a given fluctuation
whose initial wavelength $\lambda\sim a/k$ is within the Hubble radius. The
fluctuations oscillates till the wavelength becomes of the order
of the horizon scale. When the wavelength crosses the Hubble radius, the
fluctuation ceases to oscillate and gets frozen in.

Let us know study the evolution of the fluctuation in a more quantitative
way. To do so, we perform the following redefinition
$$
\delta\chi_{\bf k}=\frac{\delta\sigma_{\bf k}}{a}
$$
and we work in conformal time $\d\tau=\d t/a$. For the time
being, we solve the problem for a pure de Sitter expansion
and we take the scale 
factor
 exponentially growing as $a\sim e^{Ht}$; 
the corresponding
conformal factor reads (after choosing properly the integration constants) 
$$
a(\tau)=-\frac{1}{H\tau}\,\,\,\,(\tau<0).
$$
In the following we will also solve the problem in the
case of quasi de Sitter expansion.
The beginning of inflation coincides with some initial time $\tau_{\rm i}\ll 0$.
We find that Eq. (\ref{quantum})
becomes
\be
\delta\sigma^{\prime\prime}_{\bf k}+
\left(k^2-\frac{a^{\prime\prime}}{a}\right)\delta\sigma_{\bf k}=0.
\label{qq}
\ee
We obtain an equation which is very `close' to the equation for a 
Klein-Gordon scalar field in flat space-time, the only difference being
a negative time-dependent mass term $-a^{\prime\prime}/a=-2/\tau^2$.
Eq. (\ref{qq}) can be obtained from an action of the type
\be
\delta S_{\bf k}=\int\,\d\tau\,\left[\frac{1}{2}
\delta\sigma^{\prime 2}_{\bf k}-\frac{1}{2}
\left(k^2-\frac{a^{\prime\prime}}{a}\right)\delta\sigma^2_{\bf k}
\right],
\label{action}
\ee
which is the canonical action for a simple harmonic oscillator with
canonical commutation relations 

\be
\delta\sigma^*_{\bf 
k}\delta\sigma^\prime_{\bf k}
-\delta\sigma_{\bf k}\delta\sigma^{*\prime}_{\bf k}=-i.
\ee
Let us study the behavior of this equation on sub-Hubble and super-Hubble
scales. Since 
$$
\frac{k}{aH}=-k\,\tau,
$$
on sub-Hubble scales $k^2\gg a^{\prime\prime}/a$   Eq. (\ref{qq})
reduces to 
$$
\delta\sigma^{\prime\prime}_{\bf k}+
k^2\,\delta\sigma_{\bf k}=0,
$$
whose solution is a plane wave
\be
\delta\sigma_{\bf k}=\frac{e^{-ik\tau}}{\sqrt{2k}}\,\,\,\,(k\gg aH).
\label{q1}
\ee
We find again that  fluctuations with wavelength within the horizon
oscillate exactly like in flat space-time. 
This does not come as a surprise. In the 
ultraviolet regime, that is for wavelengths much smaller than the Hubble radius
scale, one expects that approximating the space-time as flat
is a good approximation.

On super-Hubble scales, 
$k^2\ll a^{\prime\prime}/a$ Eq. (\ref{qq})
reduces to 
$$
\delta\sigma^{\prime\prime}_{\bf k}-
\frac{a^{\prime\prime}}{a}\delta\sigma_{\bf k}=0,
$$
which is satisfied by 
\be
\delta\sigma_{\bf k}=B(k)\,a \,\,\,\,(k\ll aH).
\label{x2}
\ee
where $B(k)$ is a constant of integration. Roughly matching  the (absolute
values of the) solutions
$(\ref{q1})$ and $(\ref{x2})$ at $k=aH$ ($-k\tau=1$), we can determine the
(absolute value of the) constant $B(k)$
$$
\left|B(k)\right|a=\frac{1}{\sqrt{2k}}\Longrightarrow
\left|B(k)\right|=\frac{1}{a\sqrt{2k}}=\frac{H}{\sqrt{2k^3}}.
$$
Going back to the original variable 
$\delta\chi_{\bf k}$, we obtain that the quantum fluctuation of the 
$\chi$
field on super-Hubble scales is constant and approximately
equal to
\begin{center}
\begin{tabular}{|p{13 cm}|}
\hline
$$
\left|\delta\chi_{\bf k}\right|\simeq \frac{H}{\sqrt{2k^3}}\,\,\,\,
({\rm ON}\,\,{\rm SUPER-HUBBLE}\,\,{\rm SCALES}).
$$
\\
\hline
\end{tabular}
\end{center}
In fact we can do much better, since Eq. (\ref{qq}) has an {\it exact} 
solution:
\be
\label{sigma}
\delta\sigma_{\bf k}=\frac{e^{-ik\tau}}{\sqrt{2k}}\left(
1-\frac{i}{k\tau}\right).
\ee
This solution reproduces all what we have found by qualitative arguments
in the two extreme regimes $k\ll aH$ and $k\gg aH$. The reason why we have
performed the matching procedure is to show that the latter can be
very useful to determine the behavior of the solution on super-Hubble scales
when the exact solution is not known.

\subsection{Quantum fluctuations of a generic massive scalar field during
a de Sitter stage}

So far, we have solved the equation for the quantum perturbations of 
a generic massless  field, that is  neglecting the mass squared term 
$m_\chi^2$. 
Let us know discuss
the solution when such a mass term is present. Eq. (\ref{qq})
becomes
\be
\delta\sigma^{\prime\prime}_{\bf k}+
\left[k^2+M^2(\tau)\right]\delta\sigma_{\bf k}=0,
\label{qqq}
\ee
where
$$
M^2(\tau)=\left(m_\chi^2-2H^2\right)a^2(\tau)=\frac{1}{\tau^2}
\left(\frac{m^2}{H^2}-2\right).
$$
Eq. (\ref{qqq}) can be recast in the form
\be
\delta\sigma^{\prime\prime}_{\bf k}+
\left[k^2-\frac{1}{\tau^2}\left(\nu_\chi^2-\frac{1}{4}\right)
\right]\delta\sigma_{\bf k}=0,
\label{qqqq}
\ee
where 
\be
\nu_\chi^2=\left(\frac{9}{4}-\frac{m_\chi^2}{H^2}\right).
\label{zz}
\ee
The generic
solution to Eq. (\ref{qqq}) for $\nu_\chi$ {\it real} is
$$
\delta\sigma_{\bf k}=\sqrt{-\tau}\left[c_1(k)\,H^{(1)}_{\nu_\chi}(-k\tau)+
c_2(k)\,H_{\nu_\chi}^{(2)}(-k\tau)\right],
$$
where $H^{(1)}_{\nu_\chi}$ and $H_{\nu_\chi}^{(2)}$ are the Hankel's 
functions of the
first and second kind, respectively.
If we impose that in the ultraviolet regime $k\gg aH$  $(-k\tau\gg 1$)
the solution matches the plane-wave solution $e^{-ik\tau}/\sqrt{2k}$
that we expect in flat space-time and knowing that
$$
H^{(1)}_{\nu_\chi}(x\gg 1)\sim \sqrt{\frac{2}{\pi x}}\,e^{i\left(x-
\frac{\pi}{2}\nu_\chi-\frac{\pi}{4}\right)}\,\,\,\, ,
H_{\nu_\chi}^{(2)}(x\gg 1)\sim \sqrt{\frac{2}{\pi x}}\,
e^{-i\left(x-
\frac{\pi}{2}\nu_\chi-\frac{\pi}{4}\right)},
$$ 
we set $c_2(k)=0$ and $c_1(k)=
\frac{\sqrt{\pi}}{2}\,e^{i\left(\nu_\chi+\frac{1}{2}\right)\frac{\pi}{2}}$. 
The exact solution becomes
\be
\delta\sigma_{\bf k}=\frac{\sqrt{\pi}}{2}\,
e^{i\left(\nu_\chi+\frac{1}{2}\right)\frac{\pi}{2}}\,
\sqrt{-\tau}\,H^{(1)}_{\nu_\chi}(-k\tau).
\label{exactm}
\ee
On super-Hubble scales, since $H^{(1)}_{\nu_\chi}(x\ll 1)\sim
\sqrt{2/\pi}\, e^{-i\frac{\pi}{2}}\,2^{\nu_\chi-\frac{3}{2}}\,
(\Gamma(\nu_\chi)/\Gamma(3/2))\, x^{-\nu_\chi}$, 
the fluctuation (\ref{exactm}) becomes
$$
\delta\sigma_{\bf k}=e^{i\left(\nu_\chi-\frac{1}{2}\right)\frac{\pi}{2}}
2^{\left(\nu_\chi-\frac{3}{2}\right)}\frac{\Gamma(\nu_\chi)}{\Gamma(3/2)}
\frac{1}{\sqrt{2k}}\,(-k\tau)^{\frac{1}{2}-\nu_\chi}.
$$
Going back to the old variable $\delta\chi_{\bf k}$, we find that
on super-Hubble scales, the fluctuation with nonvanishing mass
is not exactly constant, but
it acquires a tiny dependence upon the time
\begin{center}
\begin{tabular}{|p{13 cm}|}
\hline
$$
\left|\delta\chi_{\bf k}\right|\simeq \frac{H}{\sqrt{2k^3}}
\left(\frac{k}{aH}\right)^{\frac{3}{2}-\nu_\chi}
\,\,\,\,
({\rm ON}\,\,{\rm SUPER-HUBBLE}\,\,{\rm SCALES})
$$
\\
\hline
\end{tabular}
\end{center}
If we now define, in analogy with the definition of the slow roll
parameters $\eta$ and $\epsilon$ for the 
inflaton field,  the 
parameter 
$\eta_\chi=(m_\chi^2/3H^2)\ll 1 $, one finds
\be
\frac{3}{2}-\nu_\chi\simeq \eta_\chi.
\label{nu}
\ee

\subsection{Quantum to classical transition}

We have previously said that the quantum flactuations can be regarded
as classical when their corresponding wavelengths cross the 
Hubble radius. To  better motivate this statement, we should compute the
number of particles $n_{\bf k}$ per wavenumber ${\bf k}$ on super-Hubble
scales and check that it is indeed much larger than unity, $n_{\bf k}\gg 1$
(in this limit one can neglect the ``quantum" factor $1/2$ in the 
Hamiltonian $H_{\bf k}=\omega_{\bf k}\left(n_{\bf k}+\frac{1}{2}\right)$
where $\omega_{\bf k}$ is the energy eigenvalue).
If so, the fluctuation  can be regarded as classical. The number of
particles $n_{\bf k}$ can be estimated to be of the order of $H_{\bf 
k}/\omega_{\bf k}$,
where $H_{\bf k}$ is the Hamiltonian corresponding to the action
\be
\delta S_{\bf k}=\int\,\d\tau\,\left[\frac{1}{2}
\delta\sigma^{\prime 2}_{\bf k}+\frac{1}{2}
\left(k^2-M^2(\tau)\right)\delta\sigma^2_{\bf k}
\right].
\ee
One obtains on super-Hubble scales
$$
n_{\bf k}\simeq \frac{
\delta\sigma^{'2}_{\bf k}}{\omega_{\bf k}}\sim \left(\frac{H}{\sqrt{k^3}}\right)^2\frac{a^{'2}}{k}\sim
\left(\frac{k}{aH}\right)^{-4}\gg 1,
$$
which confirms that fluctuations on super-Hubble scales may be indeed 
considered as classical.

\subsection{The power spectrum}

Let us define now the power spectrum, a useful quantity to
characterize the properties of the perturbations. 
For a generic quantity $g({\bf x},t)$, which can expanded in
Fourier space as
$$
g({\bf x},t)=\int\,\frac{\d^3{k}}{(2\pi)^{3}}\,e^{i{\bf k}
\cdot{\bf x}}\,
g_{{\bf k}}(t),
$$
the power spectrum can be defined as
\be
\langle 0|g_{{\bf k}_1}g_{{\bf k}_2}|0\rangle
\equiv(2\pi)^3\delta^{(3)}\left({\bf k}_1+{\bf k}_2\right)\,
|{g}_{\bf k}|^2,
\ee
where $\left|0\right.\rangle$ is the vacuum quantum state of the system. This
definition leads to the  relation
\be
\Big< 0\Big|g^2({\bf x},t)\Big |\,0\Big>=\int\,\frac{\d^3{k}}{(2\pi)^{3}}\,{g}_{\bf k}g_{-\bf k}=\int\,\frac{\d^3{k}}{(2\pi)^{3}}|{g}_{\bf k}|^2=
\int\,\frac{\d k}{k}\,
{\cal P}_{g}(k),
\ee
which defines the power spectrum of the perturbations of the field $g({\bf x},t)$ as

\be
\fbox{$\displaystyle
{\cal P}_{g}(k)=\frac{k^3}{2\pi^2}|{g}_{\bf k}|^2$}.
\ee

\subsection{Quantum fluctuations of a generic scalar field in a quasi de 
Sitter stage}

So far, we have computed the time evolution and the spectrum of the 
quantum fluctuations of a generic scalar field $\chi$ 
supposing that the 
scale factor evolves
like in a pure de Sitter expansion, $a(\tau)=-1/(H\tau)$. However, 
during
inflation the Hubble rate is not exactly constant, but changes with time
as $\dot H=-\epsilon\,H^2$ (quasi de Sitter expansion),
In this subsection, we will solve for the perturbations
in a quasi de Sitter expansion. Using the definition of the
conformal time, one can show that the scale factor for small values
of $\epsilon$
becomes
$$
a(\tau)=-\frac{1}{H}\frac{1}{\tau^{(1+\epsilon)}}.
$$
Eq. (\ref{qqq}) has now a squared mass term
$$
M^2(\tau)=m_\chi^2a^2-\frac{a^{\prime\prime}}{a},
$$
where
\be
\label{ap}
\frac{a^{\prime\prime}}{a}\simeq  \frac{1}{\tau^2}\left(2+3\epsilon\right).
\ee
Taking $m_\chi^2/H^2=3\eta_\chi$ and expanding for small values
of $\epsilon$ and $\eta_\chi$ we get Eq. 
(\ref{qqqq}) with
\be
\label{vv}
\nu_\chi\simeq \frac{3}{2}+\epsilon-\eta_\chi.
\ee
Armed with these results, we may  compute the  power spectrum of the
fluctuations of the scalar field $\chi$. Since we have seen that fluctuations are (nearly) 
frozen in on super-Hubble scales,
a way of characterizing the perturbations is to compute
the spectrum on scales larger than the Hubble radius 
\be
\fbox{$\displaystyle
{\cal P}_{\delta\chi}(k)\equiv\frac{k^3}{2\pi^2}
\,\left|\delta\chi_{\bf k}\right|^2=\left(\frac{H}{2\pi}\right)^2
\left(\frac{k}{aH}\right)^{3-2\nu_\chi}.$}
\label{spectrum}
\ee
We may also define the {\it spectral index} $n_{\delta\chi}$ 
of the fluctuations
as
\begin{center}
\begin{tabular}{|p{13 cm}|}
\hline
$$
n_{\delta\chi}-1=
\frac{d {\rm ln} \,{\cal P}_{\delta\phi}}{d {\rm ln} \,k}=3-2\nu_\chi=
2\eta_\chi-2\epsilon.
$$
\\
\hline
\end{tabular}
\end{center}
The power spectrum of fluctuations of the scalar
field $\chi$  is therefore
{\it nearly flat}, that is is nearly independent from the wavelength
$\lambda=\pi/k$: the amplitude of the 
fluctuation on super-Hubble scales does not (almost) depend upon the 
time at which the fluctuations crosses the Hubble radius and becomes frozen
in. The small tilt of the power spectrum arises from the fact that
the scalar field $\chi$ is massive and because 
during inflation the Hubble rate is not exactly constant, but
nearly constant, where `nearly' is quantified by the slow-roll
parameters $\epsilon$. Adopting the  traditional 
terminology,
we may say that the spectrum of perturbations is blue if 
$n_{\delta\chi}>1$
(more power in the ultraviolet)
and red if $n_{\delta\chi}<1$ (more power in the infrared).
The power spectrum of the perturbations
of a generic scalar field $\chi$ generated during a period
of slow roll inflation may be either blue or red. This
depends upon the relative magnitude between $\eta_\chi$ and $\epsilon$.
For instance, in chaotic inflation with a quadric potential
$V(\phi)=m^2\phi^2/2$, one can easily
compute 

$$
n_{\delta\chi}-1=
2\eta_\chi-2\epsilon=\frac{2}{3H^2}\left(m_\chi^2-m^2\right),
$$
which tells us that the spectrum is blue (red) if $m_\chi^2>m_\phi^2$
($m_\chi^2<m^2$). 

\vskip 0.2cm

{\it \underline{Comment}:} We might have computed the
spectral index of the spectrum ${\cal P}_{\delta\chi}(k)$ by first
solving the equation for the perturbations of the field $\chi$
in a di Sitter stage, with $H=$ constant and therefore $\epsilon=0$, and
then taking into account the time-evolution of the Hubble rate
introducing  the subscript in $H_{\bf k}$ whose time variation is determined 
by  Eq. (\ref{z}). Correspondingly, $H_{\bf k}$ 
is the value of the Hubble rate  when a given wavelength $\sim 
k^{-1}$ crosses
the horizon (from that point on the fluctuations remains
frozen in). The power spectrum in such an approach would read

\be
{\cal P}_{\delta\chi}(k)=\left(\frac{H_{\bf k}}{2\pi}\right)^2
\left(\frac{k}{aH}\right)^{3-2\nu_\chi}
\label{bb}
\ee
with $3-2\nu_\chi\simeq 2 \eta_\chi$. Using Eq. (\ref{z}), one finds

$$
n_{\delta\chi}-1=
\frac{\d {\rm ln} \,{\cal P}_{\delta\chi}}{\d {\rm ln} \,k}=
\frac{\d {\rm ln} \,H_{\bf k}^2}{\d {\rm ln} \, k}+3-2\nu_\chi=
2\eta_\chi-2\epsilon,
$$
which 
reproduces our previous findings.

\vskip 0.2cm   

{\it \underline{Comment}:} Since on super-Hubble scales 

$$
\delta\chi_{\bf k}\simeq \frac{H}{\sqrt{2k^3}}
\left(\frac{k}{aH}\right)^{\eta_\chi-\epsilon}\simeq
\frac{H}{\sqrt{2k^3}}\left[1+\left(\eta_\chi-\epsilon\right)
{\rm ln}\,\left(\frac{k}{aH}\right)\right],
$$
we discover that 

\begin{equation}
\label{e}
\left|\delta\dot{\chi}_{\bf k}\right|\simeq \left|
H\left(\eta_\chi-\epsilon\right)
\,\delta\chi_{\bf k}\right|\ll \left|H\,\delta\chi_{\bf k}\right|,
\end{equation}
that is on super-Hubble scales the time variation of the 
perturbations can be safely neglected.

\section{Quantum fluctuations during inflation}

As we have mentioned in the previous section, the linear theory of the
cosmological perturbations represent a cornerstone of modern cosmology
and is used to describe the formation and evolution of structures
in the universe as well as the anisotropies of the CMB. The seeds
 were generated during inflation and
stretched over astronomical scales because of the rapid superluminal
expansion of the universe during the (quasi) de Sitter epoch.

In the previous section we have already seen
that perturbations of a generic scalar field $\chi$ are generated
during a (quasi) de Sitter expansion. The inflaton
field is a scalar field and, as such, we conclude that 
inflaton fluctuations will be generated as well. However, 
the inflaton is special from the point of view
of perturbations. The reason is very simple. By assumption, the
inflaton field dominates the energy density of the universe during
inflation. Any perturbation in the inflaton field means a perturbation
of the stress energy-momentum tensor

$$
\delta\phi\Longrightarrow \delta T_{\mu\nu}.
$$
A perturbation in the stress energy-momentum tensor implies,
through Einstein's equations of motion, a perturbation of the metric

$$
\delta T_{\mu\nu}\Longrightarrow \left[
\delta R_{\mu\nu}-\frac{1}{2}\delta\left(g_{\mu\nu}R\right)\right]
=8\pi G\delta T_{\mu\nu}\Longrightarrow \delta g_{\mu\nu}.
$$
On the other hand, a pertubation of the metric induces a back reaction
on the evolution of the inflaton perturbation through the 
perturbed Klein-Gordon equation of the inflaton field

$$
\delta g_{\mu\nu}\Longrightarrow \delta\left(-\partial_\mu\partial^\mu\phi+
\frac{\partial V}{\partial\phi}\right)=0\Longrightarrow\delta\phi.
$$
This logic chain makes us conclude that the perturbations of the
inflaton field and of the metric are tightly coupled to each other
and have to be studied together

\begin{center}
\begin{tabular}{|p{13.0 cm}|}
\hline
$$
\delta\phi\Longleftrightarrow\delta g_{\mu\nu}.
$$
\\
\hline
\end{tabular}
\end{center}
As we will see shortly, this
relation is stronger than one might thought because of the issue
of gauge invariance.

Before launching ourselves into the problem of finding the evolution
of the quantum perturbations of  the inflaton field when they are coupled 
to gravity, 
let us give  a heuristic 
explanation of why we expect that during inflation such fluctuations are
indeed present.

If we take Eq. (\ref{nabla}) and split the  inflaton field as
its classical value $\phi_0$ plus the quantum fluctuation $\delta\phi$,  
$\phi({\bf x},t)=\phi_{0}(t)+\delta\phi({\bf x},t)$, the quantum perturbation
$\delta\phi$ satisfies the equation of motion
\begin{equation}
\delta\ddot{\phi}+3H\,\delta{\dot\phi}-\frac{\na^2\delta\phi}{a^2}
+V''\,
\delta\phi=0.
\label{aa}
\end{equation}
Differentiating  Eq. (\ref{poi}) with respect to  time and taking $H$ constant
(de Sitter expansion)  we find
\begin{equation}
({\phi}_0)^{\cdot\cdot\cdot}+3H\ddot{\phi}_0 +V''\,\dot{\phi}_0=0.
\end{equation}
Let us consider for simplicity the limit $k/a\ll H$ 
and let us disregard
the gradient term. Under this condition we see that
$\dot{\phi}_0$ and $\delta\phi$ solve the same equation. The solutions
have therefore to be related to each other by a constant of proportionality
which depends upon space only, that is 

\be
\label{old}
\delta\phi=-\dot{\phi}_0\,\delta t({\bf x}).
\ee
This tells us that $\phi({\bf x},t)$ will have the form
$$
\fbox{$\displaystyle
\phi({\bf x},t)=\phi_0\left({\bf x},t-\delta t({\bf x})\right).$}
$$
This equation indicates that the inflaton field does not acquire
the same value at a given time $t$ in all the space. On the contrary,
when the inflaton field is rolling down its potential, it acquires
different values from one spatial point ${\bf x}$ to the other. The inflaton
field is not homogeneous and fluctuations are present. These fluctuations, 
in turn, will induce fluctuations in the metric.

\subsection{The metric fluctuations}

The mathematical tool do describe the linear evolution
of the cosmological perturbations is obtained  by 
perturbing at  the first-order the 
FRW metric $g^{(0)}_{\mu\nu}$, 

\begin{equation}
g_{\mu\nu}\quad = \quad  g^{(0)}_{\mu\nu}(t) \,+\, 
\delta g_{\mu\nu}(\mathbf{x},t)\,; \qquad  \delta g_{\mu\nu} \,\ll
\,g^{(0)}_{\mu\nu}\,.
\end{equation}
The metric perturbations can be decomposed according to their spin
with respect to a local rotation of the spatial
coordinates on hypersurfaces of constant time. This leads to

\begin{itemize}

\item {\it scalar perturbations},

\item {\it vector perturbations},

\item{\it tensor perturbations}.

\end{itemize}
Tensor perturbations or gravitational waves 
have spin 2 and are the ``true'' degrees of 
freedom of the gravitational fields  in the sense that they can
exist even in the vacuum. Vector perturbations are spin 1 modes arising from
rotational velocity fields and are also called vorticity modes. Finally,
scalar perturbations have spin 0. 

Let us make  a simple exercise to count how many scalar degrees of freedom
are present. Take a space-time of dimensions $D=n+1$, of which $n$
coordinates are spatial coordinates. The symmetric metric tensor $g_{\mu\nu}$
has $\frac{1}{2}(n+2)(n+1)$ degrees of freedom. We can perform $(n+1)$
coordinate transformations in order to eliminate $(n+1)$ degrees of freedom,
this leaves us with $\frac{1}{2}n(n+1)$ degrees of freedom.
These $\frac{1}{2}n(n+1)$ degrees of freedom contain scalar, vector
and tensor modes. According to Helmholtz's theorem we can always decompose
a vector $U_{i}$ $(i=1,\cdots,n)$ as $U_{i}=\partial_i v +v_i$, where
$v$ is a scalar (usually called potential flow) which is curl-free,
$v_{[i,j]}=0$, 
and $v_i$ is a real vector (usually called vorticity) which is divergence-free,
$\nabla\cdot v=0$. This means that the real vector (vorticity) modes
are $(n-1)$. Furthermore, a generic traceless tensor $\Pi_{ij}$
can always be decomposed as $\Pi_{ij} =\Pi^S_{ij}+\Pi_{ij}^V+
\Pi_{ij}^T$, where $\Pi^S_{ij}=\left(-\frac{k_{i} k_j}{k^2}+
\frac{1}{n}\delta_{ij}\right)\Pi$, $\Pi^V_{ij}=(-i/2k)\left(k_{i}\Pi_j
+k_j\Pi_{i}\right)$ $(k_{i}\Pi_{i}=0)$  and $k_{i}\Pi^T_{ij}=0$. This means that the
true symmetric, traceless and transverse tensor degrees of freedom
are $\frac{1}{2}(n-2)(n+1)$. 

The number of
scalar degrees of freedom are therefore

$$
\frac{1}{2}n(n+1)-(n-1)-\frac{1}{2}(n-2)(n+1)=2,
$$
while the degrees of freedom  
of true vector modes are $(n-1)$ and the number of degrees of freedom of 
true
tensor modes (gravitational waves) are $\frac{1}{2}(n-2)(n+1)$. In four
dimensions $n=3$, meaning that one expects 2 scalar degrees of freedom,
2 vector degrees of freedom and 2 tensor degrees of freedom.
As we shall see, to the 2 scalar degrees of freedom from the 
metric, one has to add an another
one, the inflaton field perturbation $\delta\phi$. However, since
Einstein's equations will tell us that the two scalar degrees of freedom
from the metric are equal during inflation, we expect a total number
of scalar degrees of freedom equal to 2.

At the linear order, the scalar, vector and tensor perturbations evolve
independently (they decouple) and it is therefore possible to analyze
them separately. Vector perturbations are not excited
during inflation because there are no rotational velocity fields
during the inflationary stage. We will analyze the generation
of tensor modes (gravitational waves) in the following. For the
time being we want to focus on the scalar degrees of freedom of the metric.

Considering only the scalar degrees of freedom of the perturbed
metric, the most generic perturbed metric reads

\begin{equation}
g_{\mu\nu}\,=\, a^2 \left(
\begin{array}{c c}
- (1 \,+\, 2\,\Phi) & \partial_i B \\
\partial_i B & \left( 1 \,-\, 2\,\Psi\right)\delta_{ij} \,+\, D_{ij}
E \\
\end{array}
\right),
\end{equation}
while the line-element can be written as 
\begin{equation}
\d s^2 \,=\, a^2 \big[ -(1 + 2\,\Phi)\d\tau^2 \,+\, 2 \,\partial_i B
\,\d\tau\,\d x^i \,+\, \left((1 - 2\,\Psi)\delta_{ij} \,+\,
D_{ij}E\right) \,\d x^i\,\d x^j \big],
\end{equation}
where $D_{ij}\,=\left(\partial_i
\partial_j \,-\, \frac{1}{3}\,\delta_{ij}\,\na^2\right)$.

We now want to determine the inverse $g^{\mu\nu}$ of the metric
at the linear order 
\begin{equation}
\label{inv} g^{\mu\alpha}\,g_{\alpha\nu}\,=\,
\delta^{\mu}_{\nu}.
\end{equation}
We have therefore to solve the equations
\begin{equation}
\label{metricper}
 \left( g^{\mu\alpha}_{(0)} \,+\, 
{g^{\mu\alpha}} \right)\left( g^{(0)}_{\alpha\nu} \,+\, 
{g_{\alpha\nu}} \right) \,=\, \delta^{\mu}_{\nu}\,,
\end{equation}
where $g^{\mu\alpha}_{(0)}$ is simply the unperturbed FRW metric. 
Since 
\begin{equation}
g_{(0)}^{\mu\nu}\,=\, \frac{1}{a^2}\left(
\begin{array}{c c}
-1 & 0 \\
0 & \delta^{ij}
\end{array}
\right),
\end{equation}
we can write in general
\begin{eqnarray}
g^{00}\,&=& \,\frac{1}{a^2} \left( \,-1 \,+\, X \right)\,;\nonumber \\
 g^{0i}\,&=& \,\frac{1}{a^2} \,\partial^i Y \,;\nonumber\\
g^{ij}\,&=& \,\frac{1}{a^2}\,\left( \left( 1 \,+\, 2\,Z
\right)\delta^{ij} \,+
 \,D^{ij} K \right).
 \end{eqnarray}
Plugging these expressions into 
Eq. (\ref{metricper}) we find for   $\mu = \nu = 0$ 
\begin{equation}
( - 1 \,+\, X )( -1 \,- \,2\,\Phi ) \,+\, \partial^i Y \,\partial_i B
\,=\, 1.
\end{equation}
Neglecting the terms $ - \,2\,\Phi \cdot X$ e $\partial^i Y \cdot
\partial_i B$ because they are second-order in the 
perturbations, we find
\begin{equation}
1 \,-\,X \,+\, 2\,\Phi \,=\, 1 \qquad \Rightarrow \qquad X\,=\,
2\,\Phi\,.
\end{equation}
Analogously, 
the components $\mu = 0 ,\, \nu = i$ of Eq. 
(\ref{metricper}) give
\begin{equation}
( - 1 \,+\,2\,\Phi)( \partial_i B ) \,+\, \partial^j Y \left[ ( 1
\,-\, 2\,\Psi ) \delta_{ji} \,+\, D_{ji} E \right] \,=\, 0.
\end{equation}
At the first-order, we obtain 
\begin{equation}
- \partial_i B \,+\, \partial_i Y \,=\, 0 \qquad \Rightarrow
\qquad Y = B \,.
\end{equation}
Finally, the components
$\mu = i$, \, $\nu = j$ give
\begin{equation}
\partial^i B \,\partial_j B \,+\, \left( ( 1 \,+\,2\,Z )\delta^{ik}
\,+\, D^{ik} K \right)\,\left( ( 1 - 2\,\Psi) \delta_{kj} \,+\,
D_{kj} E \right) \,=\, \delta^i_j.
\end{equation}
Neglecting the second-order terms, we obtain
\begin{equation}
 ( 1 \,-\, 2\,\Psi \,+\, 2\,Z )\delta^i_j \,+\, D^i_j E \,+\,
D^i_j K \,=\, \delta^i_j 
 \Rightarrow Z \,=\,\Psi \,; \qquad K\,= \,-\,E\,.
\end{equation}
The metric $g^{\mu\nu}$ finally reads
\begin{equation}
g^{\mu\nu}\,=\, \frac{1}{a^2} \left(
\begin{array}{c c}
-1 \,+\, 2\,\Phi & \partial^i B \\
\partial^i B & ( 1 \,+\, 2\,\Psi )\delta^{ij} \,-\,D^{ij}E
\end{array}
\right).
\end{equation}

\subsection{Perturbed affine connections and Einstein's tensor}
\label{Sec:pp}
In this subsection we provide the reader with the perturbed affine connections
and Einstein's tensor. 
First, let us list the unperturbed affine connections
\begin{equation}
\Gamma^0_{00}\,=\, \Aa, \qquad
\Gamma^i_{0j}\,=\,\Aa\,\delta^i_j, \qquad
\Gamma^0_{ij}\,=\,\Aa\,\delta_{ij},
\end{equation}
\begin{equation}
\Gamma^i_{00}\quad=\quad\Gamma^0_{0i}\quad=\quad\Gamma^i_{jk}\quad=\quad
0 \,.
\end{equation}
The expression for the affine connections in terms
of the metric is 
\begin{equation}
\label{conness} \Gamma^\alpha_{\beta\gamma}\,=\,
\frac{1}{2}\,g^{\alpha\rho}\left( \frac{\partial
g_{\rho\gamma}}{\partial x^{\beta}} \,+\, \frac{\partial
g_{\beta\rho}}{\partial x^{\gamma}} \,-\, \frac{\partial
g_{\beta\gamma}}{\partial x^{\rho}}\right),
\end{equation}
which implies
\begin{eqnarray}
\deu{\Gamma^\alpha_{\beta\gamma}}& \,=\ &
\frac{1}{2}\,\deu{g^{\alpha\rho}}\left( \frac{\partial
g_{\rho\gamma}}{\partial x^{\beta}} \,+\, \frac{\partial
g_{\beta\rho}}{\partial x^{\gamma}} \,-\, \frac{\partial
g_{\beta\gamma}}{\partial x^{\rho}}\right) \nonumber\\
& + &\frac{1}{2}\,g^{\alpha\rho}\left( \frac{\partial
\deu{g_{\rho\gamma}}}{\partial x^{\beta}} \,+\, \frac{\partial
\deu{g_{\beta\rho}}}{\partial x^{\gamma}} \,-\, \frac{\partial
\deu{g_{\beta\gamma}}}{\partial x^{\rho}}\right),
\end{eqnarray}
or in components
\begin{equation}
\deu{\Gamma^0_{00}} \,=\, \Phi^{\prime},
\end{equation}
\begin{equation}
\deu{\Gamma^0_{0i}} \,=\, \partial_i\, \Phi \,+\,
\frac{a^{\prime}}{a}\partial_i\,B ,
\end{equation}
\begin{equation}
\deu{\Gamma^i_{00}} \,=\, \frac{a^{\prime}}{a}\,\partial^i B \,+\,
\partial^i B^{\prime} \,+\, \partial^i \Phi,
\end{equation}
\be
\deu{\Gamma^0_{ij}}=
-\,2\,\frac{a^{\prime}}{a}\,\Phi\,\delta_{ij} \,-\,
\partial_i \partial_j B \,-\,
2\,\frac{a^{\prime}}{a}\,\psi\,\delta_{ij} \nonumber\\
 -\, \Psi^{\prime}\,\delta_{ij} \,-\,
\frac{a^{\prime}}{a}\,D_{ij} E \,+\, \frac{1}{2}\,D_{ij}E^{\prime},
\ee
\begin{equation}
\deu{\Gamma^i_{0j}} \,=\, - \,\Psi^{\prime}\delta_{ij} \,+\,
\frac{1}{2}\,D_{ij}E^{\prime},
\end{equation}
\be
\deu{\Gamma^i_{jk}} = \partial_j \Psi \,\delta_k^i \,-\,
\partial_k \Psi\, \delta_j^i \,+\, \partial^i \Psi \,\delta_{jk}
\,-\, \frac{a^{\prime}}{a}\,\partial^i B
\,\delta_{jk}
   +\, \frac{1}{2}\,\partial_j D^i_k E \,+\, \frac{1}{2}\,\partial_k D^i_j E
  \,-\,
  \frac{1}{2}\,\partial^i D_{jk} E.
\ee
We may now compute the Ricci scalar
defines as 
\begin{equation}
R_{\mu\nu}\,=\, \partial_\alpha\,\Gamma^\alpha_{\mu\nu} \,-\,
\partial_{\mu}\,\Gamma^\alpha_{\nu\alpha} \,+\,
\Gamma^\alpha_{\sigma\alpha}\,\Gamma^\sigma_{\mu\nu} \,-\,
\Gamma^\alpha_{\sigma\nu} \,\Gamma^\sigma_{\mu\alpha}.
\end{equation}
Its variation at the first-order reads
\begin{eqnarray}
\deu{R_{\mu\nu}}\,&
=&\,\partial_\alpha\,\deu{\Gamma^\alpha_{\mu\nu}} \,-\,
\partial_{\mu}\,\deu{\Gamma^\alpha_{\nu\alpha}} \,+\,
\deu{\Gamma^\alpha_{\sigma\alpha}}\,\Gamma^\sigma_{\mu\nu}
\,+\,\Gamma^\alpha_{\sigma\alpha}\,\deu{\Gamma^\sigma_{\mu\nu}} \nonumber\\
&  -&\, \deu{\Gamma^\alpha_{\sigma\nu}}
\,\Gamma^\sigma_{\mu\alpha}\,-\,\Gamma^\alpha_{\sigma\nu}
\,\deu{\Gamma^\sigma_{\mu\alpha}}.
\end{eqnarray}
The background values are given by 
\begin{equation}
R_{00}\,=\,-\,3\,\Ac \,+\,3\,\Ab, \qquad R_{0i}\,=\,0\,
\end{equation}
\begin{equation}
R_{ij}\,=\,\Big( \Ac \,+\,\Ab \Big)\,\delta_{ij},
\end{equation}
which give
\begin{equation}
\deu {R_{00}} = \frac{a^{\prime}}{a}\partial_i\partial^i B +
\partial_i\partial^i B^{\prime} + \partial_i\partial^i \Phi +
3\Psi^{\prime\prime} + 3\frac{a^{\prime}}{a}\Psi^{\prime} +
3\frac{a^{\prime}}{a}\Phi^{\prime},
\end{equation}
\begin{equation}
\deu {R_{0i}} = \frac{a^{\prime\prime}}{a}\partial_i B +
\left(\frac{a^{\prime}}{a}\right)^2\partial_i B +
2\partial_i\Psi^{\prime} + 2\frac{a^{\prime}}{a}\partial_i \Phi +
\frac{1}{2}\partial_k D^K_{i} E^{\prime},
\end{equation}
\begin{eqnarray}
\deu {R_{ij}} &=& \Big(-\frac{a^{\prime}}{a}\Phi^{\prime} -
5\frac{a^{\prime}}{a}\psi^{\prime} - 2\frac{a^{\prime\prime}}{a}\Phi
-2\left(\frac{a^{\prime}}{a}\right)^2\Phi \nonumber\\
& -&2\frac{a^{\prime\prime}}{a}\Psi -
2\left(\frac{a^{\prime}}{a}\right)^2\Psi - \Psi^{\prime\prime} +
\partial_k\partial^k\Psi -
\frac{a^{\prime}}{a}\partial_k\partial^k B \Big) \delta_{ij}\nonumber\\
& -&\partial_i\partial_j B^{\prime} +
\frac{a^{\prime}}{a}D_{ij}E^{\prime} +
\frac{a^{\prime\prime}}{a}D_{ij}E +
\left(\frac{a^{\prime}}{a}\right)^2 D_{ij}E \nonumber\\
& +& \frac{1}{2}D_{ij}E^{\prime\prime} + \partial_i\partial_j \Psi
-
\partial_i\partial_j \Phi - 2\frac{a^{\prime}}{a}\partial_i\partial_j
B\nonumber\\
& +& \frac{1}{2}\partial_k\partial_iD^k_j E +
\frac{1}{2}\partial_k\partial_j D^k_{i} E -
\frac{1}{2}\partial_k\partial^k D_{ij} E,
\end{eqnarray}
The perturbation of the scalar curvature
\begin{equation}
\label{curvat} R \,=\, g^{\mu\alpha}\,R_{\alpha\mu}\,,
\end{equation}
for which the first-order perturbation is
\begin{equation}
\label{r} \deu R \,=\, \deu{g^{\mu\alpha}}\,R_{\alpha\mu} \,+\,
g^{\mu\alpha}\,\deu R_{\alpha\mu}\,.
\end{equation}
The background value is 
\begin{equation}
R \,=\, \frac{6}{a^2}\,\Ac,
\end{equation}
while from Eq.  (\ref{r}) one finds
\begin{eqnarray}
\deu R &=& \frac{1}{a^2} \Big( 
-6\frac{a^{\prime}}{a}\partial_i\partial^i B -
2\partial_i\partial^i B^{\prime} - 2\partial_i\partial^i \Phi
-6\Psi^{\prime\prime}\nonumber\\& -& 6\frac{a^{\prime}}{a}\Phi^{\prime}
-18\frac{a^{\prime}}{a}\Psi^{\prime} -
12\frac{a^{\prime\prime}}{a}\Phi +4\partial_i\partial^i\Psi +
\partial_k\partial^iD^k_{i} E \Big).
\end{eqnarray}
Finally, we may compute the 
perturbations of the Einstein tensor
\begin{equation}
G_{\mu\nu} \,=\, R_{\mu\nu} \,-\, \frac{1}{2}\,g_{\mu\nu}\,R,
\end{equation}
whose background components are  
\begin{equation}
G_{00}\,=\,3\,\Ab, \qquad G_{0i}\,=\,0,\qquad G_{ij}\,=\,\Big(
-\,2\,\Ac \,+\, \Ab \Big)\,\delta_{ij}.
\end{equation}
At first-order, one finds
\begin{equation}
\deu{G_{\mu\nu}}\,=\, \deu{R_{\mu\nu}} \,-\,
\frac{1}{2}\,\deu{g_{\mu\nu}}\,R \,-\,
\frac{1}{2}\,g_{\mu\nu}\,\deu R \,,
\end{equation}
or in components
\begin{equation}
\deu{G_{00}} =
-2\frac{a^{\prime}}{a}\thinspace\partial_i\partial^i\thinspace B -
6\frac{a^{\prime}}{a}\thinspace\Psi^{\prime} +
2\thinspace\partial_i\partial^i\thinspace\Psi +
\frac{1}{2}\thinspace\partial_k\partial^i D^K_{i} E,\\
\end{equation}
\\
\begin{equation}
\deu {G_{0i}} =  -2\frac{a^{\prime\prime}}{a}\thinspace\partial_i
B + \left(\frac{a^{\prime}}{a}\right)^2\thinspace\partial_i B +
2\partial_i\thinspace\Psi^{\prime} +
\frac{1}{2}\partial_k\thinspace D^K_{i} E^{\prime} +
2\thinspace\frac{a^{\prime}}{a}\thinspace\partial_i \Phi , \\
\end{equation}
\\
\begin{eqnarray}
\deu {G_{ij}} &= & \Bigg( 2\frac{a^{\prime}}{a}\thinspace
\Phi^{\prime} + 4\frac{a^{\prime}}{a}\thinspace \Psi^{\prime} +
4\frac{a^{\prime\prime}}{a}\thinspace \Phi
-2\left(\frac{a^{\prime}}{a}\thinspace\right)^2 \Phi \nonumber\\
& +& 4\frac{a^{\prime\prime}}{a}\thinspace\Psi
-2\left(\frac{a^{\prime}}{a}\thinspace\right)^2 \Psi +
2\Psi^{\prime\prime} - \partial_k\partial^k\thinspace\Psi \nonumber\\
& + &2\frac{a^{\prime}}{a}\thinspace\partial_k\partial^k B +
\partial_k\partial^k B^{\prime} + \partial_k\partial^k\Phi
+ \frac{1}{2}\partial_k\partial^m\thinspace D^k_m\, E\Bigg)\delta_{ij}
\nonumber\\
& -&\partial_i\partial_j B^{\prime} +\partial_i\partial_j \Psi
 - \partial_i\partial_j A + \frac{a^{\prime}}{a}\thinspace
 D_{ij}E^{\prime}
 - 2\,\frac{a^{\prime\prime}}{a}\thinspace D_{ij}E\nonumber\\
 & + &\left(\frac{a^{\prime}}{a}\thinspace\right)^2 D_{ij}E +
\frac{1}{2}D_{ij}E^{\prime\prime}+ \frac{1}{2}\partial_k\partial_i
D^k_j E\nonumber\\
&+& \frac{1}{2}\partial^k\partial_j D_{ik}E
-\frac{1}{2}\partial_k\partial^k D_{ij}E - 2 \,\Aa
\partial_i\partial_j B.
\end{eqnarray}
For convenience, we also give the expressions for the perturbations
with one index up and one index down

\begin{eqnarray}
\deu {G^\mu_\nu}\,& =&\,\deu{(g^{\mu\alpha} \, G_{\alpha\nu})} \,
\nonumber\\
&=&
\deu{g^{\mu\alpha}}\,G_{\alpha\nu}\,+\,g^{\mu\alpha}\,\deu{G_{\alpha\nu}},
\end{eqnarray}
or in components
\begin{equation}
 \deu {G_0^0} \,=\frac{1}{a^2}\left[6\,\Ab \Phi \,+\,6\,\Aa\Psi^{\prime}\,+\, 2\,\Aa \La
 B\,-\, 2\,\La \Psi \,-\, \frac{1}{2}\,\partial_k \partial^i
 \,D^K_{i} E\right],
\label{00}
 \end{equation}
 \begin{equation}
 \deu {G^0_i} \,=\,\frac{1}{a^2}\left[ -2\,\Aa \partial_i \Phi \,-\, 2\,\partial_i
 \Psi^{\prime} \,-\, \frac{1}{2}\,\partial_k D^K_{i} E^{\prime}\right],
\label{0i} 
\end{equation}
  \begin{eqnarray}
\deu {G^i_j} &=& \frac{1}{a^2}\left[\Bigg( 2\,\Aa \Phi^{\prime} \,+\, 4\,\Ac \Phi \,-\,
2\,\Ab \Phi \,+\, \La \Phi \,+\, 4\,\Aa \Psi^{\prime} 
\,+\, 2\,\Psi^{\prime\prime} \right. \nonumber\\
& -&\, \La \Psi \,+\, 2\,\Aa \La B \,+\, \La B^{\prime} \,+\,
\frac{1}{2}\partial_k \partial^m D^k_m
E \Bigg)\delta_j^i \nonumber\\
& -& \partial^i\partial_j \Phi \,+\, \partial^i\partial_j \Psi \,-\,
2\,\Aa \partial^i\partial_j B \,-\, \partial^i\partial_j
B^{\prime} \,+\, \Aa D^i_j E^{\prime} \,+\, \frac{1}{2}\,D^i_j
E^{\prime\prime}
\nonumber\\
& +&\left.\, \frac{1}{2}\,\partial_k\partial^i \,D^k_j E \,+\,
\frac{1}{2}\,\partial_k\partial_j \,D^{ik} E \,-\,
\frac{1}{2}\,\partial_k\partial^k \,D^i_j E\right]. 
\label{ij}
\end{eqnarray}

\subsection{Perturbed  stress energy-momentum
tensor}
As we have seen previously, the perturbations of the metric
are induced by the perturbations of the  stress energy-momentum
tensor of the inflaton field
\begin{equation}
T_{\mu\nu}\,=\,\partial_\mu \phi\,\partial_\nu \phi \,-\,
g_{\mu\nu}\left( \frac{1}{2}\,g^{\alpha\beta}\,\partial_\alpha
\phi\,\partial_\beta \phi \,+\, V(\phi)\right)\,,
\end{equation}
whose background values are (we are not going to put the subscript $_0$ any longer for the background quantities)
\begin{eqnarray}
 T_{00}&=&\frac{1}{2}\,{\phi^{\prime}}^2 \,+\, V(\phi)\,a^2,\nonumber\\
 T_{0i}&=&\,0, \nonumber\\
T_{ij}&=&\left( \frac{1}{2}\,{\phi^{\prime}}^2 \,-\,
V(\phi)\,a^2\right)\,\delta_{ij}.
\end{eqnarray}
The perturbed stress energy-momentum tensor reads
\begin{eqnarray}
\deu{T_{\mu\nu}} \,&=&\,\partial_\mu \deu\phi \,\partial_\nu \phi
\,+\,\partial_\mu \phi \,\partial_\nu \deu\phi \,-\,
\deu{g_{\mu\nu}}\left(
\frac{1}{2}\,g^{\alpha\beta}\,\partial_\alpha \phi\,\partial_\beta
\phi \,+\, V(\phi)\right)\nonumber\\
& -&\,g_{\mu\nu}\left(
\frac{1}{2}\deu{g^{\alpha\beta}}\,\partial_\alpha
\phi\,\partial_\beta \phi \,+\, g^{\alpha\beta}\,\partial_\alpha
\deu\phi\,\partial_\beta \phi \,+\, \frac{\partial V}{\partial
\phi}\,\deu\phi \,+\, \frac{\partial V}{\partial
\phi}\deu{\phi}\right).
\end{eqnarray}
In components 
we have
\begin{equation}
\deu {T_{00}} \,=\, \deu{\phi^{\prime}}\thinspace\phi^{\prime}
\,+\, 2\,\Phi\,V(\phi)\,a^2 \,+\, a^2\thinspace\frac{\partial
V}{\partial \phi}\thinspace\deu\phi,
\end{equation}
\begin{equation}
\deu T_{0i} \,=\, \partial_i \thinspace\deu\phi \, \phi^{\prime}
\,+\, \frac{1}{2}\,\partial_i B \,{\phi^{\prime}}^2
 \,-\, \partial_i B \,V(\phi)\,a^2,
\end{equation}
\begin{eqnarray}
\deu T_{ij}  \,&=&\,  \left(\deu{\phi^{\prime}}\,\phi^{\prime}
\,-\, \Phi\thinspace{\phi^{\prime}}^2 \,-\,
a^2\thinspace\frac{\partial
V}{\partial\phi}\thinspace\delta\phi
 \,-\, \Psi \,{\phi^{\prime}}^2 \,+\, 2\,\Psi\,V(\phi)\,a^2 
\right)\delta_{ij}
 \nonumber\\
\,&+&\, \frac{1}{2}\,D_{ij}E\,{\phi^{\prime}}^2 \,-\,
 D_{ij}E\,V(\phi)\,a^2.
\end{eqnarray}
For convenience, we list the mixed components
\begin{eqnarray}
\deu{T^\mu_\nu}\,& =&\, \deu{(g^{\mu\alpha}\,T_{\alpha\nu})} \nonumber\\
& =&\, \deu{g^{\mu\alpha}}\,T_{\alpha\nu} \,+\,
g^{\mu\alpha}\,\deu{T_{\alpha\nu}},
\end{eqnarray}
or
\begin{eqnarray}
  \deu {T^0_0} &=& \Phi\,{\phi^{\prime}}^2 \,-\,
  \deu{\phi^{\prime}}\,\phi^{\prime} 
\,-\, \deu \phi\, \frac{\partial V}{\partial
  \phi}\,a^2,\nonumber\\
  \deu {T^i_0} &=& \partial^i B\,{\phi^{\prime}}^2 \,+\,
  \partial^i \deu\phi\,\phi^{\prime},\nonumber\\
  \deu {T^0_i} &=& -\,\partial^i \deu{\phi}\,\phi^{\prime},
\nonumber\\
  \deu {T^i_j} &=& \left( -\, \Phi\,{\phi^{\prime}}^2 \,+\, \deu{
  \phi^{\prime}}\,\phi^{\prime} \,-\, \deu\phi\, \frac{\partial V}{\partial
  \phi}\,a^2 \right) \delta^i_j. 
  \end{eqnarray}

\subsection{Perturbed Klein-Gordon equation}

The inflaton equation of motion is the Klein-Gordon equation of
a scalar field under the action of its potential $V(\phi)$.
The equation to
perturb is therefore 
\be
\label{kg}
 \frac{1}{\sqrt {- g}}\,\partial_\nu \left( \sqrt{-
g}\,g^{\mu\nu}\,\partial_\nu \phi\right)=\, \frac{\partial V}{\partial \phi},
\ee
which at the zero-th order gives the inflaton equation
of motion 
\begin{equation}
\label{kg_espl} \phi^{\prime\prime}\,+\,
2\,\Aa\,\phi^{\prime}\,=\, -\,\frac{\partial V}{\partial
\phi}\,a^2.
\end{equation}
The variation of Eq. (\ref{kg}) is the sum of four different
contributions corresponding to the variations of
$\frac{1}{\sqrt{- g}}$, $\sqrt{- g}$, $g^{\mu\nu}$ and $\phi$.
For the variation of $g$ we have 
\begin{equation}
\label{gi} \deu g \,=\,g\,g^{\mu\nu}\deu{g_{\nu\mu}},
\end{equation}
which give at the linear order
\begin{eqnarray}
 \deu {\sqrt{-g}}\,&=&\, -\, \frac{\deu g}{2\,\sqrt{- g}},\nonumber \\
 \deu {\frac{1}{\sqrt{- g}}} \,&=&\, \frac{\deu{\sqrt{- g}}}{g}.
\end{eqnarray}
Plugging these results into 
the expression for the variation of 
Eq. (\ref{kg_espl}) 
\begin{eqnarray}
\label{mongo}
\deu \partial_\mu\partial^\mu
\,\phi \,& =& \,-\,\deu{\phi^{\prime\prime}} \,-\, 2\,\Aa
{\deu \phi}^{\prime} \,+\, \La \deu \phi \,+\,
2\,\Phi\,\phi^{\prime\prime}
\,+\, 4\,\Aa \Phi\,\phi^{\prime} \,+\, \Phi^{\prime}\phi^{\prime} \nonumber\\
& +&\, 3\,\Psi^{\prime}\phi^{\prime} \,+\, \La B \,\phi^{\prime} \,\nonumber
\\
& =&\,\deu \phi \,\frac{\partial^2 V}{\partial \phi^2}\,a^2.
\end{eqnarray}
Using Eq.  (\ref{kg_espl}) to write 
\begin{equation}
2\,\Phi\,\phi^{\prime\prime}\,+\, 4 \,\Aa \Phi\phi^{\prime}\,=\,
2\,\Phi\,\frac{\partial V}{\partial \phi}a^2,
\end{equation}
Eq. (\ref{mongo}) becomes
\begin{eqnarray}
\label{kgg}
\,{\deu\phi}^{\prime\prime} \,+\, 2\,\Aa {\deu \phi}^{\prime}
\,&-&\, \La \deu \phi \,-\, \Phi^{\prime}\phi^{\prime} \,-\,
 3\,\Psi^{\prime}\phi^{\prime} \,-\, \La B \,\phi^{\prime} \,\nonumber\\
& =&\,-\deu \phi \,\frac{\partial^2 V}{\partial \phi^2}\,a^2 \,-\,
2\,\Phi\,\frac{\partial V}{\partial \phi}.
\end{eqnarray}
After having computed the perturbations at the linear order
of the Einstein's tensor and of the stress energy-momentum
tensor, we are ready to solve the perturbed Einstein's equations
in order to quantify the inflaton and the metric fluctuations.
We pause, however, for a moment in order to deal with the problem
of gauge invariance.

\subsection{The issue of gauge invariance}

When studying the cosmological density perturbations, 
what we are interested in is following the evolution of a space-time which
is neither homogeneous nor isotropic.  This is done by following
the evolution of the differences between the actual space-time and a
well understood reference space-time.  So we will consider small
perturbations away from the homogeneous, isotropic space-time (see
Fig.\ \ref{tau}).
\begin{figure}[h!]
    \centering
        \includegraphics[width=.45\textwidth]{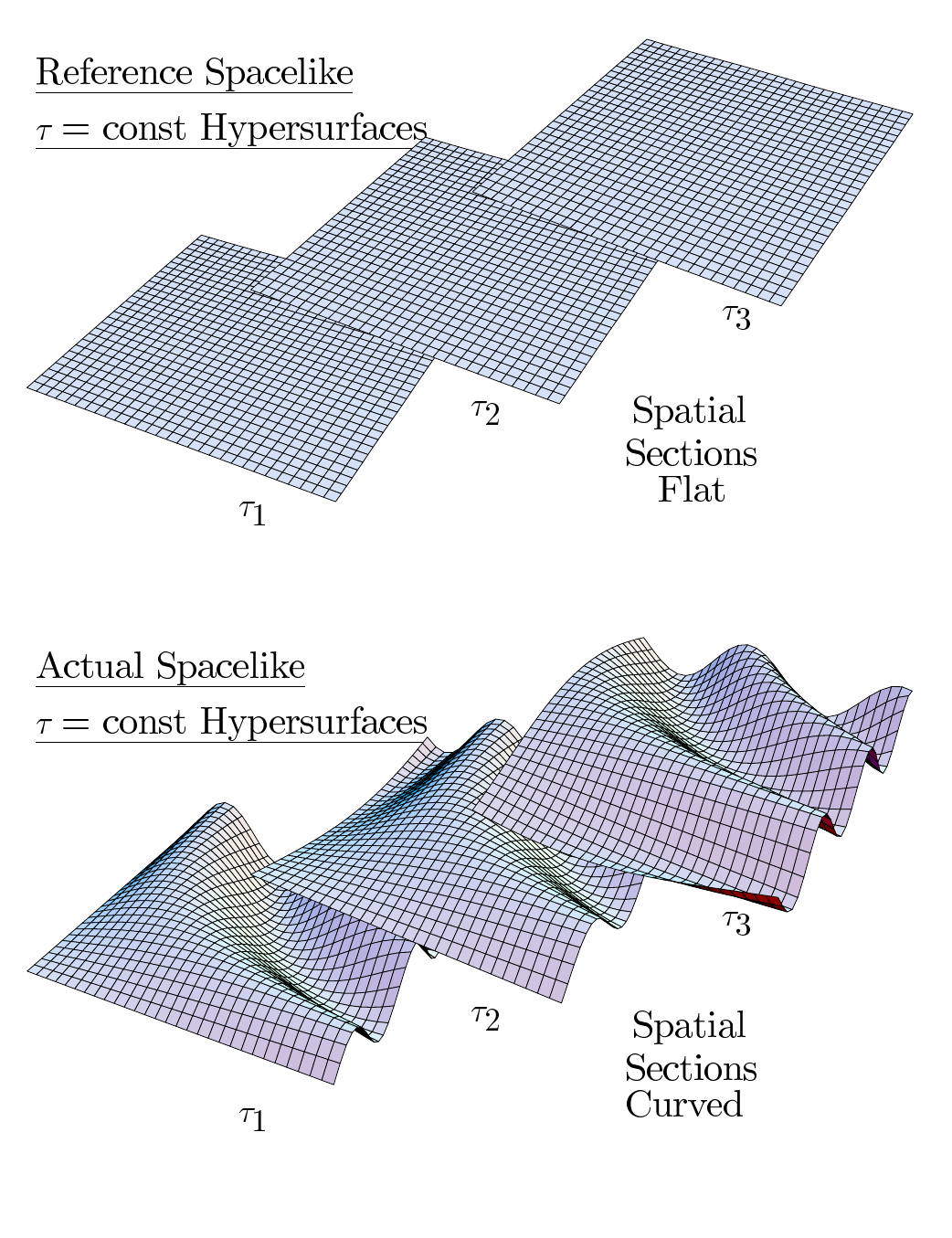}
\caption{In the reference unperturbed universe, constant-time
surfaces have constant spatial curvature (zero for a flat FRW model).
In the actual perturbed universe, constant-time surfaces have
spatially varying spatial curvature. From Ref. \cite{kolbreview}. }
    \label{tau}
\end{figure}
The reference system in our case is the
spatially flat Friedmann--Robertson--Walker  space-time, with line
element $\d s^2 = a^2(\tau) \left( -\d\tau^2 + \d{\bf x}^2\right)$.
Now, the key issue is that general relativity is a gauge theory where
the gauge transformations are the generic coordinate transformations
from a local reference frame to another. 

When we compute the perturbation of a given quantity, this is defined
to be the difference between the value that this quantity assumes on the
real physical  space-time and the value it assumes on the unperturbed 
background. Nonetheless, to perform a comparison between these two
values, it is necessary to compute the at the same space-time point.
Since the two values ``live'' on two different geometries, it is necessary to
specify a map which allows to link univocally the
same point on the two different space-times. This correspondence is called
a gauge choice and changing the map means performing a gauge transformation.

Fixing a gauge in general relativity implies choosing
a coordinate system. A choice of coordinates 
defines a {\it threading} of space-time into 
lines (corresponding to fixed spatial coordinates  ${\bf x}$) and a 
{\it slicing} into hypersurfaces (corresponding to fixed time $\tau$).
A choice of coordinates is called a {\it gauge} and one should
be aware that there is no preferred gauge

\begin{center}
\begin{tabular}{|p{13.0 cm}|}
\hline
$$
{\rm GAUGE ~~CHOICE}~~~\Longleftrightarrow~~~{\rm SLICING ~~AND ~~THREADING}
$$
\\
\hline
\end{tabular}
\end{center}
Similarly,  we can
look at the change of coordinates  either as an active transformation, in which we slightly alter 
the manifold or as a passive transformation, in which we
do not alter  the manifold, all the points remain fixed, and we just change  the coordinate
system. So this is tantamount to a relabeling of the points. From the passive point of view, in which a coordinate transformation represents
a relabeling of the points of the space, one then  compares a quantity, say the metric (or its perturbations),  at a point
$P$ (with coordinates $x^\mu$) with the new metric at the point $P'$ which has the same values
of the new coordinates as the point $P$ had in the old coordinate system, $\widetilde{x^\mu}(P') = x^\mu(P)$.
This is by the way an efficient way to  detect symmetries (isometries if one is concerned with the metric), we only need to consider infinitesimal
coordinate transformations.

From a more formal point of view, operating  an infinitesimal 
 gauge transformation on the coordinates
\begin{equation}
\label{gauge} \widetilde{x^\mu}\,=\, x^\mu \,+\, \delta x^\mu
\end{equation}
implies on a generic quantity $Q$ a transformation on its
perturbation
\begin{equation}
\widetilde{\deu Q}\,=\, \deu Q \,+\,\pounds_{\delta x} \,Q_0\,
\label{formal}
\end{equation}
where $Q_0$ is the value assumed by the quantity $Q$ on the background
and $\pounds_{\delta x}$ is the Lie-derivative of $Q$ along
the vector $\delta x^\mu$. Notice that for a scalar, the Lie derivative is just the ordinary directional derivative (and this
is as it should be since saying that a function has a certain symmetry amounts to the
assertion that its derivative in a particular direction vanishes).

Decomposing in the
usual manner the vector $\delta x^\mu$ 
\begin{eqnarray}
 \delta x^0 \,&=&\, \xi^0(x^\mu)\,; \nonumber\\
\delta x^i \,&=&\, \partial^i \beta(x^\mu) \,+\, v^i(x^\mu) 
\,; \qquad \partial_i v^i
\,=\,0\,,
\end{eqnarray}
we can easily deduce the transformation
law of a scalar quantity $f$ (like the inflaton scalar field $\phi$
and energy density $\rho$). Instead of applying the formal definition
(\ref{formal}), we find the transformation law in an alternative (and 
more pedagogical) way.
We first write $\delta f(x)=f(x)-f_0(x)$,
where $f_0(x)$ is the background value. Under a gauge transformation
we have $\widetilde{\delta f}(\widetilde{x^\mu})=
\widetilde{f}(\widetilde{x^\mu})-\widetilde{f}_0(\widetilde{x^\mu})$.
Since $f$ is a scalar we can write $\widetilde{f}(\widetilde{x^\mu})=f(x^\mu)$
(the value of the scalar function in a given physical point is
the same in all the coordinate system). On the other side, on the
unperturbed background hypersurface the background solution is the same, $\widetilde{f}_0(\widetilde{x^0})=
f_0(\widetilde{x^0})$. We have therefore

\begin{eqnarray}
\widetilde{\delta f}(\widetilde{x^\mu})&=&
\widetilde{f}(\widetilde{x^\mu})-\widetilde{f}_0(\widetilde{x^0})\nonumber\\
&=& f(x^\mu)-f_0(\widetilde{x^0})\nonumber\\
&=&f(x^\mu)-\delta x^0\,\frac{\partial f_0(x^0)}{\partial x^0}-f_0(x^0),\nonumber\\
\end{eqnarray}
from which we finally deduce

\begin{center}
\begin{tabular}{|p{13 cm}|}
\hline
$$
\widetilde{\delta f}=\delta f-f_0^\prime\,\xi^0.
$$
\\
\hline
\end{tabular}
\end{center}
For the spin zero perturbations of  the metric, we can proceed analogously.
We use the following trick. Upon a coordinate transformation
$x^\mu\rightarrow \widetilde{x^\mu}=x^\mu+\delta x^\mu$, the line
element is left invariant, $\d s^2=\widetilde{\d s^2}$. This implies, for instance,
that $a^2(\widetilde{x^0})\left(1+2\widetilde{\Phi}\right)\left(
\d\widetilde{x^0}\right)^2=
a^2(x^0)\left(1+2\Phi\right)(\d x^0)^2$. Since 
$a^2(\widetilde{x^0})\simeq a^2(x^0)+2 a\, a^\prime\,\xi^0$ and
$\d\widetilde{x^0}=\left(1+\xi^{0\prime}\right)\d x^0+
\frac{\partial x^0}{\partial x^i}\,\d x^i$, we obtain 
$1+2 \Phi=1+2\widetilde{\Phi}+2\H\xi^0+2\xi^{0\prime}$. A similar procedure
leads to the following transformation laws

\begin{center}
\begin{tabular}{|p{13 cm}|}
\hline
\begin{eqnarray}
\label{trasf}
\widetilde{\Phi}\,&=&\, \Phi \,-\, \xi^{0\prime} \,-\, \Aa \xi^0\,;\nonumber\\
 \widetilde{B} \,&=&\, B \,+\, \xi^0 \,+\, \beta^{\prime} \nonumber\\
 \widetilde{\Psi} \,&=&\, \Psi \,-\, \frac{1}{3}\,\na^2 \beta \,+\,
\Aa \xi^0 \,;\nonumber\\
  \widetilde{E}\,&=&\, E \,+\, 2\,\beta\,\nonumber.
\end{eqnarray}
\\
\hline
\end{tabular}
\end{center}
The gauge problem stems from the fact that a change of the map (a change
of the coordinate system) implies the variation of the perturbation
of a given quantity which may therefore assume different values (all of
them on a equal footing!) according to the gauge choice. 
To eliminate this ambiguity, one has therefore
a double choice: 

\begin{itemize}

\item Identify those combinations
representing gauge invariant quantities; 

\item choose a given gauge
and perform the calculations in that gauge. 

\end{itemize}
Both options have  advantages and drawbacks. Choosing a gauge may 
render the computation technically simpler with the danger, however, 
of including gauge artifacts, {\it i.e.} gauge freedoms which are
not physical. Performing a gauge-invariant computation may be
technically more involved, but has the advantage of treating only physical
quantities. 

Let us first indicate some gauge-invariant quantities. They are the so-called
gauge invariant potentials or Bardeen's \cite{Bardeen}
\begin{equation}
\Phi_{\rm GI}\,=\, \, \Phi \,-\, \frac{1}{a}\,\left[ \left( -\, B \,+\,
\frac{E^{\prime}}{2}\right) a \right]^{\prime}\,,
\label{bar1}
\end{equation}
\begin{equation}
\Psi_{\rm GI}\,=\, \,\Psi \,+\, \frac{1}{6}\,\na^2\,E \,-\, \Aa \left(
B \,-\, \frac{E^{\prime}}{2}\right)\,.
\label{bar2}
\end{equation}
Analogously, one can define a gauge invariant quantity for the
perturbation of the inflaton field. Since $\phi$ is a scalar field
$\widetilde{\delta \phi}=\left(\delta \phi-\phi^\prime\,\xi^0\right)
$ and therefore 
\begin{center}
\begin{tabular}{|p{13 cm}|}
\hline
$$
\deu{\phi_{\rm GI}} \,=\,  \,\deu{\phi} \,-\,
\phi^{\prime}\left(\frac{E^{\prime}}{2} \,-\, B \right).
$$
\\
\hline
\end{tabular}
\end{center}
is gauge-invariant.
Analogously, one can define a gauge-invariant energy-density perturbation

\begin{center}
\begin{tabular}{|p{13 cm}|}
\hline
$$
\deu{\rho_{{\rm GI}}} \,=\,  \,\deu{\rho} \,-\,
\rho^{\prime}\left(\frac{E^{\prime}}{2} \,-\, B \right).
$$
\\
\hline
\end{tabular}
\end{center}
We now want to pause to introduce in details 
some gauge-invariant quantities
which play  a major role in the computation of the
density perturbations. 
In the following we will be interested only in the 
coordinate  transformations on constant time hypersurfaces
and therefore gauge invariance will be equivalent to independence from  the 
slicing.

\subsection{The comoving curvature perturbation}

The intrinsic spatial curvature on hypersurfaces on constant
conformal  time $\tau$ and for a flat universe is given by

$$
^{(3)}R=\frac{4}{a^2}\na^2\,\Psi.
$$
The quantity $\Psi$ is usually referred to as the {\it curvature perturbation}.
We have seen, however, that the the curvature potential
$\Psi$ is {\it not} gauge invariant, but is defined only on a 
given slicing.
Under a transformation on constant time hypersurfaces
$\tau\rightarrow \tau+\xi^0$ (change of the slicing)

$$
\Psi\rightarrow \Psi+\H\,\xi^0.
$$
We now consider the {\it comoving slicing} which is defined to be the
slicing orthogonal to the world lines of 
comoving observers. The latter are    are free-falling
and the expansion defined by them is isotropic. In practice, what
this means is that there is no flux of energy measured by these
observers, that is $T_{0i}=0$. During inflation this means
that these observers measure $\delta\phi_{\rm com}=0$ since 
$T_{0i}$ goes like $\partial_i\delta\phi({\bf x},\tau)\phi^\prime(\tau)$.

Since $\delta\phi\rightarrow \delta\phi-\phi^\prime\delta\xi^0$ for a 
transformation on constant time hypersurfaces, this means that

$$
\delta\phi\rightarrow\delta\phi_{\rm com}=\delta\phi-\phi^\prime\,\xi^0=0
\Longrightarrow \xi^0=\frac{\delta\phi}{\phi^\prime},
$$
that is $\xi^0=\frac{\delta\phi}{\phi^\prime}$ is the time-displacement
needed to go from a generic slicing with generic $\delta\phi$ to the
comoving slicing where $\delta\phi_{\rm com}=0$.
At the same time the curvature pertubation $\Psi$ transforms
into

$$
\Psi\rightarrow\Psi_{\rm com}= \Psi+\H\,\xi^0=\Psi+
\H\frac{\delta\phi}{\phi^\prime}.
$$
The quantity

\begin{center}
\begin{tabular}{|p{13.0 cm}|}
\hline
$$
{\cal R}=\Psi+
\H\frac{\delta\phi}{\phi^\prime}=\Psi+H\frac{\delta\phi}{\dot\phi}
$$
\\
\hline
\end{tabular}
\end{center}
is the {\it comoving curvature perturbation}. This quantity is gauge invariant
by construction and is related to the gauge-dependent
curvature perturbation $\Psi$ on a generic slicing to the inflaton
perturbation $\delta\phi$ in that gauge. By construction, the meaning of
${\cal R}$ is that it represents the gravitational potential on 
comoving hypersurfaces where $\delta\phi=0$

$$
{\cal R}=\left.\Psi\right|_{\delta\phi=0}.
$$

\subsection{The curvature perturbation on spatial slices of uniform energy 
density}

We now consider the {\it slicing of uniform energy density } which 
is defined to be the
the slicing where there is no perturbation in the energy density,
 $\delta\rho=0$.

Since $\delta\rho\rightarrow \delta\rho-\rho^\prime\,\xi^0$ for a 
transformation on constant time hypersurfaces, this means that

$$
\delta\rho\rightarrow\delta\rho_{\rm unif}=\delta\rho-\rho^\prime\,\xi^0=0
\Longrightarrow \xi^0=\frac{\delta\rho}{\rho^\prime},
$$
that is $\xi^0=\frac{\delta\rho}{\rho^\prime}$ is the time-displacement
needed to go from a generic slicing with generic $\delta\rho$ to the
slicing of uniform energy density where $\delta\rho_{\rm unif}=0$.
At the same time the curvature pertubation $\Psi$ transforms
into

$$
\Psi\rightarrow\Psi_{\rm unif}= \psi+\H\,\xi^0=\Psi+
\H\frac{\delta\rho}{\rho^\prime}.
$$
The quantity

\begin{center}
\begin{tabular}{|p{13.0 cm}|}
\hline
$$
\zeta=\Psi+
\H\frac{\delta\rho}{\rho^\prime}=\Psi+H\frac{\delta\rho}{\dot\rho}
$$
\\
\hline
\end{tabular}
\end{center}
is the {\it curvature perturbation on slices of uniform energy density}. 
This quantity is gauge invariant
by construction and is related to the gauge-dependent
curvature perturbation $\Psi$ on a generic slicing and to the energy density
perturbation $\delta\rho$ in that gauge. By construction, the meaning of
$\zeta$ is that it represents the gravitational potential on 
slices of uniform energy density

$$
\zeta=\left.\Psi\right|_{\delta\rho=0}.
$$
Notice that, using the energy-conservation equation
$\rho^\prime+3\H(\rho+P)=0$, the  curvature perturbation on 
slices of uniform energy density
can be also written as

$$
\zeta=\Psi
-\frac{\delta\rho}{3(\rho+P)}.
$$
During inflation $\rho+P=\dot{\phi}^2$. Furthermore, on super-Hubble scales
from what we have learned in the previous section (and will be rigorously
shown in the following) the inflaton fluctuation $\delta\phi$ is
frozen in and $\delta\dot\phi=($slow roll parameters$)\times H\,\delta\phi$.
This implies that 
$\delta\rho=\dot\phi\delta\dot\phi+V^\prime\delta\phi\simeq
V^\prime\delta\phi\simeq -3H
\dot\phi\delta\phi$, leading to 

$$
\zeta\simeq 
\Psi+\frac{3H\dot\phi}{3\dot{\phi}^2}\delta\phi=\Psi+H\frac{\delta\phi}{\dot\phi}=
{\cal R}~~~~~({\rm ON 
~SUPER-HUBBLE~SCALES})
$$
The comoving curvature pertubation and the curvature perturbation
on uniform energy density slices are equal on super-Hubble scales during inflation.

\subsection{Scalar field perturbations in the spatially flat
gauge}

We now consider the {\it spatially flat gauge}  which 
is defined to be the
the slicing where there is no curvature $\Psi_{\rm flat}=0$.

Since $\psi\rightarrow \Psi+\H\,\xi^0$ for a 
transformation on constant time hypersurfaces, this means that

$$
\Psi\rightarrow\Psi_{\rm flat}= \Psi+\H\,\xi^0=0\Longrightarrow
\xi^0=-\frac{\Psi}{\H},
$$
that is $\xi^0=-\Psi/\H$ is the time-displacement
needed to go from a generic slicing with generic $\Psi$ to the
spatially flat gauge  where $\Psi_{\rm flat}=0$.
At the same time the fluctuation of the inflaton field  transforms
a

$$
\delta\phi\rightarrow\delta\phi -\phi^\prime\,\xi^0=\delta\phi+
\frac{\phi^\prime}{\H}\,\Psi.
$$
The quantity

\begin{center}
\begin{tabular}{|p{13.0 cm}|}
\hline
$$
Q=\delta\phi+
\frac{\phi^\prime}{\H}\,\,\Psi=\delta\phi+\frac{\dot\phi}{H}\Psi\equiv
\frac{\dot\phi}{H}\,{\cal R}
$$
\\
\hline
\end{tabular}
\end{center}
is the inflaton perturbation on spatially flat gauges. This quantity is often referred to as the Sasaki or Mukhanov
variable \cite{q1,q2}.
This quantity is gauge invariant
by construction and is related to the inflaton perturbation
$\delta\phi$ on a generic slicing and to  the curvature perturbation
$\Psi$ in that gauge. By construction, the meaning of
$Q$ is that it represents the inflaton potential on spatially flat 
slices 

$$
Q=\left.\delta\phi\right|_{\Psi=0}.
$$
Notice that $\delta\phi=-\phi^\prime\delta\tau=-\dot\phi\delta t$
on flat slices, where $\delta t=\xi^0$ is the time displacement going from
flat to comoving slices. This relation makes somehow rigorous the expression 
(\ref{old}). Analogously, going from flat to comoving slices one has
${\cal R}=-H\,\delta t$.

\subsection{Adiabatic and isocurvature perturbations}
Let us comment now on the type of perturbation we may have. Arbitrary cosmological perturbations can be decomposed into:

\begin{itemize} 

\item {\it adiabatic or curvature perturbations} are 
along the same trajectory in phase-space
of the  background solution. Given a generic  scalar
quantity $X$, its perturbations can be described 
by a unique perturbation in expansion with respect
to the background

$$
H\,\delta t=H\,\frac{\delta X}{\dot X}~~~~~{\rm FOR ~ EVERY~} X.
$$
In particular, this holds for the energy density and the pressure

$$
\frac{\delta\rho}{\dot \rho}=\frac{\delta P}{\dot P}
$$
which implies that $P=P(\rho)$. This explains why 
they are called adiabatic. They are called curvature perturbations because
a given time displacement $\delta t $ causes the same relative
change $\delta X/\dot X$ for all quantities. In other words the perturbations
is democratically shared by all components of the universe.

\item {\it isocurvature perturbations}  perturb the solution away from  the
background solution

$$
\frac{\delta X}{\dot X}\neq\frac{\delta Y}{\dot Y}~~{\rm FOR ~ SOME~}
X~{\rm AND}~ Y.
$$
One can specify a  generic isocurvature perturbation $\delta X$ by 
giving  its value on uniform-density slices, related to its value on a different
slicing by the gauge-invariant definition 

$$
H\,\left.\frac{\delta X}{\dot X}\right|_{\delta\rho=0}=
H\left(\frac{\delta X}{\dot X}-\frac{\delta \rho}{\dot \rho}\right).
$$
For a set of fluids with energy density $\rho_i$, the isocurvature 
perturbations are conventionally defined by the gauge invariant quantities

$$
\fbox{$\displaystyle
S_{ij}=3H\left(\frac{\delta \rho_i}{\dot{\rho}_i}-
\frac{\delta \rho_j}{\dot{\rho}_j}\right)=3\left(\zeta_i-\zeta_j\right).$}
$$
One simple example of isocurvature perturbations is the baryon-to-photon
ratio 

\be
S=
\delta(n_{\rm b}/n_\gamma)=\frac{\delta n_{\rm b}}{n_{\rm b}}-
\frac{\delta n_\gamma}{n_\gamma}.
\ee

\vskip 0.2cm

{\it 1. \underline{Comment}:}

From the definitions above, it follows that
the cosmological perturbations generated during inflation are of the adiabatic
type {\it if} the inflaton field is the only field driving inflation.
However, if inflation is driven by more than one field, isocurvature 
perturbations are expected to be generated (and they
might even be cross-correlated to the adiabatic ones \cite{b1,b2,b3}). In the following we will give one example of the utility of generating isocurvature perturbations.

{\it 2. \underline{Comment}:} The perturbations generated during inflation
are  nearly {\it Gaussian}, {\it i.e.} the two-point correlation functions
(like the power spectrum) suffice to define all the higher-order
even correlation functions, while the odd correlation functions (such
as the three-point correlation function) vanish. This conclusion is drawn 
by the very same fact that cosmological perturbations are studied {\it 
linearizing} Einstein's and Klein-Gordon equations. This turns out
to be a good approximation because we know that the inflaton potential
needs to be very flat in order to drive inflation and the interaction 
terms in the inflaton potential might be present, but they are small. 
Non-Gaussian features are therefore suppressed since the non-linearities 
of the inflaton potential are suppressed too. The same argument applies
to the metric perturbations; non-linearities appear only at the 
second-order in deviations from the homogeneous background solution and 
are therefore small. This expectation is confirmed
by a direct  computation of the cosmological perturbations 
generated during inflation up to second-order in deviations from the
     homogeneous background solution which fully account for 
the inflaton self-interactions as well as for the second-order 
fluctuations of
the background metric \cite{ac}.
\end{itemize}

\subsection{The next steps}

After all these technicalities, it is useful to rest for a moment and to
go back to  physics. Up to now we have learned that during inflation
quantum fluctuations of the inflaton field are generated and their wavelengths
are stretched on large scales by the rapid expansion of the universe.
We have also seen that the quantum fluctuations of the inflaton field
are in fact impossible to disentangle from the metric perturbations.
This happens not only because they are tightly coupled 
to each other through Einstein's equations, but also because
of the issue of gauge invariance. Take, for instance, the
gauge invariant quantity $
Q=\delta\phi+
\frac{\phi^\prime}{\H}\,\,\Psi$. We can always go to a gauge where
the fluctuation is entirely in the curvature potential $\Psi$, $Q=
\frac{\phi^\prime}{\H}\,\,\Psi$, or entirely in the inflaton
field, $Q=\delta\phi$. However, as we have stressed at the end
of the previous section, once ripples in the curvature 
are frozen in
on super-Hubble scales during inflation, it is in fact gravity that acts
as a messenger communicating to baryons and photons the small
seeds of perturbations once a given scale reenters the horizon
after inflation. This happens thanks to Newtonian physics; 
a small perturbation in the gravitational potential $\Psi$ induces
a small perturbation of the energy density $\rho$ through Poisson's
equation $\na^2\Psi=4\pi G_{\rm N}\delta\rho$. Similarly, 
if perturbations
are adiabatic/curvature  perturbations and, as such, treat democratically
all the components, a ripple in the curvature is communicated
to photons as well, giving rise to a nonvanishing $\delta T/T$.

These considerations make it clear that the next steps of these lectures will
be

\begin{itemize}

\item Compute the curvature perturbation generated during inflation
on super-Hubble scales. As we have seen we can either compute the comoving
curvature perturbation  ${\cal R}$ 
or the curvature on uniform energy density hypersurfaces $\zeta$. They
will tell us about the fluctuations of the gravitational potential.

\item See how the fluctuations of the gravitational
potential are transmitted to photons, baryons and matter in general.

\end{itemize}
We now intend to address the first point. As stressed previously, we are 
free
to follow two alternative roads: either pick up  a gauge and compute
the gauge-invariant curvature in that gauge or perform
a gauge-invariant calculation. We take
both options.

\subsection{Computation of the  curvature perturbation using 
the longitudinal gauge}

The longitudinal (or conformal newtonian) 
gauge is a convenient  gauge to compute the
cosmological perturbations. It is defined by performing a coordinate 
transformation such that $B=E=0$. This leaves behind two 
degrees of freedom in the scalar perturbations, $\Phi$ and $\Psi$.
As we have previously seen, these two degrees of
freedom fully account for the scalar perturbations in the metric.

First of all, we take the non-diagonal part ($i\neq j$) of the  $(ij)$-Einstein equation.
Since the stress energy-momentum tensor does not have
any non-diagonal component (no stress), we have

$$
\partial_i\partial_j\left(\Psi-\Phi\right)=0\Longrightarrow \Psi=\Phi
$$
and we can now work only with one variable, let it be $\Psi$.
The $(0i)$-component of Einstein equation gives 

\begin{equation}
\Psi'+\H\,\Psi=4\pi G_{\rm N}\phi'\,\delta\phi=\epsilon\H^2\frac{\delta\phi}{\phi'},
\label{oo}
\end{equation}
while the $(00)$- and the diagonal part $(ii)$-component  ($i=j$) component of Einstein equations 
give respectively

\begin{eqnarray}
\label{v}
3\H\left(\Psi'+\H \Psi\right)-\na^2\Psi&=&-
4\pi G_{\rm N}\left(\phi'\delta\phi'-{\phi'}^2\Psi+
a^2V^\prime\delta\phi\right),\\
\label{vvv}
\left(2\frac{{a''}}{a}-\left(\frac{{a'}}{a}\right)^2\right)\Psi+
3\H\Psi'+\Psi''&=&4\pi G_{\rm N}\left(\phi'\delta\phi'-{\phi'}^2\Psi
-a^2V^\prime\delta\phi\right).
\end{eqnarray}
If we now use the fact  that $a''/a=(\H'+\H^2)$,  sum 
the two equations above  and 
use  the background Klein-Gordon
equation to eliminate $V^\prime$, we arrive at the equation for the
gravitational potential


\begin{equation}
\label{masterconf}
{\Psi}^{\prime\prime}_{\bf k}+2\left(\H-
\frac{\phi^{\prime\prime}}{\phi^\prime}\right)
\Psi^\prime_{\bf k}+
2\left(\H^\prime-\H\frac{\phi^{\prime\prime}}{\phi^\prime}\right)
\Psi_{\bf k}+k^2\,
\Psi_{\bf k}=0
\end{equation}
and in terms of the slow-roll parameters $\epsilon$ and $\eta$

\begin{equation}
\label{masterslowroll}
{\Psi}^{\prime\prime}_{\bf k}+2\H\left(\eta-\epsilon\right)
\Psi^\prime_{\bf k}+
2\H^2\left(\eta-2\epsilon\right)
\Psi_{\bf k}+k^2\,
\Psi_{\bf k}=0.
\end{equation}
Using the same logic leading to Eq. (\ref{e}),  we can infer that on super-Hubble scales
the gravitational potential $\Psi$ is nearly constant (up to a  mild 
logarithmic
time-dependence proportional to slow-roll parameters), that is
$\dot{\Psi}_{\bf k}\sim($slow-roll parameters$)\times$$\Psi_{\bf k}$.
This is hardly 
surprising, we know that fluctuations are frozen in on super-Hubble scales.

Using Eq. (\ref{oo}), we can therefore relate the fluctuation 
of the gravitational potential $\Psi$ to the
fluctuation of the inflaton field $\delta\phi$ on super-Hubble scales

\begin{equation}
\label{ee}
\Psi_{\bf k}\simeq \epsilon\,H\frac{\delta\phi_{\bf k}}{\dot\phi}~~~
({\rm  ON ~ SUPER-HUBBLE~ SCALES}).
\end{equation}
This gives us the chance to compute the gauge-invariant
comoving curvature perturbation ${\cal R}_{\bf k}$

\begin{equation}
{\cal R}_{\bf k}=\Psi_{\bf k}+H\,\frac{\delta\phi_{\bf k}}{\dot\phi}
=\left(1+\epsilon\right)H
\frac{\delta\phi_{\bf k}}{\dot\phi}\simeq H\frac{\delta\phi_{\bf k}}{\dot\phi}.
\end{equation}
The power spectrum of the the comoving curvature perturbation 
${\cal R}_{\bf k}$ then reads on super-Hubble scales

$$
{\cal P}_{{\cal R}}=\frac{k^3}{2\pi^2}\frac{H^2}{\dot{\phi}^2}\left|
\delta\phi_{\bf k}\right|^2=\frac{k^3}{4\overline{M}_{\rm Pl}^2\epsilon\,\pi^2}
\left|
\delta\phi_{\bf k}\right|^2.
$$
What is left to evaluate is the time evolution of $\delta\phi_{\bf k}$.
To do so, we consider the  perturbed Klein-Gordon equation (\ref{kgg})
in the longitudinal gauge (in cosmic time)

$$
\delta\ddot{\phi}_{\bf k}+3H\delta\dot{\phi}_{\bf k}+\frac{k^2}{a^2}
\delta\phi_{\bf k}+V^{\prime\prime}\delta\phi_{\bf k}=-2\Psi_{\bf k}V^\prime
+4\dot{\Psi}_{\bf k}\dot{\phi}.
$$
Since on super-Hubble scales $\left|4\dot{\Psi}_{\bf k}\dot{\phi}\right|\ll
\left|\Psi_{\bf k}V^\prime\right|$, using Eq. (\ref{ee}) and
the relation $V^\prime\simeq
-3H\dot{\phi}$, 
we can rewrite the
perturbed Klein-Gordon equation on super-Hubble scales as

$$
\delta\dot{\phi}_{\bf k}+3H\delta\dot{\phi}_{\bf k}+\left(V^{\prime\prime}
-6\epsilon H^2\right)\delta\phi_{\bf k}=0.
$$
We now introduce as usual the field $\delta\chi_{\bf k}=\delta\phi_{\bf k}/a$
and go to conformal time $\tau$. The perturbed Klein-Gordon equation on 
super-Hubble scales becomes, using Eq. (\ref{ap}),  

\bea
\delta\chi_{\bf 
k}^{\prime\prime}&-&\frac{1}{\tau^2}\left(\nu^2-\frac{1}{4}
\right)\delta\chi_{\bf k}=0,\nonumber\\
\nu^2&=&\frac{9}{4}+9\epsilon-3\eta, \nonumber\\
\nu&\simeq&\frac{3}{2}-\eta+3\epsilon.
\eea
Using what we have learned in the previous section, we conclude that

$$
\left|\delta\phi_{\bf k}\right|\simeq \frac{H}{\sqrt{2k^3}}
\left(\frac{k}{aH}\right)^{\frac{3}{2}-\nu}
\,\,\,\,
({\rm ON}\,\,{\rm SUPER-HUBBLE}\,\,{\rm SCALES})
$$
which justifies our initial assumption that both the inflaton
perturbation and the gravitational potential are nearly constant
on super-Hubble scale. 

We may now compute the power spectrum of the comoving curvature
perturbation on super-Hubble scales

\begin{center}
\begin{tabular}{|p{15 cm}|}
\hline
$$
{\cal P}_{{\cal 
R}}(k)=\frac{4\pi G_{\rm N}}{\epsilon}\left(\frac{H}{2\pi}\right)^2
\left(\frac{k}{aH}\right)^{n_{{\cal R}}-1}=
\frac{1}{2\overline{M}_{\rm Pl}^2\epsilon}\left(\frac{H}{2\pi}\right)^2
\left(\frac{k}{aH}\right)^{n_{{\cal R}}-1}\equiv A^2_{\cal R}
\left(\frac{k}{aH}\right)^{n_{{\cal R}}-1},
$$
\\
\hline
\end{tabular}
\end{center}
where we have defined the {\it spectral index} $n_{{\cal R}}$ of the comoving
curvature perturbation
as
\begin{center}
\begin{tabular}{|p{13 cm}|}
\hline
$$
n_{{\cal R}}-1=
\frac{\d {\rm ln} \,{\cal P}_{{\cal R}}}{\d {\rm ln} \,k}=3-2\nu=
2\eta-6\epsilon.
$$
\\
\hline
\end{tabular}
\end{center}
\vskip 0.2cm
We conclude that inflation is responsible for the generation of 
adiabatic/curvature perturbations with an almost scale-independent
spectrum.

From the curvature perturbation we can easily deduce the
behavior of the gravitational potential $\Psi_{\bf k}$ from Eq. (\ref{oo}).
The latter is solved by 

$$
\Psi_{\bf k}=\frac{A(k)}{a}+\frac{4\pi G_{\rm N}}{a}\,\int^t\,\d t^\prime\,a(t^\prime)
\,\dot\phi(t^\prime)\,
\delta\phi_{\bf k}(t^\prime)\simeq\frac{A(k)}{a}+\epsilon\,{\cal R}_{\bf k}.
$$
We find that during inflation and on super-Hubble scales the gravitational
potential is the sum of a decreasing function plus a nearly constant in
time piece
proportional to the curvature perturbation. Notice in particular that
in an exact de Sitter stage, that is $\epsilon=0$, the gravitational
potential is not sourced and any initial condition  in the
gravitational potential is washed out as $a^{-1}$ during the inflationary
stage.

\subsection{Computation of the  curvature perturbation using 
the flat gauge}
 We might have computed the spectrum ${\cal P}_{{\cal R}}(k)$ by first
solving the equation for the perturbation $\delta\phi_{\bf k}$ in the flat slice
in a de Sitter stage, with $H=$ constant ($\epsilon=\eta=0$) and
then taking into account the time-evolution of the Hubble rate
and of $\phi$ introducing  the subscript in $H_{\bf k}$ and $\dot{\phi}_{\bf k}$. The 
time variation of the latter  is determined 
by 
\be
\frac{\d {\rm ln} \,\dot{\phi}_{\bf k}}{\d {\rm ln} \,k}=
\left(\frac{\d {\rm ln} \,\dot{\phi}_{\bf k}}{\d t}\right)
\left(\frac{\d t}{\d {\rm ln} \,a}
\right)\left(\frac{\d {\rm ln} \,a}{\d {\rm ln} \,k}\right)=
\frac{\ddot{\phi}_{\bf k}}{\dot\phi_{\bf k}}
\times \frac{1}{H}\times 
1=-\delta=\epsilon-\eta.
\label{zzz}
\ee
Correspondingly, $\dot{\phi}_{\bf k}$ 
is the value of the time derivative of the inflaton field
  when a given wavelength $\sim 
k^{-1}$ crosses
the horizon (from that point on the fluctuations remains
frozen in). The curvature 
perturbation in such an approach would read (remember that in the flat slide the gravitational potential is zero)

$$
{\cal R}_{\bf k}= \frac{H_{\bf k}}{\dot\phi_{\bf k}}\,\delta\phi_{\bf k}\simeq
\frac{H_{\bf k}^2}{2\pi\dot\phi_{\bf k}},
$$
which 
reproduces our previous findings.
Correspondingly

$$
n_{{\cal R}}-1=
\frac{\d {\rm ln} \,{\cal P}_{{\cal R}}}{\d {\rm ln} \,k}=
\frac{\d {\rm ln} \,H_{\bf k}^4}{\d {\rm ln} \, k}-
\frac{\d {\rm ln} \,\dot{\phi}_{\bf k}^2}{\d {\rm ln} \, k}=
-4\epsilon+(2\eta-2\epsilon)=2\eta-6\epsilon.
$$
During inflation the curvature
perturbation is generated on super-Hubble scales with a spectrum which
is nearly scale invariant,  that is  nearly independent from the wavelength
$\lambda=\pi/k$: the amplitude of the 
fluctuation on super-Hubble scales does not (almost) depend upon the 
time at which the fluctuations crosses the Hubble radius  and then becomes frozen
in after a few Hubble times. The small tilt of the power spectrum arises from the fact that
the inflaton field  is massive, giving rise to a nonvanishing $\eta$
 and because 
during inflation the Hubble rate is not exactly constant, but
nearly constant, where `nearly' is quantified by the slow-roll
parameters $\epsilon$.

\subsection{A proof of time-independence of the comoving curvature
perturbation for adiabatic modes: linear level}
From what have found so far, we may conclude that
on super-Hubble scales the comoving curvature perturbation ${\cal R}$
and the uniform-density gauge curvature $\zeta$ satisfy on super-Hubble scales
the relation

$$
{\cal R}_{\bf k}\simeq{\zeta}_{\bf k}.
$$
We now describe an  argument showing that, in the presence of only  the  adiabatic mode, the curvature perturbation remains {\it exactly} constant
on super-Hubble scales.
The  general argument follows 
from energy-momentum conservation \cite{bookLL}. 

Let us consider a generic fluid with energy-momentum tensor $T^{\mu\nu}=(\rho+P)u^\mu u^\nu+g^{\mu\nu}P$. The four-velocity $u^\mu$ is subject to the constraint $u^\mu u_\nu=-1$. Since
it can be decomposed as

\be
u^\mu=\frac{1}{a}\left(\delta^\mu_0+ v^\mu\right),
\ee
we get
\be
v^0=-\Psi.
\ee
Similarly, we obtain
\bea
u_0&=&a(-1-\Phi),\nonumber\\
U_{i}&=&a v_i.
\eea
Notice that, since we will work on super-Hubble scales we have only taken
the gravitational potentials in the metric. The associated perturbation of the 
energy-momentum tensor is

\bea
\delta T^0_0&=&-(\delta\rho+\delta P)+(\bar\rho+\bar P)(1-\Psi)(-1-\Phi)+\delta P\simeq
-\delta\rho,\nonumber\\
\delta T^i_0&\simeq &0,\nonumber\\
\delta T^i_j&=&\delta P\delta^i_j,
\eea
The associated continuity equation

\be
\boldsymbol{\nabla}_\mu T^{\mu}_{\nu}=\partial_\mu T^{\mu}_\nu+\Gamma^\mu_{\,\,\mu\lambda}T^\lambda_\nu-
\Gamma^\lambda_{\,\,\mu\nu}T^\mu_\lambda
\ee
gives

\bea
&&\partial_0T^0_0+\partial_i T^i_0+\Gamma^\mu_{\,\,\mu\lambda}T^\lambda_0-
\Gamma^\lambda_{\,\,\mu 0}T^\mu_\lambda\nonumber\\
&&=\partial_0 T^0_0+ \Gamma^\mu_{\,\,\mu \lambda}T^\lambda_0
-\Gamma^\lambda_{\,\,\mu 0}T^\mu_\lambda\nonumber\\
&&=\partial_0 T^0_0+ \Gamma^\mu_{\,\,\mu 0}T^0_0
-\Gamma^\lambda_{\,\,0 0}T^0_\lambda
-\Gamma^\lambda_{\,\,i 0}T^i_\lambda\nonumber\\
&&=\partial_0 T^0_0+ \Gamma^0_{\,\,0 0}T^0_0
+\Gamma^i_{\,\,i 0}T^0_0-\Gamma^0_{\,\,0 0}T^0_0-
\Gamma^j_{\,\,i 0}T^i_j.
\eea
This expression, using the Christoffel symbols in subsection \ref{Sec:pp} gives

$$
\delta\dot\rho=-3H\left(\delta\rho+\delta P\right)+3\dot\Psi
\left(\bar\rho+\bar P\right).
$$
We
write $\delta P=\delta P_{\rm nad}+c_s^2\delta\rho$, where 
$\delta P_{\rm nad}$ is the non-adiabatic component
of the  perturbation of the pressure and $c_s^2=\delta P_{\rm ad}/\delta\rho$
is the adiabatic one. In the uniform-density gauge 
$\Psi=\zeta$ and $\delta\rho=0$ and therefore $\delta P_{\rm ad}=0$. 
The energy conservation
equation implies
$$
\fbox{$\displaystyle
\dot\zeta=\frac{H}{\bar P+\bar \rho}\,\delta P_{\rm nad}.$}
$$
If perturbations are adiabatic, the curvature on  uniform-density gauge 
is constant on super-Hubble scales

$$
\fbox{$\displaystyle
\dot\zeta=0\,\,{\rm FOR}\,\,{\rm ADIABATIC}\,\,{\rm MODES}\,\,{\rm ON}\,\,{\rm SUPER-HUBBLE}\,\,{\rm SCALES}.$}
$$
The same holds for
the comoving curvature ${\cal R}$ as the latter and $\zeta$ are
equal on super-Hubble scales.
Let us check that this is true, {\it e.g.} in the longitudinal gauge. There we have

\be
{\cal R}_{\bf k}\simeq \frac{H}{\dot\phi}\,\delta\phi_{\bf k}\propto
\frac{H^2}{2\pi\dot\phi}(-\tau)^{\eta-3\epsilon}.
\ee
Therefore 

\bea
\frac{1}{H}\frac{{\rm d}\ln {\cal R}_{\bf k} }{{\rm d}t}&=&\frac{1}{H}\frac{{\rm d}\ln H }{{\rm d}t}-\frac{1}{H}\frac{{\rm d}\ln \dot\phi }{{\rm d}t}
+(\eta-3\epsilon)\frac{1}{H}\frac{{\rm d}\ln (-\tau) }{{\rm d}t}\nonumber\\
&=&-2\epsilon +(\eta-\epsilon)+(\eta-3\epsilon)\left(\frac{-1}{H\tau}\right)\frac{{\rm d} (-\tau) }{{\rm d}t}\nonumber\\
&=&-2\epsilon +(\eta-\epsilon)+(\eta-3\epsilon)\left(\frac{1}{H\tau}\right)\frac{1}{a}\nonumber\\
&=&-2\epsilon +(\eta-\epsilon)+(\eta-3\epsilon)\left(\frac{1}{H\tau}\right)(-H\tau)\nonumber\\
&=&-2\epsilon +(\eta-\epsilon)-(\eta-3\epsilon)\nonumber\\
&=&0.
\eea

\subsection{A proof of time-independence of the comoving curvature
perturbation for adiabatic modes: all orders}

We prove now that the comoving curvature perturbation is conserved at all orders in perturbation theory for adiabatic models on scales larger than the horizon \cite{proof}.  To do so, at momenta $k\ll Ha$ the universe looks like a collection of separate almost homogeneous universes. We choose a threading of spatial
coordinates comoving with the fluid

\be
u^\mu=\frac{\d x^\mu}{\d t},\,\,v^i=\frac{u^i}{u^0}=\frac{\d x^i}{\d t}=0.
\ee
The rate of the expansion is 

\be
\Theta={\boldsymbol \nabla}_\mu
 u^\mu=\frac{1}{{\cal N}}\partial_0\,e^{3\alpha},
 \ee
where $g_{00}={\cal N}^2$, $g_{\ij}=e^{2\alpha}\gamma_{ij}$, with ${\rm det} \gamma_{ij}=1$.
The energy conservation equation

\be
u_\nu{\boldsymbol \nabla}_\mu T^{\mu\nu}=0\Rightarrow
\frac{\d}{\d \tau}\rho+(\rho+P)\Theta=0,
\ee
where $\d t/\d\tau=u^0=1/{\cal N}$. Therefore, we obtain

\be
\dot\rho+3(\rho+P)\dot\alpha=0.
\ee
Upon defining 

\be
a(t)e^{-\Psi}=e^\alpha,
\ee
we obtain
\be
3\left(\frac{\dot a}{a}-\dot\Psi\right)=3\dot\alpha=-\frac{\dot\rho}{\rho+P}.
\ee
This implies that the number of e-folds of expansion along an integral curve of the four-velocity
comoving with the fluid is 

\be
\label{deltaNNN}
N(t_2,t_1,x^i)=\frac{1}{3}\int_{\tau_1}^{\tau_2}\d\tau\,\Theta=
\frac{1}{3}\int_{t_1}^{t_2}\d t\,{\cal N}\,\Theta=-\frac{1}{3}\int_{t_1}^{t_2}\d t\,\left.\frac{\dot\rho}{\rho+P}\right|_{x^i}.
\ee
This implies that

\be
\label{deltaN}
\Psi(t_2,x^i)-\Psi(t_1,x^i)=-N(t_2,t_1,x^i)+\ln\frac{a(t_2)}{a(t_1)},
\ee
that is the change in $\Psi$ from one slice to another equals the difference of the actual number of e-folds and the background. In particular, in a flat slice

\be
N(t_2,t_1,x^i)=\ln\frac{a(t_2)}{a(t_1)},
\ee
From (\ref{deltaN}) we find therefore

\be
-\Psi(t_2,x^i)+\Psi(t_1,x^i)=-\frac{1}{3}\int_{\rho(t_1,x^i)}^{\rho(t_2,x^i)}\,\frac{\d\rho}{\rho+P}
-\ln\frac{a(t_2)}{a(t_1)}.
\ee
If the perturbation are adiabatic, that is if $P=P(\rho)$, then we conclude that

\be
\fbox{$\displaystyle
\zeta(x^i)=\Psi(t, x^i)-\frac{1}{3}\int_{\rho(t)}^{\rho(t,x^i)}\,\frac{\d\rho}{\rho+P}$}
\ee
is constant and this holds at any order in perturbation theory. This is the non-linear generalization of the comoving curvature perturbation.

Consider now two different slices $A$ and $B$ which coincide at $t=t_1$. From (\ref{deltaN}) we have that 

\be
-N_A(t_2,t_1,x^i)+N_B(t_2,t_1,x^i)=\Psi_A(t_2,x^i)-\Psi_B(t_2,x^i).
\ee
Now, choose the slice $A$ such that it is flat at $t=t_1$ and ends on a uniform energy slice at $t=t_2$ and $B$ to be flat both at $t_1$ and $t_2$

\be
\fbox{$\displaystyle
-\Psi_A(t_2,x^i)=N_A(t_2,t_1,x^i)-N_0(t_2,t_1)\equiv \delta N$},
\ee
since $B$ is flat. This means that $-\Psi_A(t_2,x^i)$
is the difference in the number of e-folds (from $t = t_1$ to $t = t_2$) between the uniform-density
slicing and the flat slicing. Therefore, by choosing
the initial slice at the $t_1$ to be the flat slice and the 
  slice at generic time $t$ to have uniform energy density, the curvature
perturbation on that slice  is the difference in the number of e-folds between the uniform energy density slice and the flat slice from $t_1$ to $t$

\be
\fbox{$\displaystyle
-\zeta=\delta N=\delta N(\phi({\bf x},t))\Rightarrow \zeta=\frac{\partial N}{\partial \phi}\delta\phi=
\frac{\partial N}{\partial t}\frac{\delta\phi}{\dot\phi}=H\frac{\delta\phi}{\dot\phi}.$}
\ee
This is indeed the easiest way of computing the comoving curvature perturbation and is dubbed the  $\delta N$ formalism.
In general 

\be
\fbox{$\displaystyle
\zeta(x^i)=-\delta N-\frac{1}{3}\int_{\rho(t)}^{\rho(t,x^i)}\,\frac{\d\rho}{\rho+P}$}
\label{deltaNNNN}
\ee
where $\delta N$ must be interpreted ad the  amount of expansion along the world line of a comoving observer from a spatially
flat $\Psi = 0$ slice at time $t_1$ to a generic slice at time $t$.

\subsection{Gauge-invariant computation of the curvature
perturbation}

In this subsection we would like to show how the computation of the
curvature perturbation can be performed in a gauge-invariant way. 
We first rewrite Einstein's equations in terms of Bardeen's potentials
(\ref{bar1}) and (\ref{bar2})

\begin{eqnarray}
 \deu{G^0_0}\,&=&\,\frac{2}{a^2}\Bigg( -
3\,\mathcal{H}\left(\mathcal{H}\,\Phi_{\rm GI} \,+\, \Psi_{\rm GI} ^{\prime}\right)
\,+\, \na^2 \Psi_{\rm GI}  \,+\, 3\,\mathcal{H}\left( -
\mathcal{H}^{\prime} \,+\,\mathcal{H}^2
\right)\left(\frac{E^{\prime}}{2}\,-\, B  \right)\Bigg)\,,\\
 \deu{G^0_i}\,&=&\, \frac{2}{a^2}\,\partial_i \,\Bigg(
\mathcal{H}\,\Phi_{\rm GI}  \,+\, \Psi_{\rm GI} ^{\prime} \,+\, \left(
\mathcal{H}^{\prime} \,-\, \mathcal{H}^2\right)\left(
\frac{E ^{\prime}}{2}\,-\, B \right)\Bigg)\,, \\
 \deu{G^i_j} \,&=&\, -\, \frac{2}{a^2}\,\Bigg( \left( \left(
2\,\mathcal{H}^{\prime} \,+\, 2\,\mathcal{H}^2\right)\Phi_{\rm GI}  \,+\,
\mathcal{H}\,\Phi_{\rm GI} ^{\prime} \,+\, \Psi_{\rm GI} ^{\prime\prime} \,+\,
2\,\mathcal{H}\,\Psi_{\rm GI} ^{\prime} \,+\, \frac{1}{2}\,\na^2
D_{\rm GI}  \right)\delta^i_j \nonumber\\
&+&\, \left(\mathcal{H}^{\prime\prime} \,-\,
\mathcal{H}\,\mathcal{H}^{\prime} \,-\,
\mathcal{H}^3\right)\left(\frac{E ^{\prime}}{2}\,-\,
B \right)\delta^i_j \,-\, \frac{1}{2}\,\partial^i\partial_j D_{\rm GI} 
\Bigg),
\end{eqnarray}
with $D_{\rm GI}  \,=\Phi_{\rm GI} - \Psi_{\rm GI} $. 
These quantities are not gauge-invariant, but using the 
gauge transformations described previously, we can easily
generalize them to gauge-invariant quantities

\begin{eqnarray}
 \deu{G^{({\rm GI})0}_0}&=&\, \deu{G^0_0} \,+\, (G^0_0)^{\prime}\left(
\frac{E^{\prime}}{2} \,-\, B\right)\,, \\
 \deu{G^{({\rm GI})0}_i}&=&\, \deu{G^0_i} \,+\, \left(G^0_i \,-\, \frac{1}{3}
\,T^k_k \right)\partial_i\, \left( \frac{E^{\prime}}{2} \,-\,
B\right)\,,\\
\deu{G^{({\rm GI})i}_j}&=&\, \deu{G^i_j} \,+\, (G^i_j)^{\prime}\left(
\frac{E^{\prime}}{2} \,-\, B\right)\,
\label{f1}
\end{eqnarray}
and

\begin{eqnarray}
 \deu{T^{({\rm GI})0}_0}&=&\, \deu{T^0_0} \,+\, (T^0_0)^{\prime}\left(
\frac{E^{\prime}}{2} \,-\, B\right)=-\delta\rho^{({\rm GI})}\,, \\
 \deu{T^{({\rm GI})0}_i}&=&\, \deu{T^0_i} \,+\, \left(T^0_i \,-\, \frac{1}{3}
\,T^k_k \right)\partial_i\, \left( \frac{E^{\prime}}{2} \,-\,
B\right)=\left(\bar\rho+\bar P\right)a^{-1} v_i^{({\rm GI})}\,,\\
\deu{T^{({\rm GI})i}_j}&=&\, \deu{T^i_j} \,+\, (T^i_j)^{\prime}\left(
\frac{E^{\prime}}{2} \,-\, B\right)=\delta P^{({\rm GI})},
\label{f2}
\end{eqnarray}
where we have written the stress energy-momentum tensor as $T^{\mu\nu}=
\left(\rho+P\right)u^\mu u^\nu +P\eta^{\mu\nu}$ with $u^\mu=(1,v^i)$. Barred quantities
are to be intended as background quantities. 
Einstein's equations can now be written in a gauge-invariant
way

\begin{eqnarray}
\label{equ}
 &-& 3\,\mathcal{H}\left( \mathcal{H}\,\Phi_{\rm GI} \,+\,
\Psi_{\rm GI}^{\prime}\right) \,+\, \na^2 \,\Psi_{\rm GI}  \\
& =&\, 4\,\pi\,G_{\rm N} \left( - \Phi_{\rm GI} \,{\phi^{\prime}}^2 \,+\,
\deu{\phi^{({\rm GI})}}\,\phi^{\prime} \,+\,
\deu{\phi^{({\rm GI})}}
\,\frac{\partial V}{\partial \phi}\,a^2\right)\,, \nonumber\\
&& \partial_i \,\left( \mathcal{H}\,\Phi \,+\,
\Psi_{\rm GI}^{\prime}\right)\,=\,4\,\pi\,G_{\rm N}
\left(\partial_i\,\deu{\phi^{({\rm GI})}}\,\phi^{\prime}\right)\,,\nonumber\\
&& \left(\left(2\,\mathcal{H}^{\prime}\,+\,\mathcal{H}^2\right)\Phi_{\rm GI}
\,+\, \mathcal{H}\,\Phi_{\rm GI}^{\prime} \,+\, \Psi_{\rm GI}^{\prime\prime} \,+\,
2\,\mathcal{H}\,\Psi_{\rm GI}^{\prime} \,+\, \frac{1}{2}\,\na^2 D_{\rm GI}
\right)\delta^i_j \,-\, \frac{1}{2}\partial^i\partial_j D_{\rm GI} , \nonumber\\
& =&\, -\,4\,\pi\,G_{\rm N} \left( \Phi_{\rm GI}\,{\phi^{\prime}}^2 \,-\,
\deu{\phi^{({\rm GI})}}\,\phi^{\prime} \,+\,
\deu{\phi^{({\rm GI})}}\,\frac{\partial V}{\partial
\phi}\,a^2\right)\delta^i_j\,.
\label{sys}
\end{eqnarray}
Taking $i\neq j$ from the third equation, we find $D_{\rm GI}=0$, that is $\Psi_{\rm GI}=\Phi_{\rm GI}$
and from now on we can work with only the variable $\Phi_{\rm GI}$. Using
the background relation

\begin{equation}
2 \Ab \,-\, \Ac \,=\, 4\,\pi\,G_{\rm N}\,{\phi^{\prime}}^2
\end{equation}
we can rewrite the system of Eqs. (\ref{sys})
in the form 

\begin{eqnarray}
\label{pippo}
 \na^2\,\Phi_{\rm GI} \,-\, 3\,\mathcal{H}\,\Phi_{\rm GI}^{\prime}
\,-\,\left(\mathcal{H}^{\prime} \,+\, 2\,\mathcal{H}^2\right)\Phi_{\rm GI}
\,&=&\,4\,\pi\,G_{\rm N} \left( \deu{\phi^{({\rm GI})}}\,\phi^{\prime} \,+\,
\deu{\phi^{({\rm GI})}}\,\frac{\partial V}{\partial
\phi}\,a^2\right)\,;\nonumber\\
 \Phi_{\rm GI}^{\prime} \,+\, \mathcal{H}\,\Phi_{\rm GI} \,&=&\, \,4\,\pi\,G_{\rm N}
\left(\deu{\phi^{({\rm GI})}}\,\phi^{\prime}\right)\,;\nonumber\\
 \Phi_{\rm GI}^{\prime\prime} \,+\, 3\,\mathcal{H}\,\Phi_{\rm GI}^{\prime} \,+\,
\left( \mathcal{H}^{\prime} \,+\, 2\,\mathcal{H}^2\right)\Phi_{\rm GI}
\,&=&\,4\,\pi\,G_{\rm N} \left(\deu{\phi^{({\rm GI})}}\,\phi^{\prime} \,-\,
\deu{\phi^{({\rm GI})}}\,\frac{\partial V}{\partial \phi}\,a^2\right)\,.
\end{eqnarray}
Subtracting the first equation from the third, using the second 
equation to express $\deu{\phi^{({\rm GI})}}$ as a function
of $\Phi_{\rm GI}$ and $\Phi_{\rm GI}^\prime$ and using the Klein-Gordon equation one finally
finds the 

\begin{equation}
\Phi_{\rm GI}^{\prime\prime}\,+\, 2\left(\mathcal{H}\,-\,
\frac{\phi^{\prime\prime}}{\phi^{\prime}}\right)\Phi_{\rm GI}^{\prime}
\,-\, \na^2 \,\Phi_{\rm GI} \,+\, 2\left(\mathcal{H}^{\prime} \,-\,
\mathcal{H}\,\frac{\phi^{\prime\prime}}{\phi^{\prime}}\right)\Phi_{\rm GI}
\,=\, 0 \,,
\label{nn}
\end{equation}
for the gauge-invariant potential $\Phi_{\rm GI}$.
We now introduce the gauge-invariant quantity

\begin{eqnarray}
u \, &\equiv& \,a\,\deu{\phi^{({\rm GI})}} \,+\, z\,\Psi_{\rm GI} \,,\\
z \, &\equiv& a\,\frac{\phi^{\prime}}{\mathcal{H}}=
a\frac{\dot\phi}{H}.
 \end{eqnarray}
Notice that the variable $u$ is equal to 
$-a\,Q$, the gauge-invariant inflaton perturbation on 
spatially flat gauges. 
Eq. (\ref{nn}) becomes

\begin{equation}
 \label{muk}
 u^{\prime\prime} \,-\, \na^2 \,u \,-\,
 \frac{z^{\prime\prime}}{z}\,u \,=\,0\,,
 \end{equation}
while the two remaining equations of the system (\ref{pippo})
can be written as

\begin{eqnarray}
\label{rest1}
  \na^2 \,\Phi_{\rm GI} \,&=&\, 4\,\pi\,G_{\rm N}\,\frac{\mathcal{H}}{a^2}\left(
 z\,u^{\prime} \,-\, z^{\prime}u \right)\,,\\
  \left( \frac{a^2
 \,\Phi_{\rm GI}}{\mathcal{H}}\right)^{\prime}\,&=&\,4\,\pi\,G_{\rm N}\,z\,u\,,
\label{rest2} 
\end{eqnarray}
which allow to determine the variables $\Phi$ and $\deu{\phi^{({\rm GI})}}$.

We have now to solve Eq. (\ref{muk}). First, we have to evaluate
$z^{\prime\prime}/z$ in terms of the slow-roll
parameters

$$
\frac{z^\prime}{\H z}=\frac{a^\prime}{\H a}+
\frac{\phi^{\prime\prime}}{\H \phi^\prime}-\frac{\H^\prime}{\H^2}=
\epsilon+\frac{\phi^{\prime\prime}}{\H \phi^\prime}.
$$
We then deduce that

$$
\delta\equiv 1-
\frac{\phi^{\prime\prime}}{\H \phi^\prime}=1+\epsilon
-\frac{z^\prime}{\H z}.
$$
Keeping the slow-roll parameters constat in time (as we have mentioned,
this corresponds to expand all quantities to first-order in the
slow-roll parameters), we find

$$
0\simeq \delta^\prime=\epsilon^\prime(\simeq 0)-
\frac{z^{\prime\prime}}{\H z}+\frac{z^\prime\H^\prime}{z\H^2}+
\frac{\left(z^\prime\right)^2}{\H z^2},
$$
from which we deduce

$$
\frac{z^{\prime\prime}}{z}\simeq 
\frac{z^\prime\H^\prime}{z\H}+\frac{\left(z^\prime\right)^2}{z^2}.
$$
Expanding in slow-roll parameters we find

$$
\frac{z^{\prime\prime}}{z}\simeq \left(1+\epsilon-\delta\right)
\left(1-\epsilon\right)\H^2+\left(1+\epsilon-\delta\right)^2\H^2\simeq
\H^2\left(2+2\epsilon-3\delta\right).
$$
If we set

$$
\frac{z^{\prime\prime}}{z}=\frac{1}{\tau^2}\left(\nu^2-\frac{1}{4}\right),
$$
this corresponds to

$$
\nu\simeq \frac{1}{2}\left[1+4
\frac{\left(1+\epsilon-\delta\right)(2-\delta)}{(1-\epsilon)^2}\right]^{1/2}
\simeq\frac{3}{2}+\left(2\epsilon-\delta\right)
\simeq\frac{3}{2}+3\epsilon-\eta.
$$
On sub-Hubble scales $(k\gg aH)$, the solution of equation (\ref{muk})
is obviously $u_{\bf k}\simeq e^{-ik\tau}/\sqrt{2k}$. Rewriting
Eq. (\ref{rest2}) as

$$
\Phi^{\rm GI}_{\bf k}=-\frac{4\pi G_{\rm N} a^2}{k^2}\frac{\dot{\phi}^2}{H}
\left(\frac{H}{a\dot\phi}u_{\bf k}\right)^{\cdot},
$$
we infer that on sub-Hubble scales

$$
\Phi^{\rm GI}_{\bf k}\simeq i\,\frac{4\pi G_{\rm N}\dot\phi}{\sqrt{2 k^3}}\,
e^{-i\frac{k}{a}}.
$$
On super-Hubble scales $(k\ll aH)$, one obvious solution to Eq. (\ref{muk}) 
is $u_{\bf k}
\propto z$. To find the other solution, we may set $u_{\bf k}=z\,
\widetilde{u}_{\bf k}$, which satisfies the equation

$$
\frac{\widetilde{u}_{\bf k}^{\prime\prime}}{
\widetilde{u}_{\bf k}^{\prime}}=-2\frac{z^\prime}{z},
$$
which gives

$$
\widetilde{u}_{\bf k}=\int^\tau\,\frac{\d\tau^\prime}{z^2(\tau^\prime)}.
$$
On super-Hubble scales  therefore we find

$$
u_{\bf k}=c_1(k)\frac{a\dot\phi}{H}+c_2(k)\frac{a\dot\phi}{H}
\int^t\,\d t^\prime\,\frac{H^2}{a^3\dot{\phi}^2}.
\simeq c_1(k)
\frac{a\dot\phi}{H}-c_2(k)\frac{1}{3 a^2\dot\phi},
$$
where the last passage has been performed supposing a de Sitter
epoch, $H=$ constant. 
The first piece is the constant mode
$c_1(k)z$, while the second is the
decreasing mode. To find the constant $c_1(k)$, we 
apply what we have learned previously.  We know that
on super-Hubble scales  
the exact solution of equation (\ref{muk}) is
\be
u_{\bf k}=\frac{\sqrt{\pi}}{2}\,
e^{i\left(\nu+\frac{1}{2}\right)\frac{\pi}{2}}\,
\sqrt{-\tau}\,H_{\nu}(-k\tau).
\label{exactfund}
\ee
On super-Hubble scales, since $H_{\nu}(x\ll 1)\sim
\sqrt{2/\pi}\, e^{-i\frac{\pi}{2}}\,2^{\nu-\frac{3}{2}}\,
(\Gamma(\nu_\chi)/\Gamma(3/2))\, x^{-\nu}$, 
the fluctuation (\ref{exactfund}) becomes
$$
u_{\bf k}=e^{i\left(\nu-\frac{1}{2}\right)\frac{\pi}{2}}
2^{\left(\nu-\frac{3}{2}\right)}\frac{\Gamma(\nu)}{\Gamma(3/2)}
\frac{1}{\sqrt{2k}}\,(-k\tau)^{\frac{1}{2}-\nu}.
$$
Therefore

\be
c_1(k)={\rm lim}_{k\rightarrow 0}\left|\frac{u_{\bf k}}{z}\right|=
\frac{H}{a\dot\phi}\frac{1}{\sqrt{2k}}\left(
\frac{k}{aH}\right)^{\frac{1}{2}-\nu}=
\frac{H}{\dot\phi}\frac{1}{\sqrt{2k^3}}\left(
\frac{k}{aH}\right)^{\eta-3\epsilon}
\label{c1}
\ee
The last steps consist in relating the variable $u$ to the comoving
curvature ${\cal R}$ and to the gravitational potential $\Phi_{\rm GI}$. The comoving
curvature takes the form    

\begin{equation}
\label{curva} \mathcal{R}\,\equiv\,\,\Psi_{\rm GI} +
\frac{H}{\phi^{\prime}}\,\delta\phi^{({\rm GI})}=\frac{u}{z}.
\end{equation}
Since $z=a\dot\phi/H=a\sqrt{2\epsilon}\overline{M}_{\rm Pl}$, 
the power spectrum of the comoving curvature can be expressed on
super-Hubble scales as 
\begin{equation}
\mathcal{P}_\mathcal{R}(k)\,=\,
\frac{k^3}{2\,\pi^2}\,\left| \frac{u_{\bf k}}{z}\right|^2=
\frac{1}{2\overline{M}_{\rm Pl}^2\epsilon}\left(\frac{H}{2\pi}\right)^2
\left(\frac{k}{aH}\right)^{n_{{\cal R}}-1}\equiv A^2_{\cal R}
\left(\frac{k}{aH}\right)^{n_{{\cal R}}-1}
\end{equation}
with 

\be
n_{{\cal R}}-1
=3-2\nu=
2\eta-6\epsilon.
\ee
These results reproduce those found in the previous subsection.
The last step is to find the behavior of the gauge-invariant potential
$\Phi_{\rm GI}$ on super-Hubble scales. If we  recast equation (\ref{rest2}) in the form

\be
u_{\bf k}=\frac{1}{4\pi G_{\rm N}}\frac{H}{\dot\phi}
\left(\frac{a}{H}\Phi^{\rm GI}_{\bf k}\right)^{\cdot},
\label{jjj}
\ee
we can infer that on super-Hubble scales the nearly constant mode
of the gravitational potential during inflation reads

\be
\Phi^{\rm GI}_{\bf k}=c_1(k)\left[1-\frac{H}{a}\,\int^t\,\d t^\prime\,
a\left(t^\prime\right)\right]\simeq -c_1(k)\frac{\dot H}{H^2}=\epsilon\,
c_1(k)\simeq \epsilon\frac{u_{\bf k}}{z}\simeq \epsilon\,{\cal R}_{\bf k}.
\label{c2}
\ee
Indeed, plugging this solution into Eq. (\ref{jjj}), one reproduces
$u_{\bf k}=c_1(k)\frac{a\dot\phi}{H}$.

\section{Gravitational waves}

Quantum fluctuations in the gravitational fields are generated 
in a similar fashion of that of the scalar perturbations 
discussed so far. A gravitational wave
may be viewed as a ripple of space-time in the FRW background metric
 and in general the linear tensor perturbations
may be written as 

$$
\d s^2=a^2(\tau)\left[-\d\tau^2+\left(\delta_{ij}+h_{ij}\right)
\d x^i \d x^j\right],
$$
where $\left|h_{ij}\right|\ll 1$. The tensor $h_{ij}$ has six degrees of 
freedom, but, as we studied previously,
 the tensor perturbations are traceless, $\delta^{ij}h_{ij}=0$,
and transverse $\partial^i h_{ij}=0$ $(i=1,2,3)$. With these
4 constraints, there remain 2 physical degrees of freedom, or 
polarizations, 
 which are
usually indicated 
$\lambda=+,\times$. More precisely, we can write

$$
h_{ij}=h_+\,e_{ij}^+ +h_\times\,e_{ij}^\times,
$$
where $e^+$ and $e^\times$ are the polarization tensors which 
have the following properties

$$
e^\lambda_{ij}=e^\lambda_{ji},~~~ k^i e^\lambda_{ij}=0,~~~e^\lambda_{ii}=0,
$$
$$
e^\lambda_{ij}(-{\bf k})=\left[e^\lambda_{ij}({\bf k})\right]^*,~~~
\sum_{ij}\left(e^\lambda_{ij}\right)^*e^{\lambda'}_{ij}=\delta_{\lambda\lambda'}.
$$
Notice also that
the tensors $h_{ij}$ are gauge-invariant and therefore represent physical
degrees of freedom.

If the stress-energy momentum tensor is diagonal, as the one 
provided by the inflaton potential $T_{\mu\nu}=\partial_\mu\phi
\partial_\nu\phi+g_{\mu\nu}{\cal L}$, the tensor modes do not have any
source and  their action can be written as 

\begin{eqnarray}
S_h&=&\frac{1}{32\pi G_{\rm N}}\,\int\,\d^3x\d\tau\,a^2 \frac{1}{2}\left[\left(h'_{ij}\right)^2-\left(\nabla
h_{ij}\right)^2\right]=\frac{\overline{M}_{\rm Pl}^2}{4}\,\int\,\d^3x\d\tau\,a^2 \frac{1}{2}\left[\left(h'_{ij}\right)^2-\left(\nabla
h_{ij}\right)^2\right],\nonumber\\
&&\end{eqnarray}
that is the action of two (not yet canonically normalized)  independent massless scalar fields. The 
gauge-invariant tensor amplitude of each of these modes is (we drop for the time being the indices ${}_{ij}$)

$$
v_{\bf k}=\frac{a\overline{M}_{\rm Pl}}{2}\, h_{\bf k}
$$
and it satisfies  the equation

$$
v_{\bf k}^{\prime\prime}+\left(k^2-\frac{a^{\prime\prime}}{a}\right)
v_{\bf k}=0,
$$
which is the equation of motion of a massless scalar field
in a quasi-de Sitter epoch. We can therefore make use of the results
learnt previously  to  conclude that
on super-Hubble scales the tensor modes scale like

$$
\left|v_{\bf k}\right|=\left(\frac{H}{2\pi}\right)\left(\frac{k}{aH}\right)
^{\frac{3}{2}-\nu_T},
$$
where 

$$
\nu_T\simeq \frac{3}{2}-\epsilon.
$$
Since fluctuations are (nearly)
frozen in on super-Hubble scales,
a way of characterizing the tensor perturbations is to compute
the spectrum on scales larger than the horizon

\be   
{\cal P}_{T}(k)=\frac{k^3}{2\pi^2}\sum_{ij}\left|
h_{ij\,\bf k}
\right|^2=4\sum_{ij}\frac{k^3}{2\pi^2}\left|v_{ij\,\bf k}\right|^2=4\sum_{ij}\frac{k^3}{2\pi^2}\left(\left|v^+_{\bf k}\right|^2+\left|v^\times_{\bf k}\right|^2\right).
\ee
This gives the power spectrum on super-Hubble scales

\begin{center}
\begin{tabular}{|p{13 cm}|}
\hline
$$
{\cal P}_{T}(k)=\frac{8}{\overline{M}_{\rm Pl}^2}\left(\frac{H}{2\pi}\right)^2
\left(\frac{k}{aH}\right)^{n_T}\equiv A^2_{T}
\left(\frac{k}{aH}\right)^{n_T}
\label{ttt}
$$
\\
\hline
\end{tabular}
\end{center}
where  {\it spectral index} $n_{T}$ is the spectral index of the tensor 
perturbations 

\begin{center}
\begin{tabular}{|p{13 cm}|}
\hline
$$
n_T=
\frac{\d {\rm ln} \,{\cal P}_{T}}{\d {\rm ln} \,k}=3-2\nu_T=
-2\epsilon.
$$
\\
\hline
\end{tabular}
\end{center}
The tensor perturbation is almost scale-invariant. Notice that
the amplitude of the tensor modes depends only on the value
of the Hubble rate during inflation. This amounts
to saying that it depends only on the energy scale $V^{1/4}$
associated to the inflaton potential. A detection of gravitational
waves from inflation is  therefore a direct measurement
of the energy scale associated to inflation.
\subsection{The consistency relation}

The results obtained so far for the scalar and 
tensor perturbations allow to predict a {\it consistency relation}
which holds for the models of inflation addressed in these
lectures, {\it i.e.} the models of inflation driven by
one-single field $\phi$. We define tensor-to-scalar amplitude ratio to be

$$
r=\frac{ A_T^2}{A_{\cal R}^2}=
\frac{ 8 \left(\frac{H}{2\, 
\pi\,\overline{M}_{\rm Pl}}\right)^2}{(2\epsilon)^{-1}\left(
\frac{H}{2\,\pi\,\overline{M}_{\rm Pl}}\right)^2}=16\epsilon.
$$
This means that 

\begin{center}
\begin{tabular}{|p{13 cm}|}
\hline
$$
r=-8 n_T.
$$
\\
\hline
\end{tabular}
\end{center}
One-single models of inflation predict that during inflation
driven by a single scalar field, the ratio between the
amplitude of the tensor modes and that of the curvature perturbations
is equal to minus one-half of the tilt of the spectrum of tensor modes.
If this relation turns out to be
falsified by the future measurements of the CMB anisotropies, this
does not mean that inflation is wrong, but only that
inflation has not been driven by
only one field. Generalizations to two-field models of inflation
can be found for instance in Refs. \cite{b2,b3}.

 Furthermore, using the
consistency relation $r=16\epsilon$ and the definition of $epsilon$,  one deduces that \cite{lyth}

\begin{equation}
\frac{\Delta \phi}{M_{\rm Pl}}\simeq \left(\frac{r}{2\times 10^{-2}}\right)^{1/2},
\end{equation}
meaning that a future measurement of the $B$-mode of CMB polarization will imply an inflaton excursus of Planckian
values. Therefore,  future measurements of the $B$-mode polarization of the CMB will allow a 
determination of the value of the energy scale of inflation. 
This explains the utility of CMB polarization measurements 
as probes of the physics of inflation. They are fundamental to prove or disprove that inflation took place
at high-energy and to see if super-Planckian ranges were probed.

\section{Transferring the perturbation to radiation during reheating}
When the inflaton decays, the comoving curvature perturbation associated to the inflaton field
are transferred to radiation. Let us see how this works \cite{b1}.

Let us consider the system composed by the oscillating scalar field
$\phi$
and the radiation fluid. Each component has energy-momentum tensor 
$T^{\mu\nu}_{(\phi)}$ and $T^{\mu\nu}_{(\gamma)}$. The total energy 
momentum $T^{\mu\nu}=T^{\mu\nu}_{(\phi)}+T^{\mu\nu}_{(\gamma)}$ is 
covariantly conserved, but allowing for an interaction between the two fluids

\begin{eqnarray}
 \label{Qvector}
\boldsymbol{\nabla}_\mu T^{\mu\nu}_{(\phi)}&=&Q^\nu_{(\phi)}\,, \nonumber\\
\boldsymbol{\nabla}_\mu T^{\mu\nu}_{(\gamma)}&=&Q^\nu_{(\gamma)}\,, 
\end{eqnarray}
where $Q^\nu_{(\phi)}$ and $Q^\nu_{(\gamma)}$ are
 the generic energy-momentum transfer to
the scalar field and radiation sector respectively
and are  subject to the constraint
\begin{equation}
\label{Qconstraint}
Q^\nu_{(\phi)}+Q^\nu_{(\gamma)}=0 \,.
\end{equation} 
The energy-momentum transfer $Q^\nu_{(\phi)}$ and 
$Q^\nu_{(\gamma)}$ can be decomposed for convenience as  
\bea
\label{splitQ}
Q^\nu_{(\phi)}&=&\hat{Q}_{\phi}u^{\nu}+f_{(\phi)}^{\nu}
\, , \nonumber \\
Q^\nu_{(\gamma)}&=&\hat{Q}_{\gamma}u^{\nu}+f_{(\gamma)}^{\nu}\, ,
\eea
where the $f^{\nu}$'s are 
required to be orthogonal to the  the total velocity of the fluid  $u^{\nu}$.
The energy continuity equations for the scalar field and radiation 
can be obtained from $u_{\nu}\nabla_\mu T^{\mu\nu}_{(\phi)}=
u_{\nu}Q^{\nu}_{(\phi)}$ and $u_{\nu}\nabla_\mu T^{\mu\nu}_{(\gamma)}=
u_{\nu}Q^{\nu}_{(\gamma)}$ and hence from Eq.~(\ref{splitQ})
\bea
\label{conseq}
u_{\nu}\nabla_\mu T^{\mu\nu}_{(\phi)}&=& \hat{Q}_{\phi} \, ,\nonumber \\
u_{\nu}\nabla_\mu T^{\mu\nu}_{(\gamma)}&=& \hat{Q}_{\gamma}\, .
\eea  
In the case of an oscillating scalar field decaying into radiation 
the energy transfer coefficient $\hat{Q}_\phi$ is given 
by
\bea 
\label{Qreh}
\hat{Q}_\phi&=&-\Gamma \rho_\phi, \nonumber \\
\hat{Q}_\gamma&=&\Gamma \rho_\phi,
\eea
where $\Gamma$ is the decay rate of the scalar field into radiation.
 
The equations of motion for the curvature perturbations $\zeta_\phi$
and  $\zeta_\gamma$ can be obtained perturbing at first order the 
continuity energy 
equations~(\ref{conseq}) for the scalar field and  
radiation energy densities, 
including the energy transfer. Expanding the transfer coefficients 
$\hat{Q}_{\phi}$ and $\hat{Q}_{\gamma}$ up to first 
order in the perturbations around the 
homogeneous background as 
\begin{eqnarray}
\hat{Q}_{\phi}&=& Q_{\phi}+\delta Q_{\phi}\, , \\
\hat{Q}_{\gamma}&=& Q_{\gamma}+\delta Q_{\gamma }\, ,
\end{eqnarray}
Eqs.~(\ref{conseq}) give  on wavelengths larger than the 
horizon scale 

\begin{eqnarray} 
\label{pertenergyexact1}
&&{\delta\rho'}_{\phi}+3\H\left( \delta\rho_{\phi}
+\delta P_{\phi} \right)
- 3 \left( \rho_{\phi}+P_{\phi} \right)\Psi^{'}\nonumber \\ 
&&= a \, Q_{\phi}\Phi+a\,  \delta Q_{\phi}\, , \\
\label{pertenergyexact2}
&&{\delta\rho'}_{\gamma}+3\H\left( 
\delta\rho_{\gamma} +\delta P_{\gamma} \right)
- 3 \left( \rho_{\gamma}+P_{\gamma} \right) \Psi^{'}\nonumber \\
&&= a\, Q_{\gamma}\Phi+a\, \delta Q_{\gamma}\, .
\end{eqnarray}
Notice that the oscillating scalar field and radiation have fixed equations 
of state with $\delta P_\phi = 0$ and $\delta P_\gamma =  
\delta\rho_\gamma/3$ (which correspond to vanishing intrinsic 
non-adiabatic pressure perturbations).
Using the perturbed $(0-0)$-component
of Einstein's equations for super-horizon wavelengths
$\Psi^{'}+\H \Phi=-\H(\delta\rho/\rho)/2$, 
we can 
rewrite Eqs. (\ref{pertenergyexact1}) and~(\ref{pertenergyexact2})
in terms of the gauge-invariant curvature
perturbations $\zeta_\phi$ and $\zeta_\gamma$   
\begin{eqnarray}
\label{eq:zeta1phi}
\zeta^{'}_\phi&=& 
\frac{a\H}{\rho_\phi'}\bigg[\delta Q_\phi-
\frac{Q_\phi'}{\rho_\phi'} \delta \rho_\phi\nonumber\\
&+&Q_\phi \frac{\rho'}{2\rho} 
\bigg( \frac{\delta\rho_\phi}{\rho_\phi'} - 
\frac{\delta\rho}{\rho'} \bigg) \bigg]\, ,\\
\label{eq:zeta1g}  
\zeta^{'}_\gamma&=& 
\frac{a\H}{\rho_\gamma'}\left[\delta Q_\gamma-
\frac{Q_\gamma'}{\rho_\gamma'} \delta \rho_\gamma\right.\nonumber\\ 
&+&\left.Q_\gamma \frac{\rho'}{2\rho} 
\left(\frac{\delta\rho_\gamma}{\rho_\gamma'} -
\frac{\delta\rho}{\rho'} \right) \right] \, ,
\end{eqnarray}
where $\delta Q_{\gamma}=-\delta Q_{\phi}$ from the constraint
in Eq~(\ref{Qconstraint}).
If the energy transfer coefficients $\hat{Q}_\phi$ and 
$\hat{Q}_\gamma$ are given in terms 
of the decay rate $\Gamma$ as in Eq.~(\ref{Qreh}), the first order perturbation 
are respectively
\begin{eqnarray}
\label{Qgamma1}
\delta Q_\phi&=&-\Gamma \delta\rho_\phi, \\
\label{Qgamma2}
\delta Q_\gamma &=&\Gamma \delta\rho_\phi.
\end{eqnarray}   
Plugging the expressions ~(\ref{Qgamma1}-\ref{Qgamma2}) 
into Eqs.~(\ref{eq:zeta1phi}-\ref{eq:zeta1g}), the first order 
curvature perturbations for the scalar field and radiation obey on large 
scales 
\begin{eqnarray}
\label{zeta1phi'}
\zeta^{'}_\phi&=&\frac{a \Gamma}{2} \frac{\rho_\phi}{\rho_\phi'} 
\frac{\rho'}{\rho} \left( \zeta_\phi -\zeta 
\right) ,\\
\label{zeta1g'}
\zeta^{'}_\gamma&=&\frac{a}{\rho_\gamma'} \left[ \Gamma \rho' 
\frac{\rho_\phi'}{\rho_\gamma'}\left(1-\frac{\rho_\phi}{2\rho} 
\right)\left( \zeta -\zeta_\phi 
\right)
\right]. \nonumber \\
&&
\end{eqnarray}
From the total comoving curvature perturbation

\be
\zeta=\frac{\dot{\rho}_\phi}{\dot\rho}\zeta_\phi+\frac{\dot{\rho}_\gamma}{\dot\rho}\zeta_\gamma,
\,\,\rho=\rho_\phi+\rho_\gamma.
\ee
it is thus possible to find the equation of 
motion for the total curvature perturbation $\zeta$ using the evolution 
of the individual curvature perturbations in Eqs.~(\ref{zeta1phi'}) 
and~(\ref{zeta1g'})
\begin{eqnarray}
\label{zeta1'}
\zeta^{'}&=&f'\left( \zeta_\phi- \zeta_\gamma \right)
+f \zeta^{'}_\phi+(1-f)\zeta^{'}_\gamma \nonumber \\
&=& \H f (1-f)\left( \zeta_\phi-\zeta_\gamma \right)=
-\H f \left( \zeta-\zeta_\phi \right)\, ,
\end{eqnarray}
where $f=(\dot{\rho}_\phi/\dot\rho)$.
Notice that during the decay of the scalar field into the radiation fluid, 
$\rho_{\gamma}'$ in Eq.~(\ref{zeta1g'}) may vanish. 
So it is convenient to close the system of equations by using 
the two equations~(\ref{zeta1phi'}) and~(\ref{zeta1'}) for the evolution of 
$\zeta_{\phi}$ and $\zeta$. These equations say that $\zeta=\zeta_\phi$ is a fixed point: during the
reheating phase the comoving curvature perturbation stored in the inflaton field is transferred to radiation smoothly.

\section{Comoving curvature perturbation from isocurvature perturbation}
Let us give one example of how the fact that the comoving curvature perturbation is not constant
when there are isocurvature perturbation can be useful. The paradigm we will describe goes under the name of the curvaton mechanism.

Suppose that during inflation there is another  field $\sigma$, the curvaton, which is supposed to give a 
negligible contribution to the energy density and to be an almost free 
scalar field,  with a small effective mass 
$m^2_\sigma=|\partial^2 V/\partial \sigma^2| \ll H^2$.

The unperturbed curvaton field satisfies the equation of motion
\be
\label{backg}
\sigma''+2 \H \sigma'+a^2 \frac{\partial V}{\partial \sigma}=0.
\ee
It is also usually assumed that the curvaton field is very weakly coupled to 
the scalar fields driving inflation and that the curvature perturbation 
from the inflaton fluctuations is negligible. Thus, if we expand the curvaton field  up to first-order 
in the 
perturbations around the homogeneous background as $\sigma({\bf x},\tau)
=\sigma_0(\tau)+\delta\sigma({\bf x},\tau)$, the 
linear perturbations satisfy on large scales
\be
\label{sigma1}
\delta \sigma ''+2 \H \delta \sigma' +a^2 
\frac{\partial^2 V}{\partial \sigma^2}\, \delta \sigma=0.
\ee
As a result on super-Hubble scales its fluctuations 
$\delta \sigma$ will be Gaussian distributed and with a nearly 
scale-invariant spectrum given by 
\be
\calp_{\delta\sigma}^\frac12(k) \approx \frac{H_*}{2\pi},
\label{pinf}
\ee
where the subscript $*$ denotes the epoch of horizon exit $k=aH$.     
Once inflation is over the inflaton energy density will be converted to 
radiation ($\gamma$) and the curvaton field will remain approximately 
constant until $H^2 \sim m_\sigma^2$. At this epoch the curvaton field begins 
to oscillate around the minimum of its potential which can be safely 
approximated to be quadratic $V \approx \frac{1}{2} m_\sigma^2 \sigma^2$.
During this stage the energy density of the curvaton field just scales as 
non-relativistic matter $\rho_\sigma \propto a^{-3}$. 
The energy density in the oscillating field is
\be
\label{energyoscill}
\rho_\sigma(\tau,{\bf x}) \approx
m_\sigma^2 \sigma^2(\tau,{\bf x}),
\ee
and it can be expanded into a homogeneous background 
$\rho_\sigma(\tau)$ and a first-order perturbation $\delta
\rho_\sigma$ as 
\be
\label{rhocurv1}
\rho_\sigma(\tau,{\bf x})=\rho_\sigma(\tau)+\delta \rho_\sigma(\tau,{\bf x})=
m_\sigma^2 \sigma+2 m_\sigma^2\, \sigma \, \delta \sigma. 
\ee
As it follows from Eqs.~(\ref{backg}) and (\ref{sigma1}) for a 
quadratic potential the ratio $\delta\sigma/\sigma$ remains 
constant and     
the resulting relative energy density perturbation is
\be
\label{relrhocurv}
\frac{\delta\rho_\sigma}{\rho_\sigma}=2 \left(
\frac{\delta \sigma}{\sigma} \right)_* ,
\ee
Such perturbations in the energy density of the 
curvaton field produce in fact a primordial density 
perturbation well after the end of inflation.     
The primordial adiabatic density perturbation is associated with a 
perturbation in the spatial curvature $\Psi$ and it is, as we have shown,  characterized 
in a gauge-invariant manner by the curvature perturbation $\zeta$ on 
hypersurfaces of uniform total density $\rho$. We recall that at linear order the quantity 
$\zeta$ is given by the gauge-invariant formula 
\be
\label{zetatot}
\zeta= \Psi+ \H\frac{\delta\rho}{\rho'},
\ee     
and on large scales it obeys the equation of motion  
\begin{equation}
 \label{zetadot}
 \zeta' = { \H \over \rho +P} \,\delta P_{\rm nad},
\end{equation}
In the curvaton scenario the curvature perturbation is generated well after 
the end of inflation during the oscillations of the curvaton field because 
the pressure of the mixture of matter (curvaton) and radiation produced by 
the inflaton decay is not adiabatic. A convenient way to study this 
mechanism is 
to consider the curvature perturbations $\zeta_i$ associated with each 
individual energy density components, which to linear order are defined as

\begin{equation}
\label{zetai}
\zeta_{i}
 \equiv  \Psi+ \H\left(\frac{\delta\rho_{i}}
{\rho_{i}'}\right).
\end{equation}   
Therefore, during the oscillations of the curvaton field,  
the total curvature perturbation  can be written as a weighted sum of the single 
curvature perturbations 
\be
\label{zetasum}
\zeta=\frac{\dot\rho_\gamma}{\dot\rho}\zeta_\gamma+\frac{\dot\rho_\sigma}{\dot\rho}\zeta_\sigma
=
(1-f)\zeta_\gamma+f\zeta_\sigma,
\ee
where the quantity 
\begin{equation}
\label{deff}
 f = \frac{3\rho_\sigma}{4\rho_\gamma +3\rho_\sigma} 
\end{equation}
defines the relative contribution of the curvaton field 
to the total curvature 
perturbation. From now on we shall work under the approximation of  
sudden decay of the curvaton field. Under this approximation the curvaton and 
the radiation components $\rho_\sigma$ and $\rho_\gamma$ satisfy 
separately the energy conservation equations 
\bea
\label{conseqs}
\rho_\gamma'=-4 \H \rho_\gamma\, ,\nonumber \\
\rho_\sigma'=-3 \H \rho_\sigma,
\eea
and the curvature 
perturbations $\zeta_i$ remains constant on super-Hubble scales until 
the decay of the curvaton. Therefore from Eq.~(\ref{zetasum}) it follows 
that the first-order curvature pertubation evolves on large scales as
\be
\zeta'=f'(\zeta_\sigma-\zeta_\gamma)
=\H f(1-f)(\zeta_\sigma-\zeta_\gamma),
\ee 
and by comparison with Eq.~(\ref{zetadot}) one obtains the expression for the
non-adiabatic pressure perturbation at first order 
\be
\label{pressurepert}
\delta P_{\rm nad}=\rho_\sigma (1-f)(\zeta_\sigma-
\zeta_\gamma) .
\ee 
Since in the curvaton scenario 
it is supposed that the curvature perturbation 
in the radiation produced at the end of inflation is negligible, then
\be
\label{zetagamma}
\zeta_\gamma=\Psi-\frac{1}{4}\frac{\delta \rho_\gamma}
{\rho_\gamma}=0.
\ee
Similarly the value of $\zeta_\sigma$ is fixed by 
the fluctuations of  the curvaton during inflation
\be
\label{zetasigma}
\zeta_\sigma=\Psi-\frac{1}{3}
\frac{\delta\rho_\sigma}{\rho_\sigma}= 
\zeta_{\sigma {\rm I}},
\ee 
where ${\rm I}$ stands for the value of the 
fluctuations during inflation.  
From Eq.~(\ref{zetasum}) the total curvature perturbation 
during the curvaton oscillations is given by  
\be
\label{zetaoscill}
\zeta=f \zeta_\sigma. 
\ee
As it is clear from Eq.~(\ref{zetaoscill}) initially, 
when the curvaton energy density is subdominant, the 
density perturbation in the curvaton field $\zeta_\sigma$ gives a 
negligible contribution to the total curvature perturbation, 
thus corresponding to an isocurvature (or entropy) perturbation. 
On the other hand during the oscillations $\rho_\sigma \propto a^{-3}$ 
increases with respect to the energy density of radiation 
$\rho_\gamma\propto a^{-4}$, and the perturbations in the curvaton field 
are then converted into the curvature perturbation.     
Well after the decay of the curvaton, 
during the conventional radiation and 
matter dominated eras, the total curvature perturbation  
will remain constant on super-Hubble scales at a value which, 
in the sudden decay approximation, is fixed by Eq.~(\ref{zetaoscill}) at 
the epoch of curvaton decay
\be
\label{atcurvdecay}
\zeta=f_{\rm D}\, \zeta_\sigma,
\ee
where ${\rm D}$ stands for the epoch of the curvaton decay.

Going beyond the sudden decay approximation it is possible to introduce a 
transfer parameter $r$ defined as 
\be
\zeta=r\zeta_\sigma,
\ee  
where $\zeta$ is evaluated well after the epoch of the curvaton 
decay and $\zeta_\sigma$ is evaluated well before this epoch.
Numerical studies of the coupled perturbation equations has been performed 
show that the sudden decay approximation is exact when 
the curvaton dominates the energy density before it decays $(r=1)$, while 
in the opposite case   
\be
r\approx \left( \frac{\rho_\sigma}{\rho} \right)_{\rm D}.
\ee

\section{Symmetries of the de Sitter geometry}
\noindent
Before launching ourselves into the computation of the post-inflationary evolution of the cosmological perturbations, we wish to
summarize the symmetries of the de Sitter geometry to understand better the properties of the inflationary perturbations. 
De Sitter space is maximally symmetric and  it posseses 10  isometries, which is the  maximum one can obtain in four dimensions (which is  $\frac{4\cdot 5}{2} =10)$. They are identified by solving the Killing equations
\be
\nabla_{\mu} \epsilon^{(\alpha)}_{\nu} + \nabla_{\nu} \epsilon^{(\alpha)}_{\mu} = 0\,, 
\qquad \mu,\nu = 0, 1, 2, 3\, ;
\quad \alpha = 1, \dots , 10\, .
\label{Kil}
\ee
Each of the independent solutions correspond to transformations $x^{\mu} \rightarrow x^{\mu} + \epsilon^{\mu}(x)$ leaving  the line element of the de Sitter geometry 
 invariant, providing $10$ generators of the de Sitter 
isometry group, the non-compact Lie group SO(4,1).
We write the  Killing Eq. (\ref{Kil}) as follows

\be
g_{\nu\lambda}\pa_\mu \epsilon^\lambda+g_{\mu\lambda}\pa_\nu \epsilon^\lambda+\pa_\sigma g_{\mu\nu}\epsilon^\sigma=0\, ,
\ee
which, for de Sitter space, they provide

\bea
\partial_{t}\epsilon_{t} &=& 0\,,\\
\partial_{t} \epsilon_i + \partial_i\epsilon_{t} &=& 2H \epsilon_i \,,\\
\partial_i \epsilon_j + \partial_j \epsilon_i &=& 2Ha^2 \delta_{ij} \epsilon_{t}\,.
\label{cKv}
\eea
\label{KilldS}
For $ \epsilon_{t} =0$ we have the three translations,
\be
 \epsilon_{t}^{({\rm T}j)} = 0\,,\qquad  \epsilon_i^{({\rm T}j)} = a^2 \delta_i^{\ j}\,,\qquad j = 1, 2, 3\,,
\ee
and the three rotations,
\be
 \epsilon_{\tau}^{({\rm R}\ell)} = 0\,,\qquad  \epsilon_i^{({\rm R}\ell)} = a^2 \epsilon_{i\ell m}x^m\,,\qquad \ell = 1, 2, 3\,.
\ee
The spatial ${\mathbb R}^3$ sections also have four conformal Killing vectors  which satisfy
the Killing vector equations with $ \epsilon_{t} \neq 0$. They are the three special conformal transformations
of ${\mathbb R}^3$,
\be
 \epsilon_{t}^{({\rm C})} = -2H x^n\,,\qquad  \epsilon_i^{({\rm C})} = H^2a^2( \delta_i^{\ n} \delta_{jk}x^jx^k 
- 2 \delta_{ij}x^jx^n)
 - \delta_i^n\,,\qquad n = 1, 2, 3\,,
 \label{speconformal}
 \ee
 and the dilation,
 \be
  \epsilon_{t}^{({\rm D})} = 1\,,\qquad  \epsilon_i^{({\rm D})} = H a^2\,\delta_{ij} x^j\,.
 \ee
Infinitesimally,  the dilation isometry can be written as $\tau\to\lambda\tau$ and $\vx\to \lambda \x$, while special conformal transformations are written as 

\begin{equation}
\label{sc}
\tau\to \tau-2\tau({\bf b}\cdot\vx),\,\,\,\, \vx\to \vx+b^2(-\tau^2+x^2)-2\vx({\bf b}\cdot\x),
\end{equation}
where ${\bf b}$ is  an infinitesimal vector. The point is that for late time correlation functions of scalars  with a mass much smaller than $H$, when the wavelengths are much larger than the Hubble radius,  the transformation of the spatial coordinates in the isometry (\ref{sc}) becomes

\begin{equation}
\label{sc1}
\vx\to \vx+{\bf b} x^2-2\vx({\bf b}\cdot\vx),
\end{equation}
which is a special conformal transformation of the spatial coordinates of $\mathbb{R}^3$ on future constant-time hypersurfaces $\tau=0$. A free scalar $\phi(\tau,\x)$ with mass $m$ on super-Hubble scales evolves as

\begin{equation}
\label{vb}
\phi\sim \tau^\Delta,\,\,\,\,\Delta=\frac{3}{2}\left(1-\sqrt{1-\frac{4}{9}\frac{m^2}{H^2}}\right)
\end{equation}
and the  transformation in (\ref{sc}) acts the same way a conformal transformation would act on a primary field of dimension $\Delta$. The same holds for dilations and thus we conclude that the late time correlation functions, at equal time, are conformal invariant with conformal weight $\Delta$. 
Let us consider dilations.  In Fourier space dilations act on a free massless scalar field $\phi(\vx,\tau)$ on large scales as

\be
\phi_\k\rightarrow\lambda^{-3}\phi_{\k/\lambda}\, .
\ee
Indeed, consider a transformation $\vx\rightarrow \lambda \vx$. Then, in real space $\phi(\vx)\rightarrow
\phi_\lambda(\vx)=\phi(\lambda \vx)$. Expressing this in terms of the Fourier transform of
$\phi(\vx)$ gives how the rescaling acts in Fourier space

\be
\phi(\lambda \vx)=\int\d^3 k \, e^{-i\vk\cdot \lambda\vx}\phi(\vk)=\lambda^{-3}\int\d^3 p \, e^{-i{\bf p}\cdot \vx}\phi({\bf p}/\lambda)\, ,
\ee
where, in the last step, we have made a change in the variable of integration with ${\bf p}=\lambda\vk$. 
Therefore,   the two-point function is constrained to have the form

\be
\langle\phi_{\k_1}\phi_{\k_2}\rangle=(2\pi)^3\delta^{(3)}(\k_1+\k_2)\frac{F(k_1\tau)}{k_1^3}.
\ee
In such a way 

\be
\langle\phi_{\k_1}\phi_{\k_2}\rangle\rightarrow \lambda^{-6}\langle\phi_{{\k_1}/\lambda}\phi_{{\k_2}/\lambda}\rangle=
\lambda^{-6}(2\pi)^3\delta^{(3)}(\k_1/\lambda+\k_2/\lambda)\frac{F(k_1\tau/\lambda)}{(k_1/\lambda)^3}=
(2\pi)^3\delta^{(3)}(\k_1+\k_2)\frac{F(k_1\tau/\lambda)}{k_1^3}.
\ee
If perturbations become time independent when out of
the Hubble radius, the function $F$ must be a constant in
this limit and this gives a scale invariant spectrum.

\part{The post-inflationary evolution of the cosmological perturbations}
The primordial perturbations set up during inflation manifest themselves in the radiation as well as in the matter distribution. By understanding the evolution of the photon perturbations we can make therefore predictions about the expected CMB anisotropy spectrum today. The evolution is determined by Einstein equations and Boltzmann equations. Perturbations to photons
evolved completely different before and after the epoch of last scattering at $z_{\rm ls}\sim 1100$. Before recombination, photons were tightly coupled to electrons and protons; all together
they can be described by a single fluid, dubbed the baryon-photon fluid. After recombination, photons
free-streamed from the surface of last scattering to us today. This means that detecting these photons
is like taking a picture of the universe when it was about 300.000 yr old.

Inflation provides the initial conditions for the  perturbations once the latter re-enter the horizon. Let us turn again to the longitudinal gauge. On super-Hubble scales, from Eq. (\ref{00}) we have

\be
6{\mathcal H}^2\Phi_{\bf k}=-4\pi G_{\rm N}a^2\frac{\delta\rho_{\bf k}}{\bar \rho}\Rightarrow \frac{\delta\rho_{\bf k}}{\bar\rho}=-2\Phi,
\ee
on super-Hubble scales, where $\H^2=(8\pi G_{\rm N}/3)\bar\rho a^2$ defines the average
energy density.
 Recalling now that $\Psi=\Phi$ and that  $\zeta=\Psi+\H\delta\rho/\rho'$ and 
$\rho'=-3\H(\rho+P)=-3\H(1+w)\rho$, where we have defined $P/\rho=w$, we find that
on super-Hubble scales

\be
\zeta_{\bf k}=\Phi_{\bf k}-\frac{\delta\rho_{\bf k}}{3(1+w)\bar\rho}=\left(1+\frac{2}{3(1+w)}\right)\Phi_{\bf k}=\frac{5+3w}{3(1+w)}\Phi_{\bf k}.
\ee
This means that during the RD phase one has ($w=1/3$)

\be
\fbox{$\displaystyle
\Phi^{\rm RD}_{\bf k}=\frac{2}{3}\zeta_{\bf k}\,\,({\rm RD})$},
\ee
and during the MD phase 

\be
\fbox{$\displaystyle
\Phi^{\rm MD}_{\bf k}=\frac{3}{5}\zeta_{\bf k}\,\,({\rm MD})$},
\ee
In particular, notice that 

\be
\fbox{$\displaystyle
\Phi^{\rm MD}_{\bf k}=\frac{9}{10}\Phi^{\rm RD}_{\bf k}$}.
\ee
One of the last steps we wish to take is now fixing the amplitude of the density perturbation
in the CMB through inflation. As we have seen, on large scales and during matter-domination  we have at last scattering

\be
\frac{\delta\rho_{\rm m}}{\bar{\rho}_{\rm m}}=-2\Phi^{\rm MD}(\tau_{\rm ls}),
\ee
and, if the adiabatic condition holds,

\be
\frac{1}{3}\frac{\delta\rho_{\rm m}}{\bar{\rho}_{\rm m}}=\frac{1}{4}\Delta_0(\tau_{\rm ls}),
\ee
where $\Delta=\delta\rho_{\rm r}/\bar{\rho}_{\rm mr}$ and the lower index ${}_0$ stands for the monopole.
The CMB anisotropy has an oscillating structure (the famous Doppler peaks)
because the baryon-photon fluid oscillations. However,  The overall amplitude of the CMB anisotropy
can be fixed at large angular scales (super-Hubble modes) where 
there is no evolution and therefore one can match the amplitude with the theoretical prediction from  inflation.
A standard computation implies that   the observed CMB anisotropy on large scales at the last scattering epoch
should be the Sachs-Wolfe  term \cite{SW,bookD}

\be
\fbox{$\displaystyle
\left(\frac{\Delta}{4}+\Phi^{\rm MD}\right)_{\rm SW}(\tau_{\rm ls})=\left(-\frac{2}{3}+1\right)\Phi^{\rm MD}=\frac{1}{3}\Phi^{\rm MD}(\tau_{\rm ls}).$}
\ee
The 
temperature anisotropy is commonly expanded  in spherical harmonics
\be
\frac{\Delta T}{T}({\bf x}_0,\tau_0,{\bf n})=\sum_{\ell m}
a_{\ell m}({\bf x}_0)Y_{\ell m}({\bf n}),
\ee
where ${\bf x}_0$ and $\tau_0$ are our position and the present time, respectively,
 ${\bf n}$ is the
direction of observation, $\ell'$s are the
different multipoles  and
\be
\langle a_{\ell m}a^*_{\ell'm'}\rangle=\delta_{\ell\ell'}\delta_{mm'} C_\ell,
\ee
where the deltas are due to the fact that the process that created
the anisotropy is statistically isotropic. 
The $C_\ell$ are the so-called CMB power spectrum.
For homogeneity and isotropy, the $C_\ell$'s are neither a function
of ${\bf x}_0$, nor of $m$.
We get therefore that

\be
a_{\ell m}({\bf x}_0,\tau_0)=\int\frac{\d^3 k}{(2\pi)^3}e^{i{\bf k}\cdot{\bf x}_0}\int
\d\Omega\,Y_{\ell m}^*({\bf n})\Theta({\bf k},{\bf n},\tau_0),
\ee
where we have made use the orthonormality property of the spherical harmonics

\be
\int\d\Omega\,Y^*_{\ell m}({\bf n})\,Y_{\ell' m'}({\bf n})=\delta_{\ell \ell'}\delta_{mm'}.
\ee
The $C_\ell$ are given by

\bea
C_\ell&=&\int\frac{\d^3 k}{(2\pi)^3}\int\frac{\d^3 p}{(2\pi)^3}e^{i({\bf k}-{\bf p})\cdot{\bf x}_0}\int
\d\Omega\,Y_{\ell m}^*({\bf n})
\int \d\Omega'\,Y_{\ell m}({\bf n}')\Big<\Theta({\bf k},{\bf n},\tau_0)\Theta^*({\bf p},{\bf n}',\tau_0)\Big>
\nonumber\\
&=&\sum_{\ell'\ell''}(-i)^{\ell'+\ell''}(2\ell'+1)(2\ell''+1)
\int\frac{\d^3 k}{(2\pi)^3}\int\frac{\d^3 p}{(2\pi)^3}e^{i({\bf k}-{\bf p})\cdot{\bf x}_0}\nonumber\\
&\times&\int
\d\Omega\,Y_{\ell m}^*({\bf n})P_{\ell'}({\bf k}\cdot{\bf n})
\int \d\Omega'\,Y_{\ell m}({\bf n}')P_{\ell''}({\bf p}\cdot{\bf n}')
\Big<\Theta_{\ell'}({\bf k})\Theta^*_{\ell''}({\bf p})\Big>.
\eea
where we have decomposed the temperature anisotropy in multipoles as usual

\be
\Theta({\bf k},{\bf n},\tau_0)=\sum_{\ell}(-i)^\ell(2\ell+1)P_\ell({\bf k}\cdot{\bf n})\Theta_\ell(k).
\ee
In the SW limit we have

\be
\Theta^{\rm SW}_\ell({\bf k})\simeq \frac{1}{3}\Phi^{\rm MD}({\bf k},\tau_{\rm ls})j_\ell(k\tau_0),
\ee
with the spectrum of the gravitational potential defined as

\be
\Big<\Phi^{\rm MD}({\bf k},\tau_{\rm ls})\Phi^{\rm MD}({\bf p},\tau_{\rm ls})\Big>=(2\pi)^3\delta^{(3)}({\bf k}+{\bf p})
P_{\Phi^{\rm MD}}(k).
\ee
Therefore we obtain

\bea
C^{\rm SW}_\ell&=&\frac{1}{9}\int\frac{\d^3 k}{(2\pi)^3}P_{\Phi^{\rm MD}}(k)j^2_\ell(k\tau_0)
\nonumber\\
&\times&\sum_{\ell'\ell''}(-i)^{\ell'-\ell''}(2\ell'+1)(2\ell''+1)
\int
\d\Omega\,Y_{\ell m}^*({\bf n})P_{\ell'}({\bf k}\cdot{\bf n})
\int \d\Omega'\,Y_{\ell m}({\bf n}')P_{\ell''}({\bf p}\cdot{\bf n}')\nonumber\\
&=&\frac{1}{9}\int\frac{\d^3 k}{(2\pi)^3}P_{\Phi^{\rm MD}}(k)j^2_\ell(k\tau_0)
\nonumber\\
&\times&\sum_{\ell'\ell''}(-i)^{\ell'-\ell''}(2\ell'+1)(2\ell''+1)
\frac{4\pi}{(2\ell +1)}\delta_{\ell\ell'}Y_{\ell m}({\bf k})\frac{4\pi}{(2\ell +1)}\delta_{\ell\ell''}Y^*_{\ell m}({\bf k})
\nonumber\\
&=&\frac{2}{9\pi}\int\,\d k\,k^2 P_{\Phi^{\rm MD}}(k)j^2_\ell(k\tau_0)
\int\d\Omega_{\bf k}\,\left|Y_{\ell m}({\bf k})\right|^2\nonumber\\
&=&\frac{2}{9\pi}\int\,\ dk\,k^2 P_{\Phi^{\rm MD}}(k)j^2_\ell(k\tau_0),
\eea
where we have made use of the property

\be
P_{\ell}({\bf n}\cdot{\bf n}')=\frac{4\pi}{2\ell +1}\sum_{m=-\ell}^\ell\, Y_{\ell m}({\bf n})Y^*_{\ell m}({\bf n}').
\ee
If  we generically indicate by 

\be
\Big<\left|\Phi^{\rm MD}_{\bf k}\right|^2\Big> k^3=A^2\,(k\tau_0)^{n-1},
\ee 
we can perform the integration knowing that

\be
\int_0^\infty \d x\, j_\ell^2(x) x^{n-2}=2^{n-4}\frac{\Gamma(3-n)\Gamma\left(\frac{2\ell+n-1}{2}\right)}{\Gamma^2\left(\frac{4-n}{2}\right)\Gamma\left(\frac{2\ell+5-n}{2}\right)}.
\ee
We obtain
\begin{equation}
C^{\rm SW}_\ell=\frac{2^{n-3} A^2}{9\pi}\left(\frac{H_0}{2}\right)^{n-1}
\frac{\Gamma(3-n)
\Gamma(\ell+\frac{n-1}{2})}{\Gamma^2\left(\frac{4-n}{2}\right)\Gamma
\left(\ell+
\frac{5-n}{2}\right)}
\end{equation}
For $n\simeq 1$ and $100\gg \ell\gg 1$, we can approximate this expression knowing that $\Gamma(\ell)/\Gamma(\ell+2)=[\ell(\ell+1)]^{-1}$ and $\Gamma(2)/\Gamma^2(3/2)=4/\pi$, and we get

\begin{equation}
\ell(\ell+1)C^{\rm SW}_\ell=
\frac{A^2}{9\pi^2}.
\label{Clad}
\end{equation}
This result shows that inflation predicts
a very flat spectrum  for  low $\ell$. This prediction has
been confirmed by CMB anisotropy measurements.
Furthermore, 
since inflation predicts $\Phi^{\rm MD}_{\bf k}=\frac{3}{5}\zeta_{\bf k}$, we find
that 

\begin{equation}
\ell(\ell+1)C^{\rm SW}_{\ell}=\frac{A_\zeta^2}{25\pi^2}=
\frac{1}{25\pi^2}\frac{1}{2\,\overline{M}_{\rm Pl}^2\,\epsilon}\left(\frac{H}{2\pi}\right)^2.
\end{equation}
Assuming that 
\be
\ell(\ell+1)C^{\rm SW}_\ell\simeq 10^{-10},
\ee
we find
\begin{center}
\begin{tabular}{|p{13 cm}|}
\hline
$$
\left(\frac{V}{\epsilon}\right)^{1/4}\simeq 6.7\times 10^{16}\,{\rm GeV}.
$$
\\
\hline
\end{tabular}
\end{center}
Take for instance a model of chaotic inflation with quadratic 
potential $V(\phi)=\frac{1}{2}m^2\phi^2$. Using Eq. (\ref{togo})
one easily computes that when there are $\Delta N$ e-foldings to go,
the value of the inflaton field is $\phi_{\Delta N}^2=(\Delta N/2\pi G_{\rm N})$
and the corresponding value of $\epsilon$ is $1/(2 \Delta N)$. Taking
$\Delta N\simeq 60$ (corresponding to large-angle CMB anisotropies), one finds
that COBE normalization imposes $m\simeq 10^{13}$ GeV.

\part{Comments on   non-Gaussianity in the cosmological perturbations}
Up to now, we have been describing the cosmological perturbations at the linear level. Is that a correct assumption?
After all we know that gravity is non-linear, so some amount of non-linearities should be expected. This is true both during inflation and after inflation: gravity is always present. Furthermore, the inflaton potential may be characterized by self-interactions.

Non-Gaussianities (NG's) in cosmology are reviewed in Ref. 
\cite{ng} and we wish to illustrate a simple computation illustrating why NG arises due to the gravitational interactions, which are non-linear \cite{by}. We  use the  metric in the Poisson gauge    (we only focus on the scalar degrees of freedom)
\begin{equation}
\label{metric}
\d s^2=
-e^{2\Phi} \d t^2+a^2(t)e^{-2\Psi}\delta_{ij}
\d x^i \d x^j\, .
\end{equation}
 The  choice of adopting  exponentials is a technical and convenient one, as we shall see.
We consider  only perturbations with wavelengths larger than the horizon at the last scattering surface.
 A local
observer perceives  them as classical a 
background.
We  write the gravitational potential $\Phi$ as
\begin{equation}
\Phi=\Phi_\ell +\Phi_s\, ,
\end{equation}
where 

\begin{eqnarray}
\Phi_\ell&=&\int\frac{\d^3\!k}{(2\pi)^3}\, \theta\left(aH-k\right)
\, \Phi_{\k} \ e^{i\vk\cdot\vx} \, ,\nonumber\\
\Phi_s&=&\int\frac{\d^3\!k}{(2\pi)^3}\, \theta\left(k-aH\right)
\, \Phi_{\vk} \ e^{i\vk\cdot\vx} \, ,
\end{eqnarray}
An analogous definition
is adopted  for the gravitational potential $\Psi$. 
For  $\Phi_\ell$ and $\Psi_\ell$  we can neglect  the spatial gradients and can be considered  only as functions of time.
By the redefinition
\begin{eqnarray}
\label{tbar}
\d\overline{t}&=&e^{\Phi_\ell} \d t\, , \\
\label{abar}
\overline{a}&=&a\, e^{-\Psi_\ell}\, ,
\end{eqnarray}
we can therefore absorb the long modes and obtain a new  
  metric  describing a  homogeneous and 
isotropic Universe
\begin{equation}
\label{newmetric}
\d s^2=-\d\overline{t}^2 + \overline{a}^{2} \delta_{ij} \,\d x^i\,\d x^j\, .
\end{equation} 
The  Universe now is a  a collection
of regions of size of the Hubble radius 
evolving like  unperturbed patches. 

Take now a photon moving from an emission point ${\mathcal E}$ to the observer point ${\mathcal O}$.
It suffers a redshift 
given by the ration between the  
frequency $\overline{\omega}_{\mathcal E}$ and  the observed one 
$\overline{\omega}_{\mathcal O}$ 
\begin{equation}
\label{Texact}
\overline{T}_{\mathcal O}=\overline{T}_{\mathcal E}\,
\frac{\overline{\omega}_{\mathcal O}}{\overline{\omega}_{\mathcal E}}\, .
\end{equation}
This expression tells us that 
 large-scale anisotropies  
get  contributions:  the intrinsic 
inhomogeneities at the last scattering
surface  and the anisotropies in the frequencies. 
The photon  
frequency goes like    the inverse of a time period and therefore
\begin{equation}
\label{1result}
\frac{\overline{\omega}_{\mathcal E}}{\overline{\omega}_{\mathcal O}}
=\frac{\omega_{\mathcal E}}{\omega_{\mathcal O}}e^{-\Phi_{\ell
{\mathcal E}}+\Phi_{\ell
{\mathcal O}}}\, .
\end{equation}             
As for the    intrinsic 
 fluctuation, we  relate the photon energy density 
$\overline{\rho}_\gamma$ to the energy density of the non-relativistic matter 
$\overline{\rho}_{\rm m}$ by using the adiabaticity condition. 
 Using the 
energy continuity equation on large scales $\partial \overline{\rho}/  
\partial\overline{t} =-3 \overline{H} (\overline{\rho}+\overline{P})$, where 
$\overline{H}=\d \ln \overline{a}/\d \overline{t}$ 
and $\overline{P}$ is the pressure of the fluid, we have shown previously in the lectures that  
the quantity 
\begin{equation}
-\zeta \equiv \ln \overline{a} +\frac{1}{3}\int^{\overline{\rho}}\, 
\frac{d \overline{\rho'}}{\left(\overline{\rho'}+\overline{P'}\right)}\, 
\label{fun}
\end{equation}
is  conserved in  time at any order in perturbation theory.  
 At the non-linear level
the adiabaticity condition becomes to 

\begin{equation}
\frac{1}{3} \int \frac{\d \overline{\rho}_{\rm m}}{\overline{\rho}_{\rm m}} =
\frac{1}{4} \int  
\frac{\d \overline{\rho}_\gamma}{\overline{\rho}_\gamma}\, ,
\end{equation}
or
\begin{equation}
\ln \overline{\rho}_{\rm m} = \ln \overline{\rho}_\gamma^{3/4}\, .
\label{2result}
\end{equation}
The  (0-0) component of Einstein equations in the 
matter-dominated era with the  
 ``barred'' metric~(\ref{newmetric}) gives 
\begin{equation}
\label{0-0}
\overline{H}^2=\frac{8\pi G_N}{3} \overline{\rho}_{\rm m} \, .
\end{equation}     
Using Eqs.~(\ref{tbar}) and~(\ref{abar}) the 
Hubble parameter $\overline{H}$ reads
\begin{equation}
\overline{H}=\frac{1}{\overline{a}}\frac{\d\overline{a}}{\d \overline{t}}=
e^{-\Phi_\ell}
(H-\dot{\Psi}_\ell)\, ,
\end{equation} 
where $H=\d \ln a/\d t$ is the  Hubble parameter in the
``unbarred'' metric.   We then obtain
\begin{equation}
\label{3result}
\overline{\rho}_{\rm m} =\rho_{\rm m}  e^{-2 \Phi_\ell}\, ,
\end{equation}
having dropped the negligible piece   $\dot{\Psi}_\ell$  on 
large scales. Finally, we obtain
\begin{equation}
\label{4result}
\overline{T}_{\mathcal E}=T_{\mathcal E}\, e^{-2 \Phi_\ell/3}\, ,
\end{equation}
and 
\begin{equation}
\label{final22}
\overline{T}_{\mathcal O}=\left(
\frac{\omega_{\mathcal O}}{\omega_{\mathcal E}}\right) 
T_{\mathcal E}\, e^{\Phi_\ell/3}\, .
\end{equation}
It allows to find the  Sachs-Wolfe effect at all orders
\begin{equation}
\label{final2}
\frac{\delta_{\rm np} \overline{T}_{\mathcal O}}{T_{\mathcal O}}=
e^{\Phi_\ell/3}-1\, .
\end{equation}        
Eq.~(\ref{final2}) 
represents
 the extension of the 
linear Sachs-Wolfe effect. At first order one gets
\begin{equation}
\frac{\delta^{(1)} T_{\mathcal O}}{T_{\mathcal O}}=\frac{1}{3} \Phi^{(1)}\, ,
\end{equation} 
and at  second order 
\begin{equation}
\label{SW2nd}
\frac{1}{2} \frac{\delta^{(2)} T_{\mathcal O}}{T_{\mathcal O}}=
\frac{1}{6} \Phi^{(2)}+\frac{1}{18} \left( \Phi^{(1)} \right)^2\, .
\end{equation}
This result shows that the CMB anisotropies are nonlinear on large scales and that a source of NG is inevitably
sourced by gravity and that the corresponding nonlinearities are order unity in units of the linear gravitational potential.

\section{Primordial non-Gaussianity}
One of the first ways to parameterize non-Gaussianity phenomenologically was via a non-linear correction to a Gaussian primordial perturbation $\zeta_{\rm g}$,
\be
\label{equ:local}
\fbox{$\displaystyle
\zeta({\bf x}) = \zeta_{\rm g}({\bf x}) + \frac{3}{5}f_{\rm NL}^{\rm local}\, \left[\zeta_{\rm g}({\bf x})^2 - \langle \zeta_{\rm g}({\bf x})^2\rangle \right]$}\, .
\ee
This definition is local in real space and therefore called local NG.
Experimental constraints on non-Gaussianity  are often set on the parameter $f_{\rm NL}^{\rm local}$ defined via Eq.~(\ref{equ:local}). The factor of 3/5 in Eq.~(\ref{equ:local}) is historical because NG was initially parametrized by using the  Newtonian potential, $\Phi({\bf x}) = \Phi_{\rm g}({\bf x}) + f_{\rm NL}^{\rm local} \, \left[\Phi_{\rm g}({\bf x})^2 - \langle \Phi_{\rm g}({\bf x})^2\rangle \right]$, which during the matter era is related to $\zeta$ by a factor of 3/5.
Despite the absence of primordial NG in the CMB data \cite{PlanckNG}
\be
f_{\rm NL}^{\rm loc}=0.8\pm 5\,\,\,{\rm at}\,\,\, 68\% \,{\rm CL},
\ee
the detecting some level of  NG is still an important target of many current experiments in cosmology. What size do we expect as far as primordial NG is concerned?

\subsection{Non-Gaussianity in single-field models}
Let us consider now a  period of inflation and  that there are a number of light fields $\sigma^I$
which are quantum mechanically excited. As we have previously seen, 
by the $\delta N$ formalism, the comoving curvature perturbation $\zeta$ 
on a uniform
energy density hypersurface at time $t_{\rm f}$ is, on sufficiently large scales, equal to the perturbation
in the time integral of the local expansion from an initial flat hypersurface ($t = t_{*}$)
to the final uniform energy density hypersurface. On sufficiently large scales, the local expansion
can be approximated quite well by the expansion of the unperturbed Friedmann
universe. Hence the curvature perturbation at time $t_{\rm f}$ can be expressed in terms of  the values of the relevant scalar
fields $\sigma^I(t_{*},\vec{x})$ at $t_{*}$ (notice the change of an irrelevant sign with respect to the previous definition of $\zeta$
(\ref{deltaNNNN}))

\be
\zeta(t_{\rm f},\vec{x})=N_I\sigma^I+\frac{1}{2}N_{IJ}\sigma^I\sigma^J+\cdots\, , \label{zeta}
\ee
where $N_I$ and $N_{IJ}$ are the first and second derivative, respectively, of the number of e-folds 
\be
N(t_{\rm f},t_{*},\vec{x})=\int_{t_{*}}^{t_{\rm f}}\,{\rm d}t\, H(t,\vec{x})\, .
\ee
with respect to the 
field $\sigma^I$. From the expansion (\ref{zeta}) one can read off the $n$-point correlators. 
For instance, the three-point correlator of the comoving curvature perturbation, the so-called bispectrum and trispectrum respectively,  is given by
\be
B_\zeta(k_1,k_2,k_3)=
N_I N_J N_K B^{IJK}_{k_1 k_2 k_3}+ N_I N_{JK}N_{L}\left(P^{IK}_{{k}_1}P^{JL}_{{k}_2}+2\,\,{\rm permutations} \label{zeta3}
\right)
\ee
Let us consider single-field inflation. 
%
Using the fact that $N_\phi=H/\dot\phi$, one gets that 

\be
\frac{N_{\phi\phi}}{N_\phi^2}= \frac{1}{N_\phi^2}\frac{\d }{\dot \phi\, \d t} N_\phi=\left(\frac{\dot H}{\dot\phi}-\frac{H\ddot\phi}{\dot\phi^2}\right)\times\frac{1}{\dot\phi}\times
\frac{\dot\phi^2}{H^2}=\left(\frac{\dot H}{H^2}-\frac{\ddot\phi}{H\dot\phi}
\right)=(-\epsilon+\eta-\epsilon)=\eta-2\epsilon\, .
\label{phiphi}
\ee
This incomplete result makes intuitive sense since the slow-roll parameters characterize deviations of the inflaton from a free field.
To get the the full result  behavior, let us consider Eq. (\ref{zeta3})
restricting ourselves to the one-single field case. Then

\be
B_\zeta(k_1,k_2,k_3)=
N^3_\phi B^{\phi}_{k_1 k_2 k_3}+ N^2_\phi N_{\phi\phi}\left(P^{\phi}(k_1)P^{\phi}(k_2)+2\,\,{\rm permutations} \label{zeta33}
\right)\, .
\ee
At first-order we have $\delta\phi^{(1)}_{\vk}\simeq (H/2\pi)$. However at second-order there is a local correction to the amplitude of vacuum fluctuations at Hubble exit due to
first-order perturbations in the local Hubble rate $\widetilde{H}(\phi)$. This is determined by the local scalar field value due to longer
wavelength modes that have already left the horizon 

\be
\widetilde{H}(\phi)=H(\phi)+H^\prime (\phi)\int_0^{k_{\rm c}}\frac{\d^3 k}{(2\pi)^3}\delta\phi_{\vk}\, ,
\ee
where $k_{\rm c}$ is the cut-off wavenumber which selects only long wavelength perturbation at horizon crossing. Thus for a mode
$k_1\simeq k_2\gg k_3$ one can write at second-order

\be
\delta\phi^{(2)}_{\vk_1}\simeq\frac{H^\prime}{H} \int_0^{k_{\rm c}}\frac{\d^3 k^\prime}{(2\pi)^3}\delta\phi^{(1)}_{\vk^\prime}\delta\phi^{(1)}_{\vk_1-\vk^\prime}\, ,
\ee
where $k_1\simeq k_2\gg k_{\rm c}$. The bispectrum for the inflation field therefore reads in the squeezed limit

\begin{eqnarray}
B^{\phi}_{k_1 k_2 k_3}&\simeq& \langle \delta\phi^{(2)}_{\vk_1}\delta\phi^{(1)}_{\vk_2}\delta\phi^{(1)}_{\vk_3}\rangle+
\langle \delta\phi^{(1)}_{\vk_1}\delta\phi^{(2)}_{\vk_2}\delta\phi^{(1)}_{\vk_3}\rangle\simeq (2\pi)^3\delta^{(3)}(\vk_1+\vk_2+\vk_3)2
\frac{H^\prime}{H} P^\phi(k_3)P^\phi(k_1)\nonumber\\
&\simeq& -2\epsilon
\left(\frac{H}{\dot\phi}\right)
(2\pi)^3\delta^{(3)}(\vk_1+\vk_2+\vk_3) P^\phi(k_3)P^\phi(k_1)\nonumber\\
&=&-2\epsilon N_\phi
(2\pi)^3\delta^{(3)}(\vk_1+\vk_2+\vk_3) P^\phi(k_3)P^\phi(k_1)\, .
\end{eqnarray}
Using Eq. (\ref{phiphi}) we then get

\begin{eqnarray}
B_\zeta(k_1,k_2,k_3)&=&(2\pi)^3\delta^{(3)}(\vk_1+\vk_2+\vk_3)\left[-2\epsilon
N^4_\phi P^\phi(k_3)P^\phi(k_1) +2(\eta-2\epsilon)P^\zeta(k_3)P^\zeta(k_1)\right]\nonumber\\
&=&(2\eta-6\epsilon)P^\zeta(k_3)P^\zeta(k_1)\nonumber\\
&=&(n_\zeta-1)P^\zeta(k_3)P^\zeta(k_1)
\end{eqnarray}
and we have obtained a $(n_\zeta-1)$ suppression.
In fact this result is more general is valid for any type of single-field models (for instance even for those without a canonical kinetic term) as long as the inflaton is the only degree of freedom present
\be
\label{equ:CZ}
\fbox{$\displaystyle
\lim_{k_3 \to 0} \langle \zeta_{{\bf k}_1} \zeta_{{\bf k}_2} \zeta_{{\bf k}_3} \rangle = (2\pi)^3 \delta^{(3)}({\bf k}_1 + {\bf k}_2 + {\bf k}_3) \, (n_\zeta-1) \, P^\zeta(k_1) P\zeta(k_3) $} \, .
\ee
Eq.~(\ref{equ:CZ}) states that the squeezed limit of the three-point function  for single-field inflation,as well as the corresponding $f^{\rm loc}_{\rm NL}$,  is always suppressed by ($1-n_\zeta$). 
A detection of a large enough non-Gaussianity in the squeezed limit can therefore rule out single-field inflation altogether.
let us see how to get the result (\ref{equ:CZ}) in general. Let us take a mode with long wavelength $k_L$. 
The corresponding curvature perturbation $\zeta_{{\bf k}_{\rm L}}$ rescales the spatial coordinates (or changes the effective scale factor) within a given Hubble patch
\be
\d s^2 =  - \d t^2 + a^2(t)e^{-2\zeta} \d {\bf x}^2\, .
\ee
The two-point function $\langle \zeta_{{\bf k}_1} \zeta_{{\bf k}_2} \rangle$, with $k_1\simeq k_2\gg k_3=k_L$,  will depend on the value of the background fluctuations $\zeta_{{\bf k}_{\rm L}}$ already frozen outside the horizon. In position space the variation of the two-point function given by the long-wavelength fluctuations $\zeta_{\rm L}$ is at linear order
\be
\label{equ:4}
\frac{\partial}{\partial \zeta_{\rm L}} \langle \zeta(x) \zeta(0) \rangle \cdot \zeta_{\rm L} = -x \frac{\d}{\d x} \langle \zeta(x) \zeta(0) \rangle \cdot \zeta_{\rm L} \, .
\ee 
To get the three-point function one multiplies Eq.~(\ref{equ:4}) by $\zeta_{\rm L}$ and average over it.
Going to Fourier space gives Eq.~(\ref{equ:CZ}). 

\subsection{Non-Gaussianity in multiple field models}
If  we depart from single-field inflation models, where the interactions are limited by the requirement that the flatness of the potential  is not spoiled, one can construct models where non-Gaussianity can be sizable. In  models like the
 curvaton mechanism  or inhomogeneous reheating  non-Gaussian fluctuations are generated thanks to an extra field other than the inflaton.
Let us describe in some detail  the curvaton case.
We expand the curvaton field  up to first-order 
in the 
perturbations around the homogeneous background as $\sigma(\tau,{\bf x})
=\sigma_0(\tau)+\delta\sigma$, the 
linear perturbations satisfy on large scales
\be
\label{sigma2}
\delta \sigma ''+2 \H \delta \sigma' +a^2 
\frac{\partial^2 V}{\partial \sigma^2}\, \delta \sigma=0\, .
\ee
As a result on super-Hubble scales its fluctuations 
$\delta \sigma$ will be Gaussian distributed and with a nearly 
scale-invariant spectrum given by 
\be
\calp_{\delta\sigma}^\frac12(k) \approx \frac{H_*}{2\pi}
\label{pinf2}
\,, 
\ee
where the subscript $*$ denotes the epoch of horizon exit $k=aH$.     
Once inflation is over the inflaton energy density will be converted to 
radiation ($\gamma$) and the curvaton field will remain approximately 
constant until $H^2 \sim m_\sigma^2$. At this epoch the curvaton field begins 
to oscillate around the minimum of its potential which can be safely 
approximated to be quadratic $V \approx \frac{1}{2} m_\sigma^2 \sigma^2$.
During this stage the energy density of the curvaton field just scales as 
non-relativistic matter $\rho_\sigma \propto a^{-3}$. 
The energy density in the oscillating field is
\be
\label{energyoscill}
\rho_\sigma(\tau,{\bf x}) \approx
m_\sigma^2 \sigma^2(\tau,{\bf x})\, ,
\ee
and it can be expanded into a homogeneous background 
$\rho_\sigma(\tau)$ and a second-order perturbation $\delta
\rho_\sigma$ as 
\be
\label{rhocurv2}
\rho_\sigma(\tau,{\bf x})=\rho_\sigma(\tau)+\delta \rho_\sigma(\tau,{\bf x})=
m_\sigma^2 \sigma+2 m_\sigma^2\, \sigma \, \delta \sigma+m_\sigma^2 \delta\sigma^2\, . 
\ee
The ratio $\delta\sigma/\sigma$ remains 
constant and     
the resulting relative energy density perturbation is
\be
\label{relrhocurv2}
\frac{\delta\rho_\sigma}{\rho_\sigma}=2 \left(
\frac{\delta \sigma}{\sigma} \right)_* +\left(
\frac{\delta \sigma}{\sigma} \right)^2_* ,
\ee
where the $*$ stands for the value at  horizon crossing.
Such perturbations in the energy density of the 
curvaton field produce in fact a primordial density 
perturbation well after the end of inflation and a potentially large NG.     

During the oscillations of the curvaton field,  
the total curvature perturbation  can be written as a weighted sum of the single 
curvature perturbations 
\be
\label{zetasum2}
\zeta=(1-f)\zeta_\gamma+f\zeta_\sigma\, ,
\ee
where the quantity 
\begin{equation}
\label{deff2}
 f = \frac{3\rho_\sigma}{4\rho_\gamma +3\rho_\sigma} 
\end{equation}
defines the relative contribution of the curvaton field 
to the total curvature 
perturbation. Working  under the approximation of  
sudden decay of the curvaton field. Under this approximation the curvaton and 
the radiation components $\rho_\sigma$ and $\rho_\gamma$ satisfy 
separately the energy conservation equations 
\bea
\label{conseqs2}
\rho_\gamma'=-4 \H \rho_\gamma\, ,\nonumber \\
\rho_\sigma'=-3 \H \rho_\sigma\, ,
\eea
and the curvature 
perturbations $\zeta_i$ remains constant on super-Hubble scales until 
the decay of the curvaton. In the curvaton scenario 
it is supposed that the curvature perturbation 
in the radiation produced at the end of inflation is negligible. 
From Eq.~(\ref{zetasum2}) the total curvature perturbation 
during the curvaton oscillations is given by  
\be
\label{zetaoscill2}
\zeta=f \zeta_\sigma\simeq \frac{f}{3}\frac{\delta\rho_\sigma}{\rho_\sigma}\simeq \frac{f}{3}\left[
2 \left(
\frac{\delta \sigma}{\sigma} \right)_* +\left(
\frac{\delta \sigma}{\sigma} \right)^2_*
\right]\, ,
\ee
from which we deduce that 

\be
\zeta=\zeta_{\rm g}+\frac{3}{4 f}(\zeta^2_{\rm g}-\langle \zeta^2_{\rm g}\rangle)\, ,\,\,\, \zeta_{\rm g}=(2 f/3)(\delta\sigma/\sigma)_*\, ,
\ee
and therefore

\be 
\label{yaw}
f_{\rm NL}^{\rm loc}=\frac{5}{4 f}\, .
\ee
We discover that the NG can be very large if $f\ll 1$. Furthermore, the NG is of the local type. This is because it is generated
not at horizon-crossing, but when the fluctuations are already outside the horizon. 

It is nice to reproduce the same result with the $\delta N$ formalism \cite{gen}. 
In the absence of interactions, fluids characterized by well-defined equation of state  (for radiation $P_\gamma=\rho_\gamma/3$) or for the
non-relativistic curvaton ($P_\sigma=0$), there are  the conserved curvature
perturbations (notice again a change of an irrelevant sign from Eq. (\ref{deltaNNNN}))
\begin{equation}
 \label{zetai}
\zeta_i
 = \delta N +
 \frac{1}{3} \int_{\bar\rho_i}^{\rho_i}
 \frac{\d\tilde\rho_i}{\tilde\rho_i+P_i(\tilde\rho_i)} \,. 
\end{equation}
If the  curvaton decays on a uniform-total density
hypersurface corresponding to $H=\Gamma$, {\it i.e.}, when the local
Hubble rate equals the decay rate for the curvaton (assumed
constant), then on this s hypersurface we have
\begin{equation}
 \label{barrho}
\rho_\gamma(t_{\rm dec},{\bf x}) + \rho_\sigma(t_{\rm dec},{\bf x}) =
\bar\rho(t_{\rm dec}) \, .
\end{equation}
The quantity  $\zeta$ is conserved after
the curvaton decay since the total pressure is simply $P=\rho/3$.

However, the  local curvaton and radiation densities on the decay
surface are perturbed
\begin{eqnarray}
\zeta_\gamma &=& \zeta + \frac{1}{4} \ln \left( \frac{\rho_\gamma}{\bar\rho_\gamma}
\right) \,,\\
\zeta_\sigma &=& \zeta + \frac{1}{3} \ln \left( \frac{\rho_\sigma}{\bar\rho_\sigma}
\right) \,,
\end{eqnarray}
or, equivalently,
\begin{eqnarray}
\rho_\gamma &=& \bar\rho_\gamma e^{4(\zeta_\gamma-\zeta)} \,,\\
\rho_\sigma &=& \bar\rho_\sigma e^{3(\zeta_\sigma-\zeta)} \,.
\end{eqnarray}
Imposing that total density is not perturbed on the decay surface,
we obtain  the relation
\begin{equation}
 \label{nlzeta}
(1-\Omsd) e^{4(\zeta_\gamma-\zeta)} + \Omsd e^{3(\zeta_\sigma-\zeta)} = 1 \,,
\end{equation}
where $\Omsd=\bar\rho_\sigma/(\bar\rho_\gamma+\bar\rho_\sigma)$ is the
dimensionless density parameter for the curvaton at the decay
time. 

Let us consider the simplest possible scenario, where 
the  curvature perturbation in
the radiation fluid before the curvaton decays is negligible,
{\it i.e.}, $\zeta_\gamma=0$.  Eq.~(\ref{nlzeta})
reads
\begin{equation}
 \label{nlzetasimpl}
e^{4\zeta} - \left[ \Omsd e^{3\zeta_\sigma} \right] e^{\zeta} + \left[ \Omsd -1 \right] = 0 \,.
\end{equation}
At first-order  Eq.~(\ref{nlzeta}) gives
\begin{equation}
4(1-\Omsd) \zeta^{(1)} = 3 \Omsd (\zeta_\sigma^{(1)}-\zeta^{(1)}) \,,
\end{equation}
and hence we can write
\begin{equation}
 \label{defr}
\zeta^{(1)} = f \zeta_\sigma^{(1)}
 \,,
\end{equation}
where 
\begin{equation}
 \label{defrSD}
f = \frac{3\Omsd}{4-\Omsd}
= \left.\frac{3\bar\rho_\sigma}{3\bar\rho_\sigma + 4\bar\rho_\gamma}\right|_{t_{\rm dec}} \,.
\end{equation}
At second order Eq.~(\ref{nlzeta}) gives
\begin{equation}
4(1-\Omsd) \zeta^{(2)} - 16(1-\Omsd) \zeta^{(1)2}
 = 3 \Omsd ( \zeta_\sigma^{(2)} -\zeta^{(2)} ) + 9 \Omsd
 (\zeta_\sigma^{(1)}-\zeta^{(1)})^2
 \,,
\end{equation}
and hence
\begin{equation}
\zeta^{(2)} = \frac{3}{4f} \zeta^{(1)2} \,,
\end{equation}
which gives again Eq. (\ref{yaw}).

\section{Conclusions}

Along these lectures, we have learned that a stage
of inflation during the early epochs of the evolution of the
universe solves many drawbacks of the standard Big-Bang cosmology, such
as the flatness or entropy problem and the horizon problem. Luckily,
despite inflation occurs after a tiny bit after the bang, 
it leaves behind some observable predictions:

\begin{itemize}

\item {\it The universe should be flat}, that is the total density
of all components of matter should sum to the critical energy
density and $\Omega_0=1$. The current data on the CMB anisotropies
offer a spectacular confirmation of such a prediction. The universe
appears indeed to be spatially flat.

\item {\it Primordial perturbations are adiabatic}. Inflation provides the 
seeds for the cosmological perturbations. In one-single field
models of inflation, the perturbations  are {\it adiabatic} or curvature
perturbations, {\it i.e.} they are fluctuations in the total energy density
of the universe or, equivalently, scalar perturbations to the
spacetime metric. Adiabaticity implies that the spatial
distribution of each species in the universe is the same, that is the ratio
of number densities of any two of these species is everywhere
the same. Adiabatic perturbations  in excellent agreement
with the CMB data \cite{Amendola:2001ni}.

\item {\it Primordial perturbations are almost scale-independent}.
The primordial power spectrum predicted by inflation has a characteristic
feature, it is almost scale-independent, that is the spectral index
$n_\zeta$ is very close to unity. Possible deviations from exact
scale-independence arise because during inflation the inflaton is not
massless and the Hubble rate is not exactly constant.

\item {\it Primordial perturbations are nearly gaussian}. The fact that
cosmological perturbations are tiny allow their analysis in terms of 
linear perturbation theory. 
Non-gaussian features are therefore suppressed since the non-linearities 
of the inflaton potential and of the metric perturbations are suppressed.
Non-gaussian features are indeed present, but 
may appear only at the 
second-order in deviations from the homogeneous background solution and 
are therefore small. This is rigorously true only for 
one-single field models of inflation. Many-field models of inflation
may give rise to some level of NG  \cite{b4}.
If the next generation of observations we  will detect a non-negligible
amount of primordial NG,  this will rule
out one-single field models of inflation.

\item {\it Production of gravitational waves}. A stochastic
background of gravitational waves is produced during inflation
 in the very same way classical perturbations to the inflaton
field are generated. The spectrum of such gravitational waves is
again flat, {\it i.e.} scale-independent and the tensor-to-scalar
amplitude ratio $r$ is, 
at least in one-single
models of inflation, related to the  spectral index $n_T$  by the 
consistency relation $r=-8n_T$. A confirmation of such a relation
would be a spectacular proof  of one-single field models
of inflation. The detection of the primordial stochastic
background of gravitational waves from inflation is challenging, but
would not only set the energy scale of inflation, but would also give the
opportunity of discriminating among different models of inflation \cite{kmr,bacci}.

\end{itemize}

\section*{Acknowledgments}

The author would like to thank the organizers of the School,
G. Dvali, A. Perez-Lorenzana, G. Senjanovic, G. Thompson and F. Vissani
for providing
the students and the lecturers with such an excellent and stimulating
environment. He also thanks the students for their questions
and enthusiasm.

 \newpage
 

\newpage
 
\end{document}